\documentclass{aa} 

\bibpunct{(}{)}{;}{a}{}{,} 

\usepackage{graphicx}
\usepackage[varg]{txfonts}
\begin{document}

\title{Discovery and characterisation of long-period eclipsing binary stars
from Kepler K2 campaigns 1, 2 and 3\thanks{Based on observations made with the
Southern African Large Telescope (SALT)},\thanks{Based on observations
collected at the European Organisation for Astronomical Research in the
Southern Hemisphere under ESO programmes 073.C-0337(A), 089.D-0097(B),
091.C-0713(A), 091.D-0145(B), 094.A-9029(R), 178.D-0361(B), 178.D-0361(F)}}

\titlerunning{Long period eclipsing binaries from K2}
\author{P.~F.~L.~Maxted\inst{1} 
\and
R.~J.~Hutcheon\inst{1} 
}           
          
\institute{Astrophysics Group,  Keele University, Keele, Staffordshire,
ST5~5BG, UK\\
\email{p.maxted@keele.ac.uk, richard.hutcheon@btinternet.com}
}

\date{Dates to be inserted}

 
  \abstract
{The Kepler K2 mission now makes it possible to find and study a wider variety
of eclipsing binary stars than has been possible to-date, particularly
long-period systems with narrow eclipses.}
{Our aim is to characterise eclipsing binary stars observed by the Kepler K2
mission with orbital periods longer than $P\approx 5.5$ days. }
{ The \texttt{ellc} binary star model has been used to determine the geometry of
eclipsing binary systems in Kepler K2 campaigns 1, 2 and 3. The nature of
the stars in each binary is estimated by comparison to stellar evolution
tracks in the effective temperature -- mean stellar density plane. }
{43 eclipsing binary systems have been identified and 40 of these are
characterised in some detail. The majority of these systems are found to be
late-type dwarf and sub-giant stars with masses in the range 0.6\,--\,1.4
solar masses. We identify two eclipsing binaries containing red giant stars,
including one bright system with total eclipses that is ideal for detailed
follow-up observations. The bright B3V-type star HD~142883 is found to be an
eclipsing binary in a triple star system. We observe a series of frequencies
at large multiples of the orbital frequency in BW~Aqr that we tentatively
identify as tidally induced pulsations in this well-studied eccentric binary
system. We find that the faint eclipsing binary EPIC~201160323 shows rapid
apsidal motion. Rotational modulation signals are observed in 13 eclipsing 
systems, the majority of which are found to rotate non-synchronously with
their orbits. }
{The K2 mission is a rich source of data that can be used to find long period
eclipsing binary stars. These data combined with follow-up observations can be
used to precisely measure the masses and radii of stars for which such
fundamental data are currently lacking, e.g.,  sub-giant stars and
slowly-rotating low-mass stars.}

\maketitle
%

\section{Introduction}

 Apart from the Sun and a few nearby stars, detached (i.e., non-interacting)
eclipsing binaries (DEBS) provide the only means to measure accurate,
model-independent masses and radii for normal stars. Using high-quality
multi-wavelength photometry and high-resolution spectroscopy, masses and radii
for stars in DEBS can be measured to $\pm 0.5$\% or better
\citep[e.g.,][]{2015A&A...578A..25M, 2016A&A...594A..92G}. Spectral
disentangling techniques also make it possible to determine the effective
temperature (T$_{\rm eff}$) and surface composition of both stars in the
binary from the analysis of their spectra \citep{2010ASPC..435..207P}.  As a
result, DEBS provide the most stringent test available for the accuracy of
stellar evolution models for many different types of star
\citep{2010A&ARv..18...67T}.  Empirical relations between mass, density,
T$_{\rm eff}$ and metallicity  based on DEBS can be used to estimate
model-independent masses and radii for low-mass companions in SB1 eclipsing
binaries, e.g., transiting hot-Jupiter systems \citep{2011MNRAS.417.2166S}
or brown dwarf or very low mass stars in eclipsing binaries with solar-type
stars \citep{2013A&A...549A..18T}. DEBS are also useful as distance indicators
because their absolute magnitudes can be accurately estimated from the radii
of the stars combined with a calibration of the stars' surface brightness
against colour or T$_{\rm eff}$ \citep{2017ApJ...837....7G}. DEBS have been
used to investigate the systematic errors in parallax measurements for the
Gaia DR1 data release \citep{2016ApJ...831L...6S}, and to accurately measure
the distance to the Magellanic Clouds \citep{ 2013Natur.495...76P,
2014ApJ...780...59G}.

The Kepler K2 mission is providing very high quality photometry for thousands
of moderately bright stars in selected regions of the sky (``campaign
fields'') near the ecliptic plane \citep{2014PASP..126..398H}. Each campaign
field is observed almost continuously for up to 80 days, making it possible to
discover and characterise eclipsing binaries with orbital periods of weeks
that are very hard to study using light curves obtained from ground-based
instruments. Extracting high quality photometry from the K2 images is
challenging because the spacecraft is being operated using only 2 reaction
wheels. This operating mode has made it possible to extend the mission
lifetime, but does result in the pointing of the spacecraft being less stable
than during the original Kepler mission. Nevertheless, there is now a variety
of algorithms available to correct for the instrumental noise caused by this
pointing drift that make it possible to recover photometric performance better
than 100\,ppm per 6-hours at 12th magnitude, close to the performance of the
original Kepler mission \citep{ 2016AJ....152..100L, 2016MNRAS.459.2408A,
2014PASP..126..948V, 2015A&A...579A..19A, 2016A&A...594A.100B}. These
algorithms are generally optimised for the detection of the periodic shallow
eclipses in the light curves of transiting exoplanets. Eclipsing binary stars
have been found both as a by-product of these searches for transiting
exoplanets and by searches for variable stars of all types in the Kepler K2
data. To-date, the characterisation of these eclipsing binaries has not been
very detailed, being limited to estimates of the period plus, in some cases,
some basic characterisation of the eclipse properties, e.g., depth and width.
  
 At the time of writing, there are approximately 200 DEBS that have masses and
radii measured to a precision of 2\% or better \citep{2015ASPC..496..164S}.
This sample is dominated by short-period systems ($P \loa 10$\,d) in which the
components of the binary system are forced to co-rotate with the orbit. This
makes it difficult to study phenomena such as interior mixing processes that
can have subtle effects on the evolution of normal stars, but which may be
disrupted by rapid rotation, particularly for sub-giant and giant stars. 

 We have conducted our own search of the Kepler K2 data from campaigns 1, 2
and 3 and characterised the stars in these binaries in some detail using
modelling of the Kepler K2 light curve plus existing optical and infrared
photometry. Our study is motivated by the opportunity to study in detail stars
of a type for which little fundamental accurate data are currently available.
We have concentrated on bright stars with well-defined eclipses and long
orbital periods that are ideal for detailed characterisation using
high-resolution spectroscopy, but also discuss some other DEBS of interest
that we have found in our survey. The results are presented here for the
benefit of those who can share the task of characterising these binary systems
and as a useful indicator of the number and properties of long-period
eclipsing binaries that will be found in future large-scale photometric
surveys.

\section{Analysis}

 Note that where we refer to the primary and secondary stars in the following
description (star 1 and 2, respectively) these labels refer to the star
eclipsed during the deeper and shallower eclipses in the K2 light curve,
respectively, irrespective of the stars' effective temperatures, masses,
radii, etc. 

\subsection{Target selection}

 Targets were identified by visual inspection of the detrended light curves
generated by the {\sc k2sff} algorithm \citep{2014PASP..126..948V}. We
downloaded the light curve data from the Mikulski Archive for Space
Telescopes\footnote{\url {http://archive.stsci.edu/missions/hlsp/k2sff}} (MAST)
and used a simple script to plot the data for each system while making a  note
of any stars showing eclipse-like features in the light curve at least 5\%
deep and with orbital periods $P\ga 5.5$ days. We excluded stars from our list
with a strong ellipsoidal effect in the light curve, i.e., a quasi-sinusoidal
variation in flux with two maxima per orbital cycle due to the gravitational
distortion of the stars in a close binary system. We also excluded systems
fainter than Kepler magnitude ${\rm Kp} \approx 13$ unless they seemed
particularly interesting based on an initial appraisal of the light curve or
other information available. These points of interest are noted
in section \ref{notes_sec}. 

 The list of stars selected for further analysis is shown in
Table~\ref{lcinfo} together with some basic characteristics  of the light
curves. The rotation periods $P_{\rm rot}$ listed in this table were determined
as part of the detrending process described in section~\ref{detrend_sec}.

\subsection{Aperture photometry and detrending \label{detrend_sec}}
 We downloaded the target pixel files for each target from MAST and used these
data to produce light curves using synthetic aperture photometry. We first
calculated the median value for every pixel in the data cube. The pixels in
the lowest 10-percentile of this median image were then used to calculate the
background level in the individual images. We used the target aperture
specified in the target pixel file where available, otherwise we used a
circular aperture centered on the flux-weighted centroid of the median image
with a radius selected by-eye to encompass most of the flux in the star --
typically 4\,--\,8 pixels. We also calculated the flux-weighted centroid
within the target aperture for each image. 

\begin{figure*}
  \begin{center}
\includegraphics{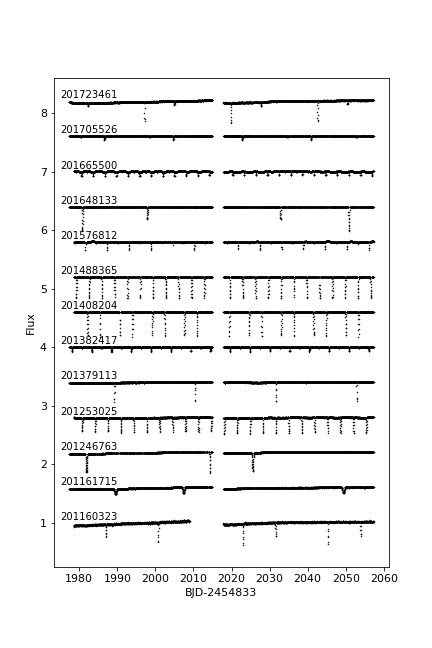} 
  \end{center}
\caption{Light curves of long-period eclipsing binaries from Kepler campaign
1. The flux is measured relative to the median out-of-eclipse level and offset
  by multiple of 0.5 units for clarity. Trends in the data due to variations
  in spacecraft pointing have been removed.}
\label{lcplot1} 
\end{figure*}

\begin{figure*}
  \begin{center}
\includegraphics{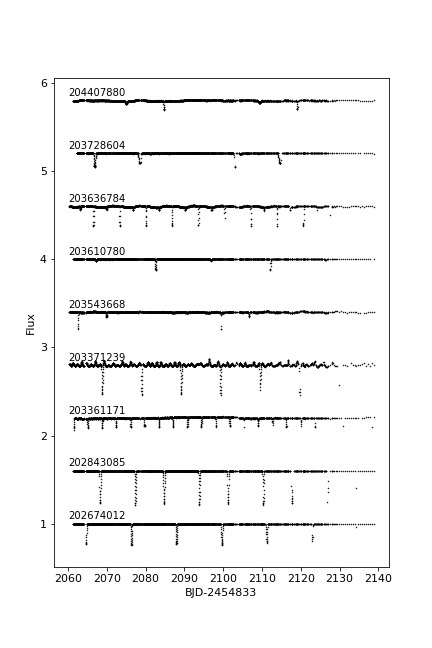} 
  \end{center}
\caption{Light curves of long-period eclipsing binaries from Kepler campaign
2. The flux is measured relative to the median out-of-eclipse level and offset
  by multiple of 0.5 units for clarity. Trends in the data due to variations 
  in spacecraft pointing have been removed.}
\label{lcplot2a} 
\end{figure*}

\begin{figure*}
  \begin{center}
\includegraphics{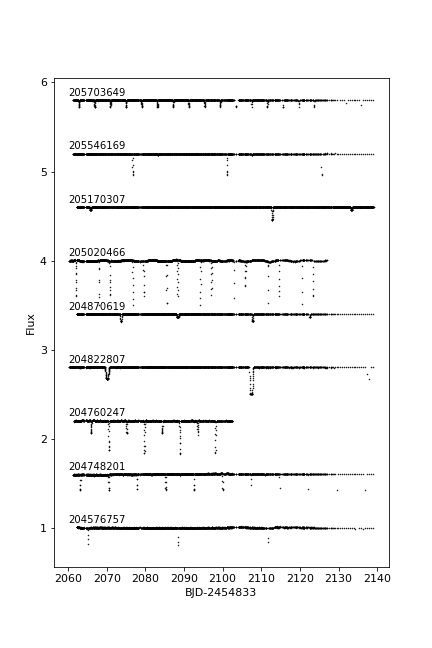} 
  \end{center}
\caption{Light curves of long-period eclipsing binaries from Kepler campaign
2. The flux is measured relative to the median out-of-eclipse level and offset
  by multiple of 0.5 units for clarity. Trends in the data due to variations 
    in spacecraft pointing have been removed.}
\label{lcplot2b} 
\end{figure*}

\begin{figure*}
  \begin{center}
\includegraphics{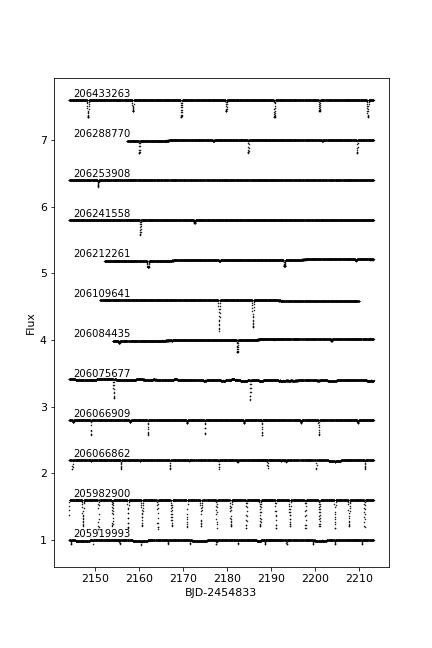} 
  \end{center}
\caption{Light curves of long-period eclipsing binaries from Kepler campaign
3. The flux is measured relative to the median out-of-eclipse level and offset
  by multiple of 0.5 units for clarity. Trends in the data due to variations 
    in spacecraft pointing have been removed.}
\label{lcplot3} 
\end{figure*}

 The light curves produced by this method clearly show instrumental noise due
to the varying position of the star on the detector. We used the {\sc k2sc}
algorithm \citep{2016MNRAS.459.2408A} to remove this instrumental noise. This
algorithm uses Gaussian processes to decompose the light curve into a trend
associated with the position of the star on the detector plus a trend with
time that represents the intrinsic variability of the star. We first detrend
the data using a squared-exponential kernel to describe the covariance
properties of the trend with time. This kernel is suitable for smooth,
aperiodic variations so we mask the eclipses for this calculation. We then
use a Lomb-Scargle periodogram \citep{1992nrfa.book.....P} to characterise
any periodic or quasi-periodic variability in the detrended light curve
between the eclipses. This variability can be due to modulation of the light
curve by star spots on one or both stars, or due to pulsations. The periods
that we judged to be significant detected by this process are noted in
Table~\ref{lcinfo} and are listed in order of power from strongest to weakest.
For the stars whose period is noted in Table~\ref{lcinfo} we repeated the
detrending using  a quasi-periodic kernel for the time trend, again with the
eclipses masked. In both cases (squared-exponential and  quasi-periodic
kernels) the trend with position determined from the data between the eclipses
was used to interpolate a correction to the data during the eclipses. f
These light curves are shown in Figs. \ref{lcplot1}, \ref{lcplot2a},
\ref{lcplot2b} and \ref{lcplot3}.

\subsection{WASP archive photometry}

 The WASP project has obtained over 580 billion photometric observations for
more than 30 million bright stars during a survey that has discovered more
than 150 transiting exoplanets since observations started in May 2004
\citep{2006PASP..118.1407P}. WASP photometry is available for many of the
systems in Table~\ref{lcinfo}, but is of much lower quality than the K2
photometry. Nevertheless, WASP photometry has enabled us to determine or
refine the orbital period for long-period binaries where only two or three
eclipses have been observed by the Kepler K2 mission.

\begin{figure*}
\includegraphics[width=0.49\textwidth]{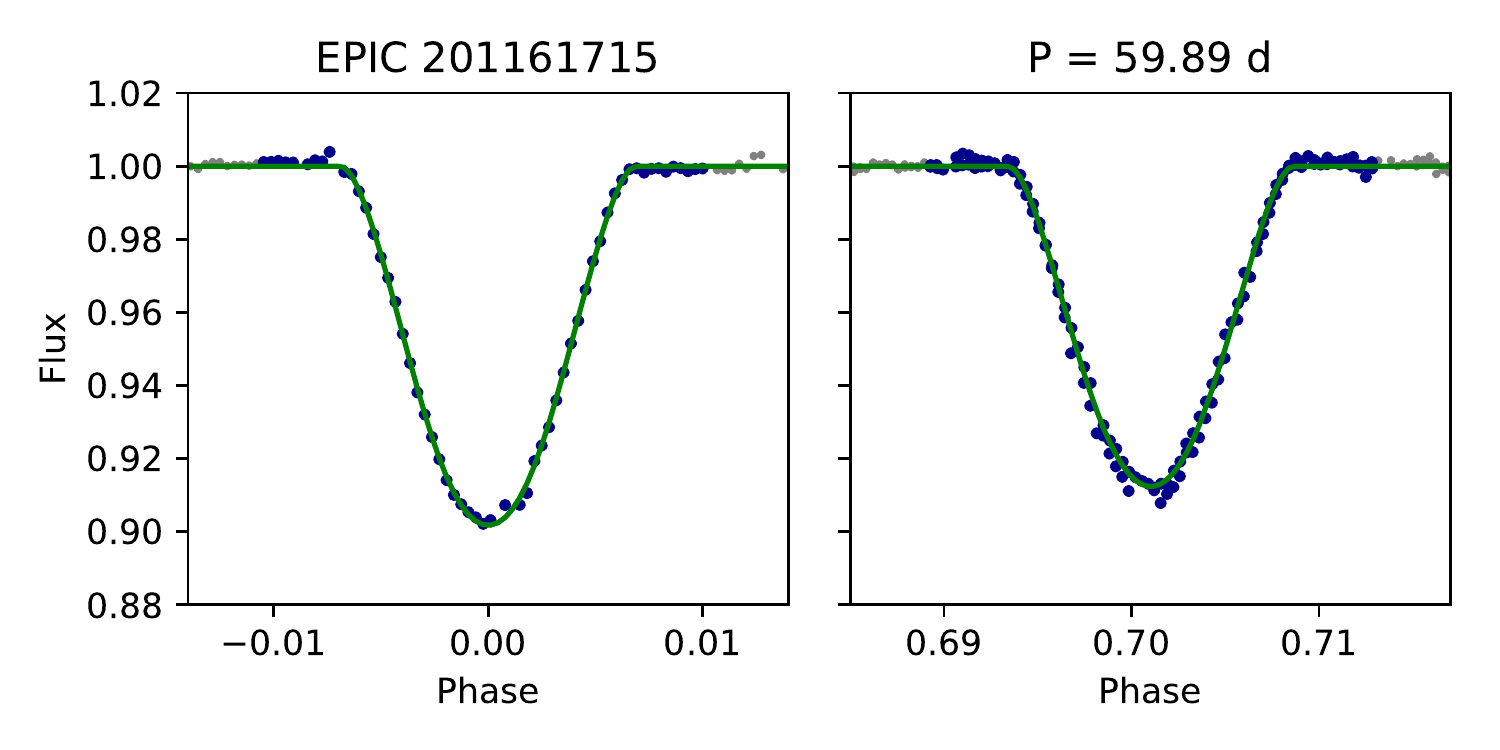} 
\includegraphics[width=0.49\textwidth]{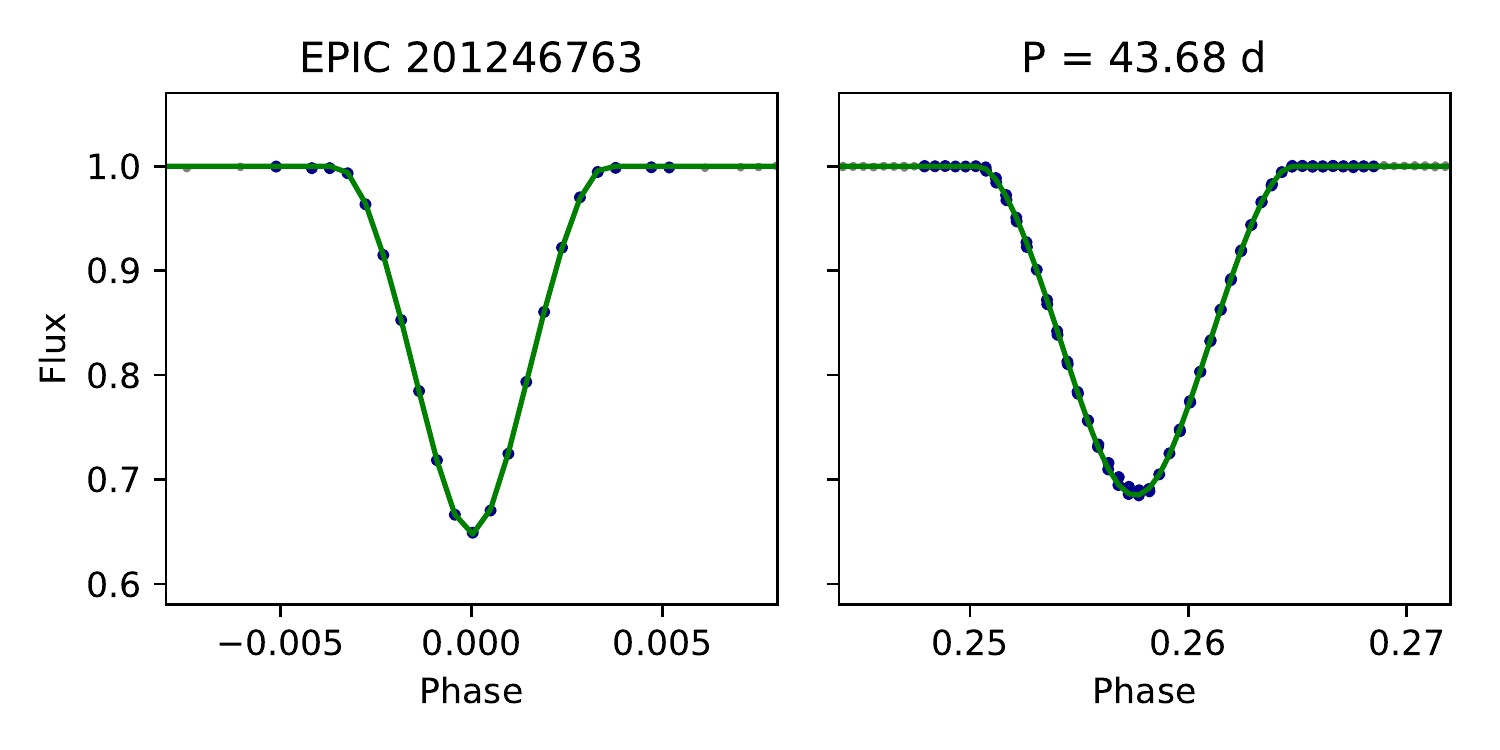} 
\includegraphics[width=0.49\textwidth]{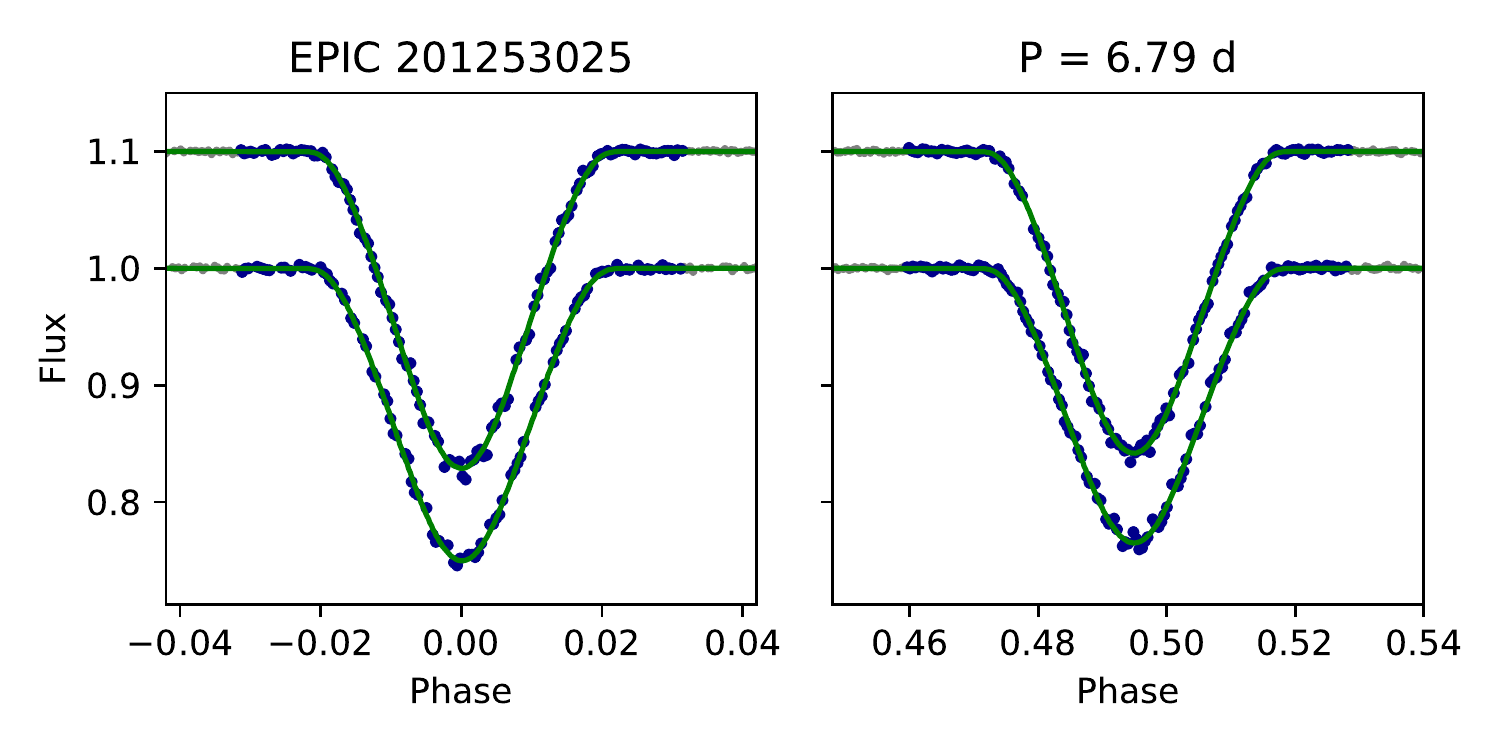} 
\includegraphics[width=0.49\textwidth]{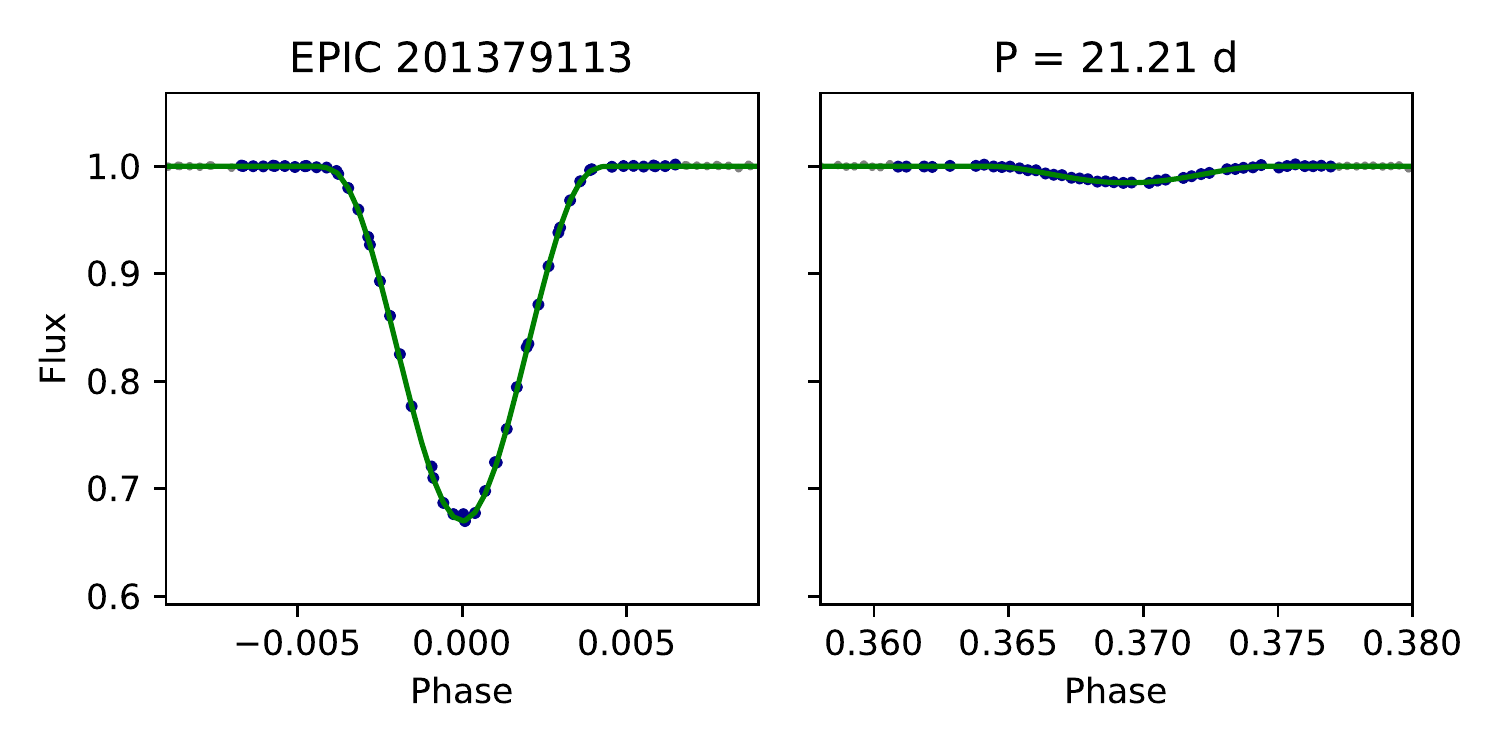} 
\includegraphics[width=0.49\textwidth]{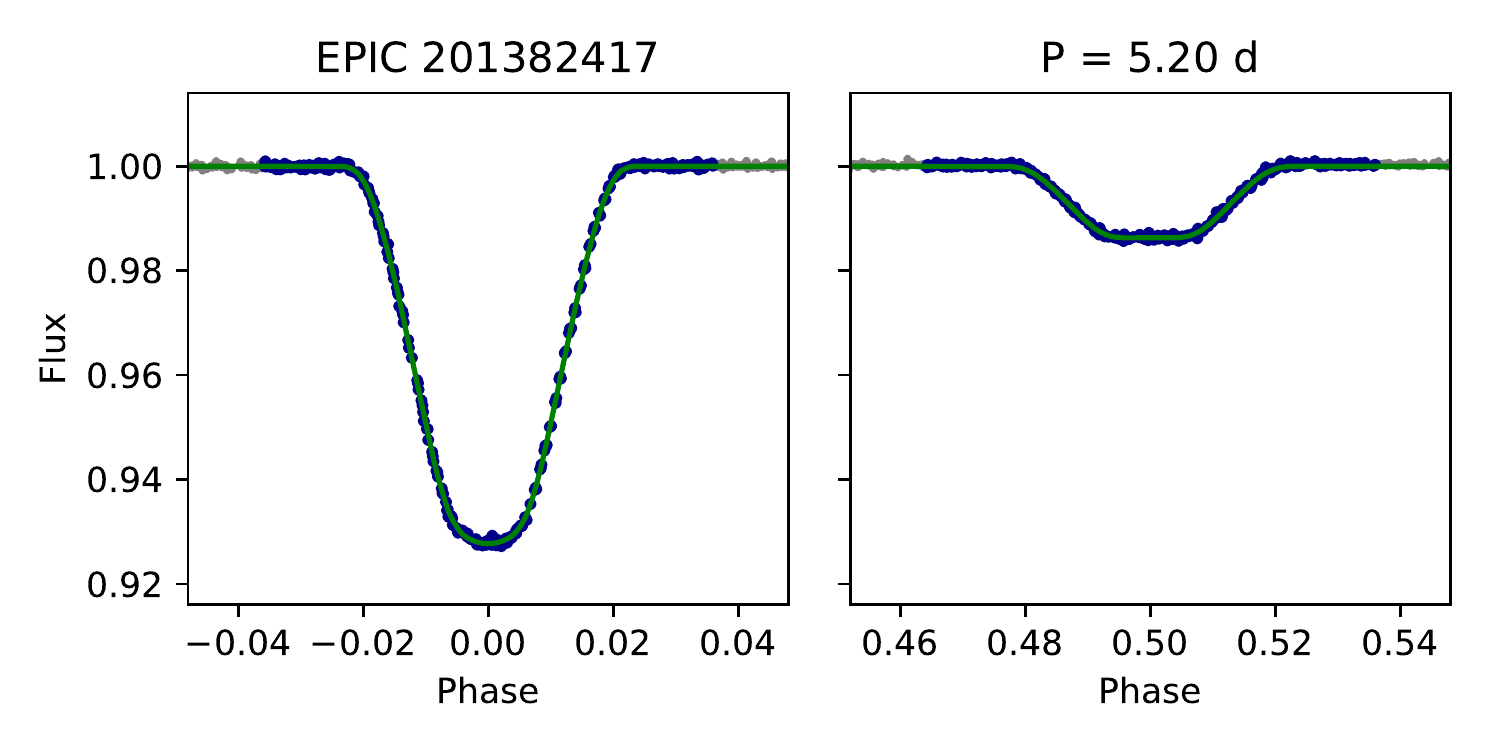} 
\includegraphics[width=0.49\textwidth]{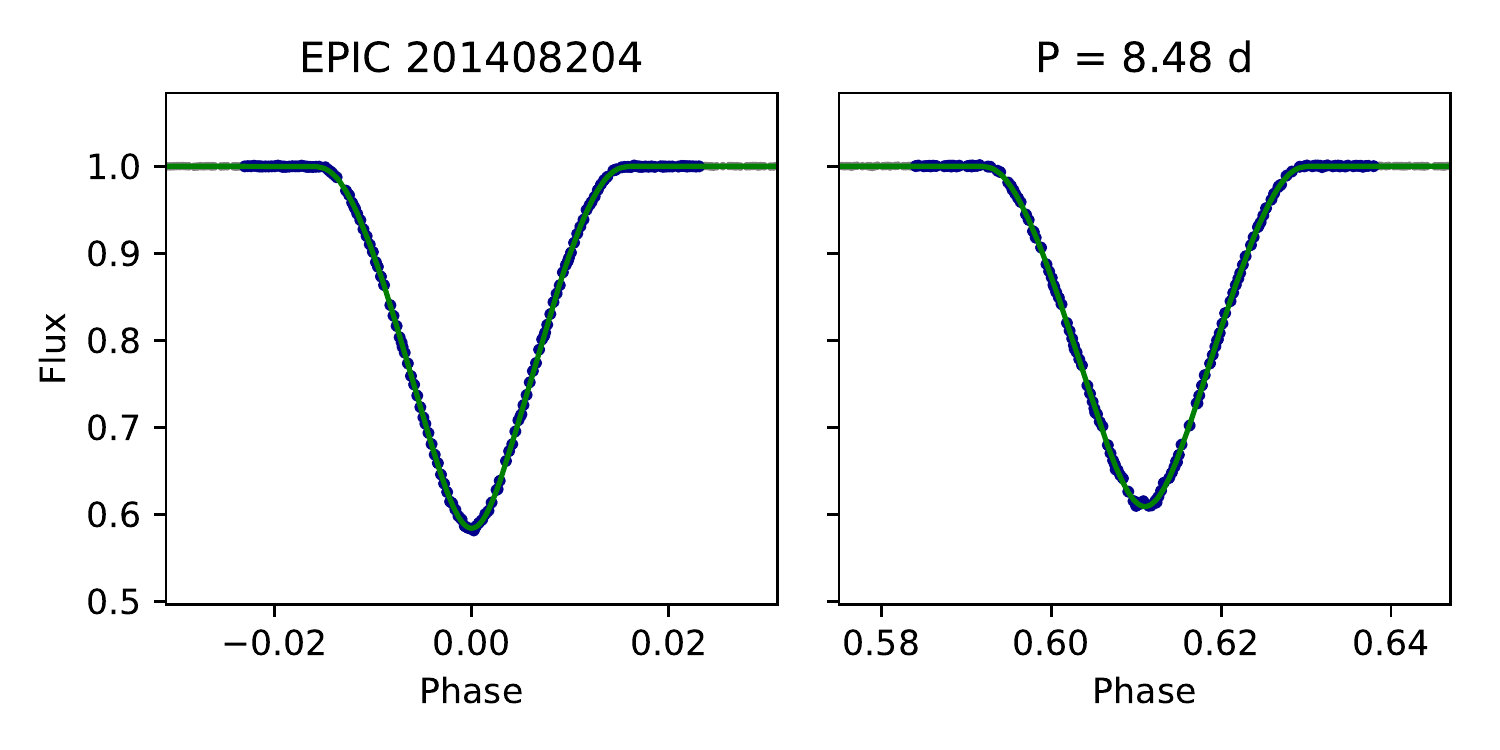} 
\includegraphics[width=0.49\textwidth]{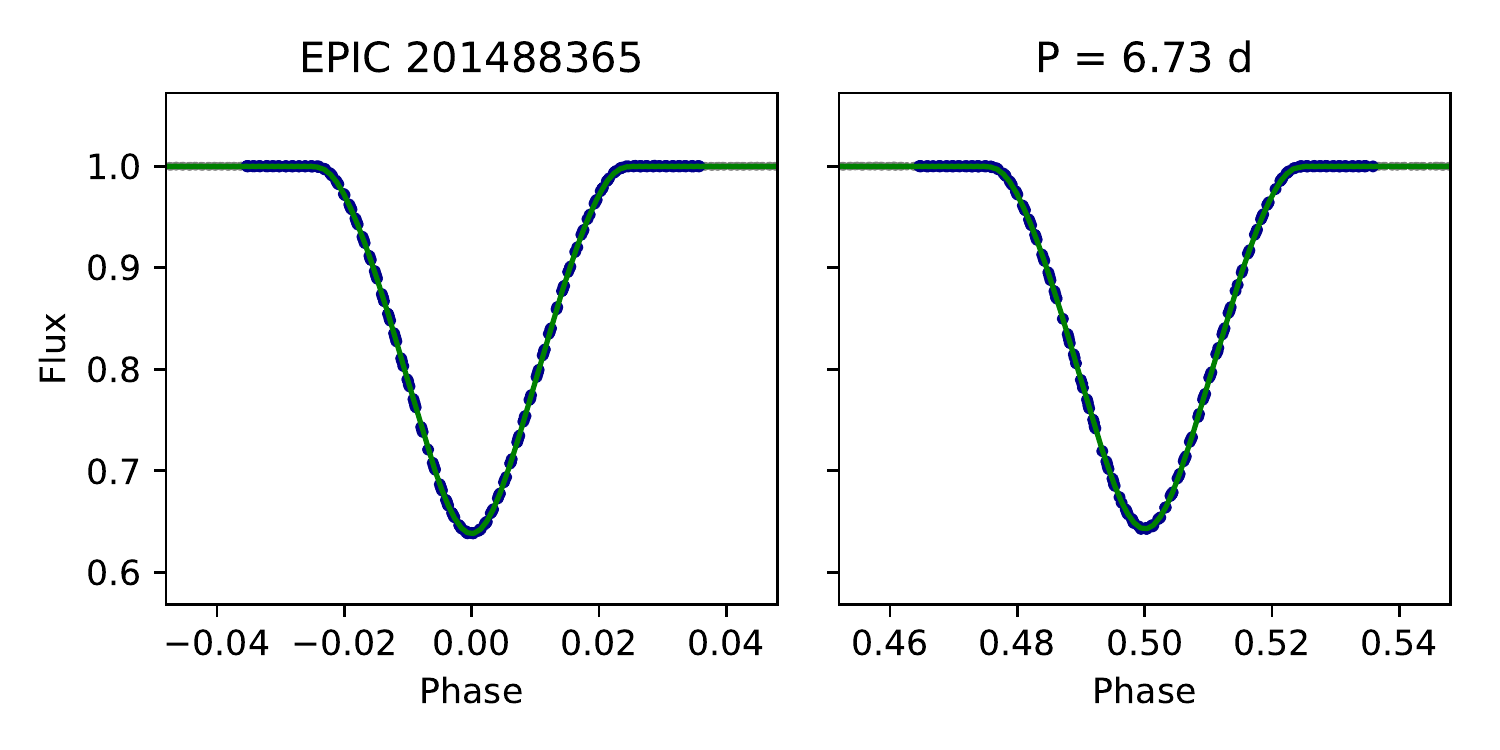} 
\includegraphics[width=0.49\textwidth]{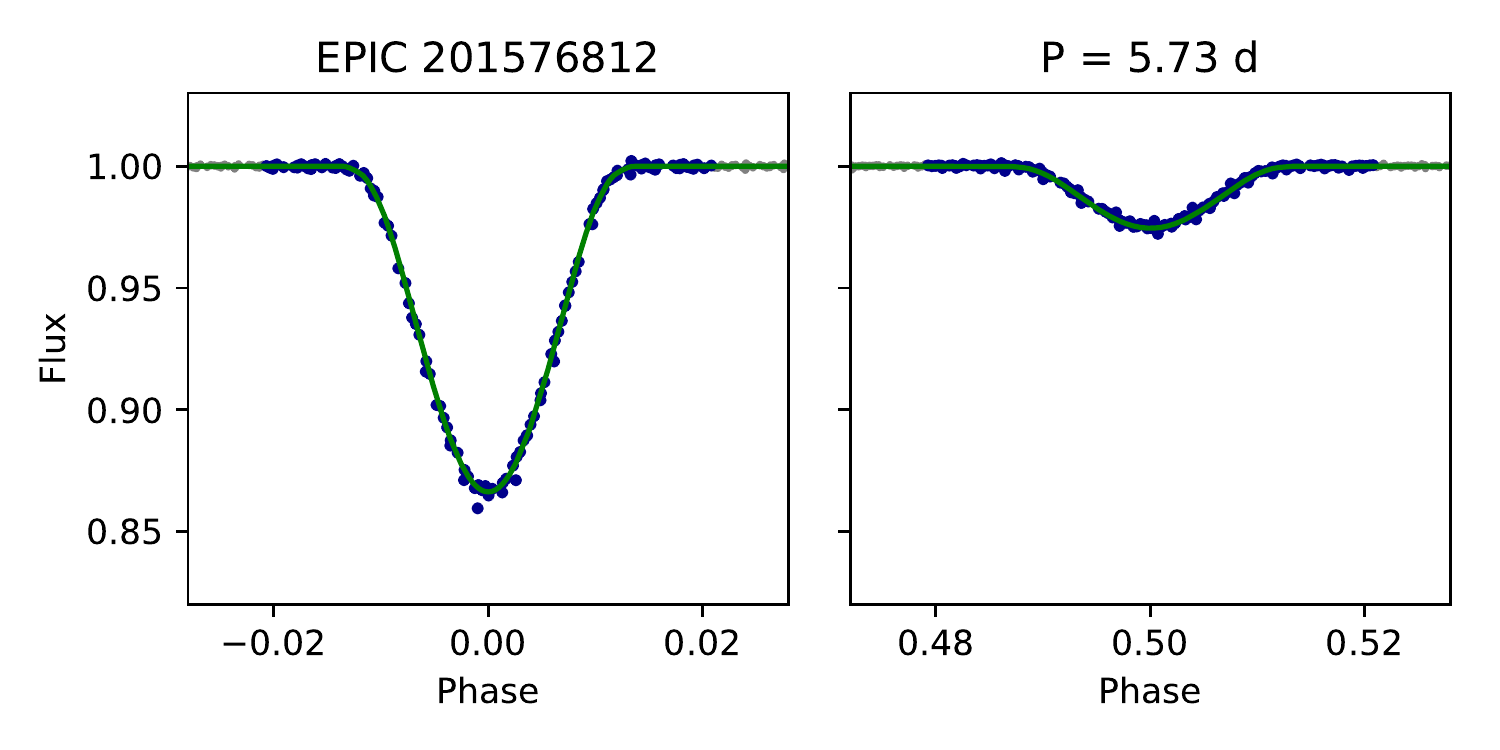} 
\includegraphics[width=0.49\textwidth]{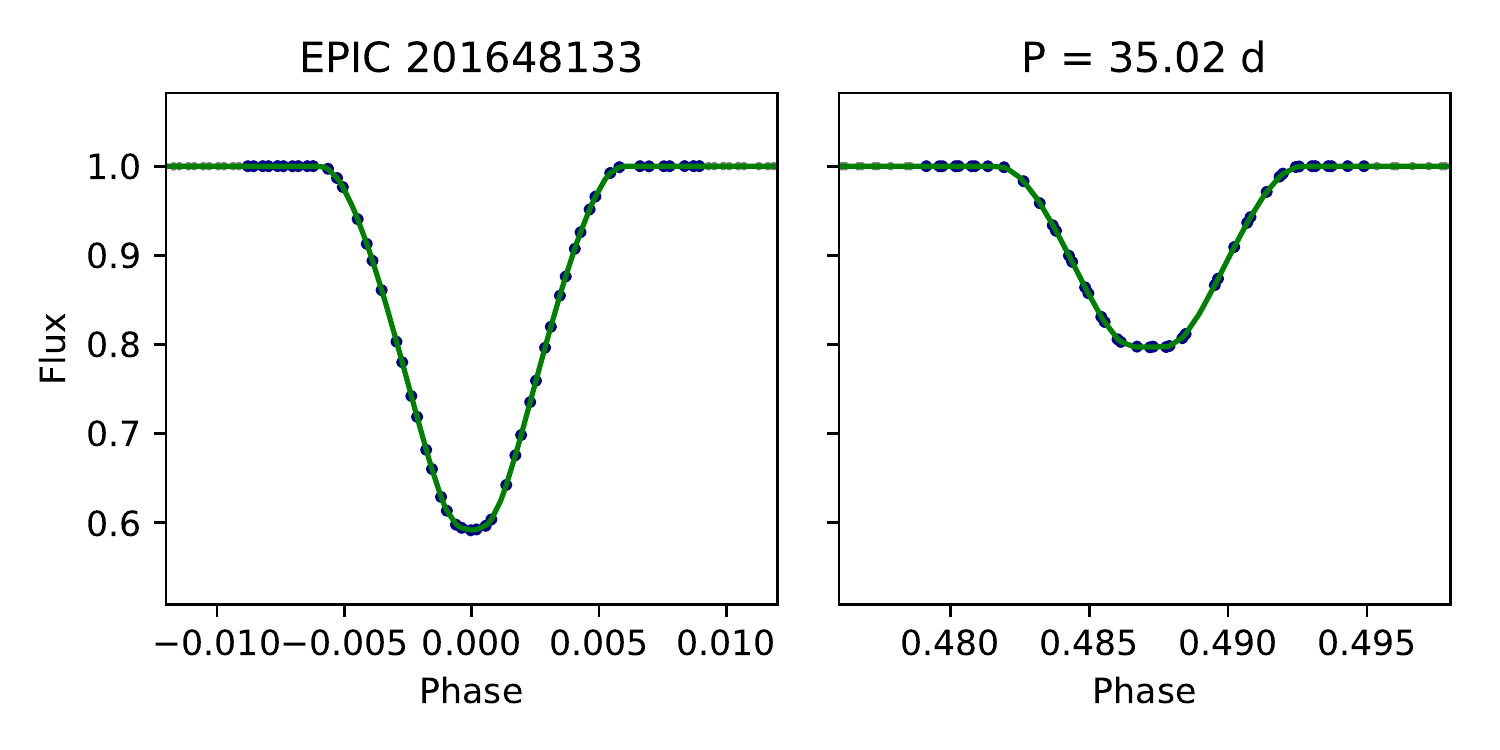} 
\includegraphics[width=0.49\textwidth]{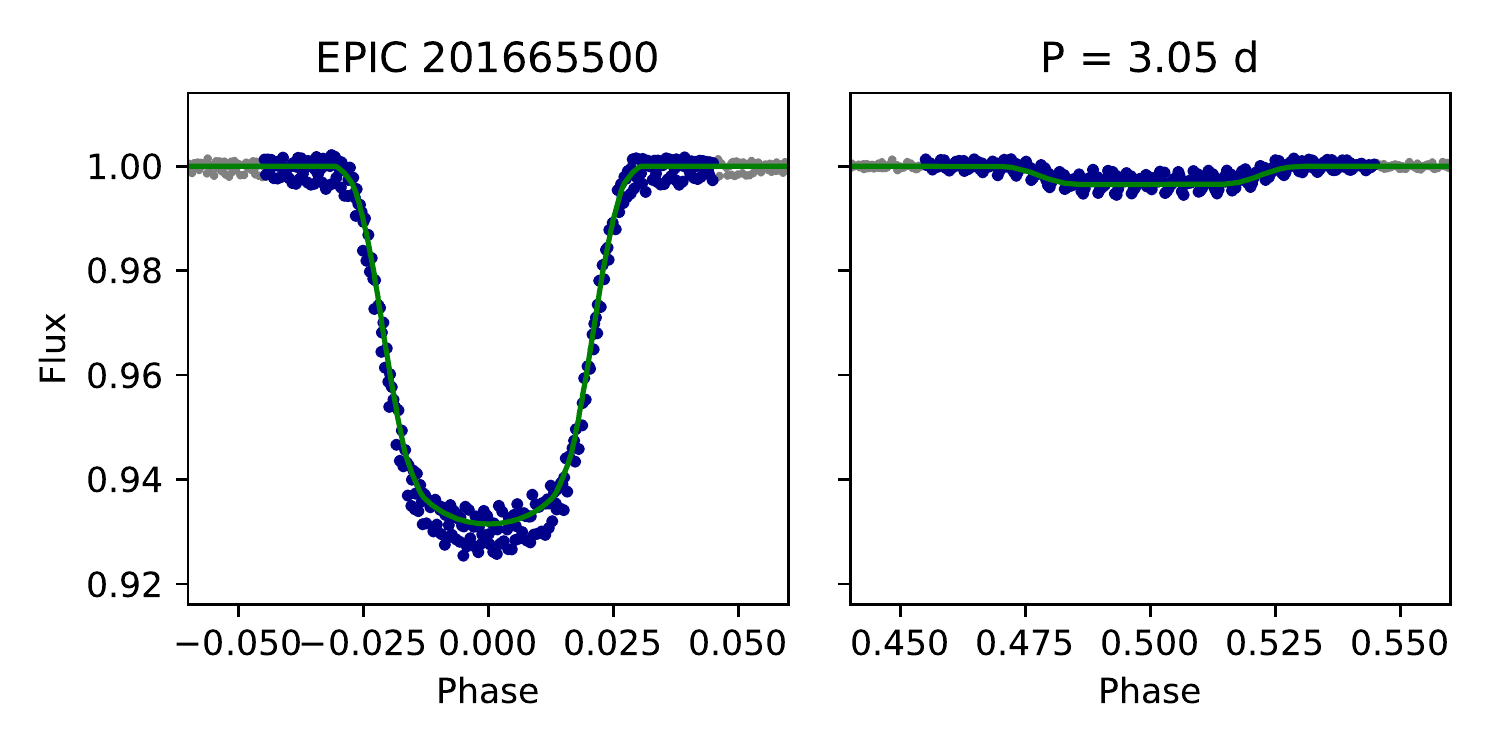} 
  \caption{K2 light curves with the best-fit \texttt{ellc} model. Data not
  included in the fit are plotted using small grey points. Data obtained after
  BJD 2456849 for 201253025  are offset vertically by 0.1 flux units. 
  \label{lcfit1}}
\end{figure*}

\begin{figure*}
\includegraphics[width=0.49\textwidth]{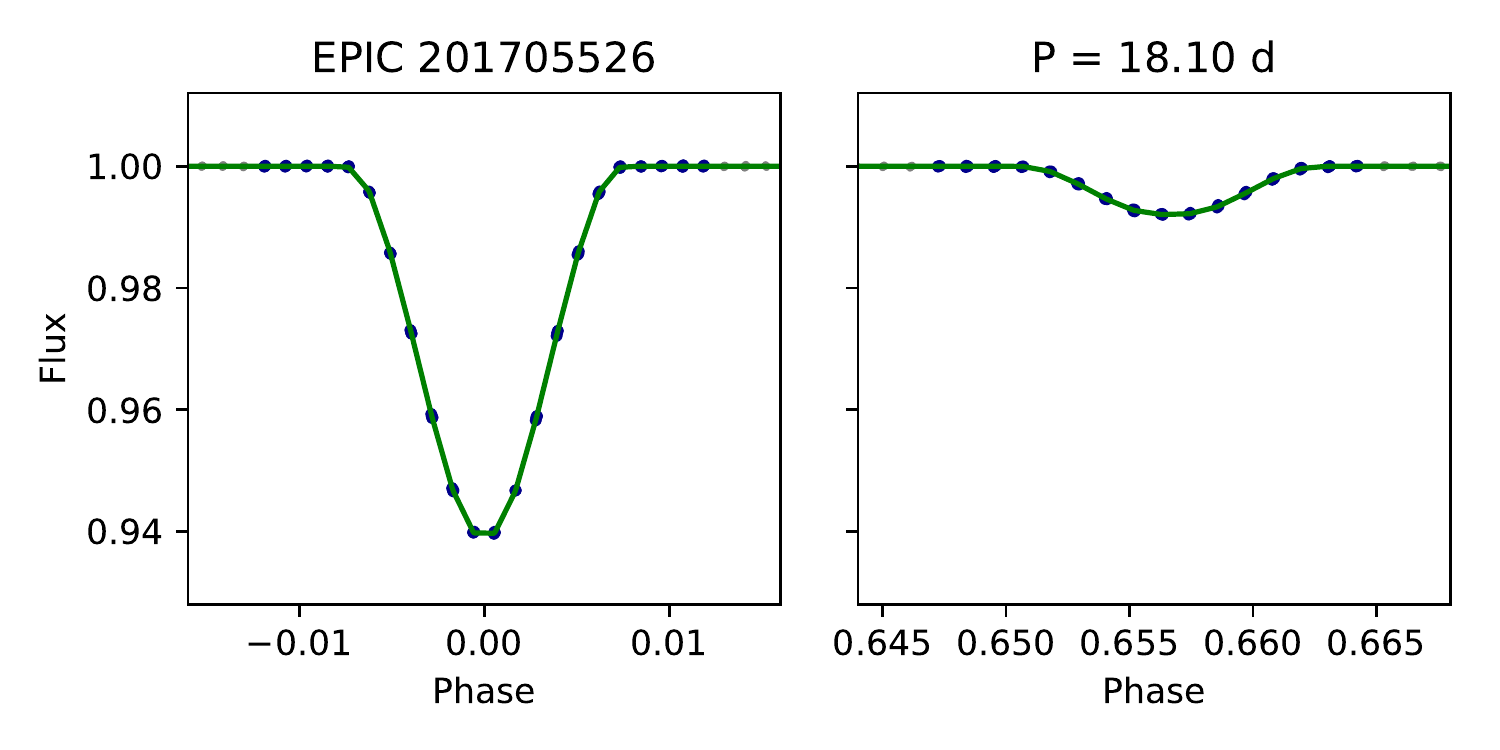} 
\includegraphics[width=0.49\textwidth]{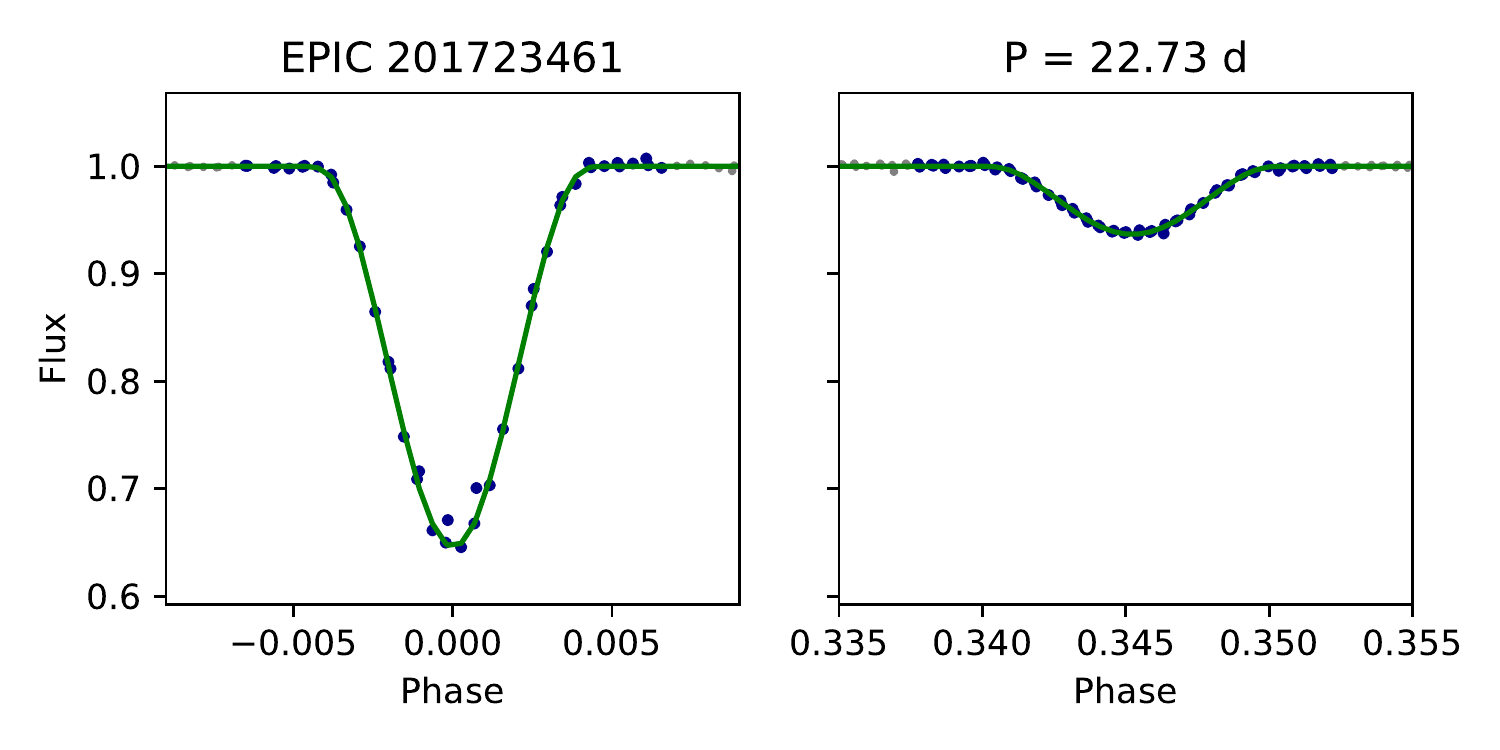} 
\includegraphics[width=0.49\textwidth]{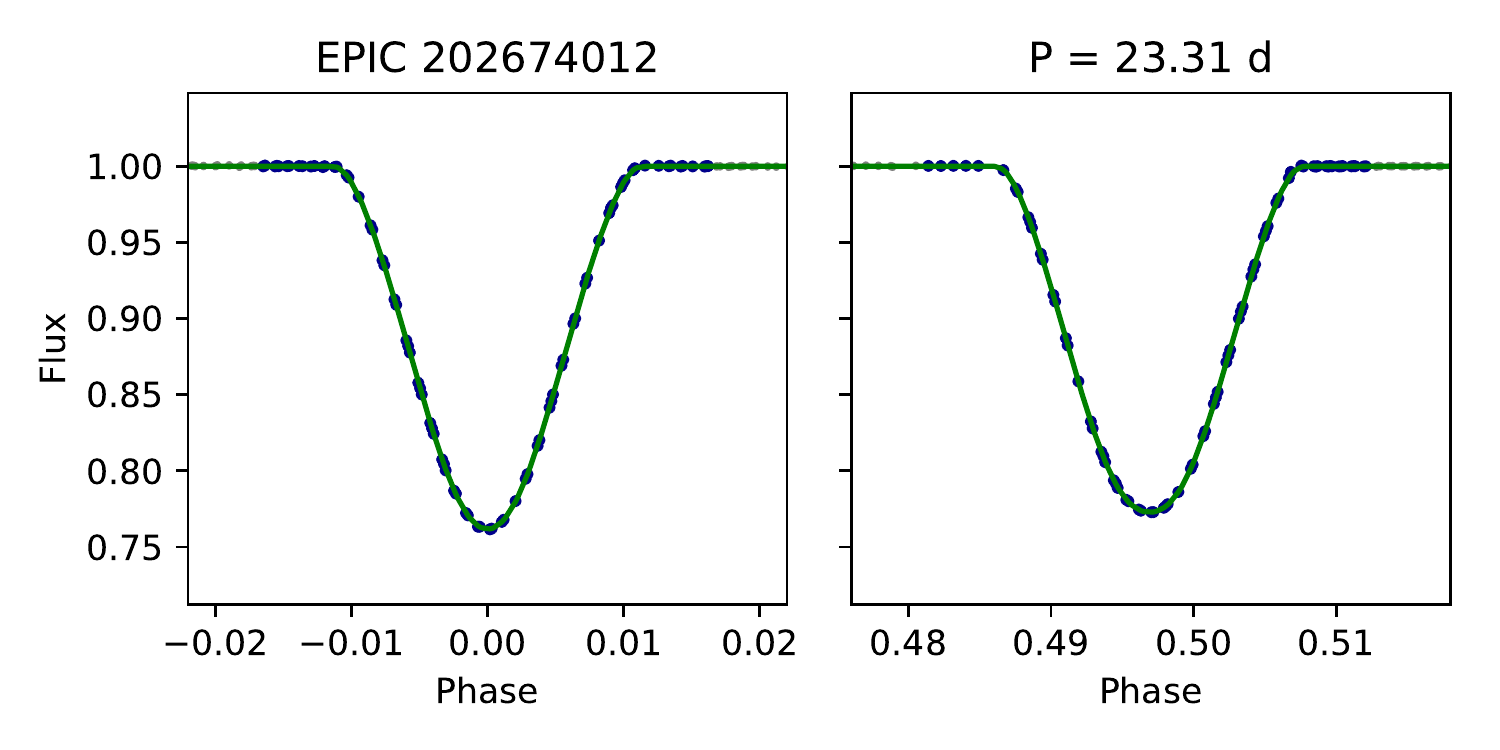} 
\includegraphics[width=0.49\textwidth]{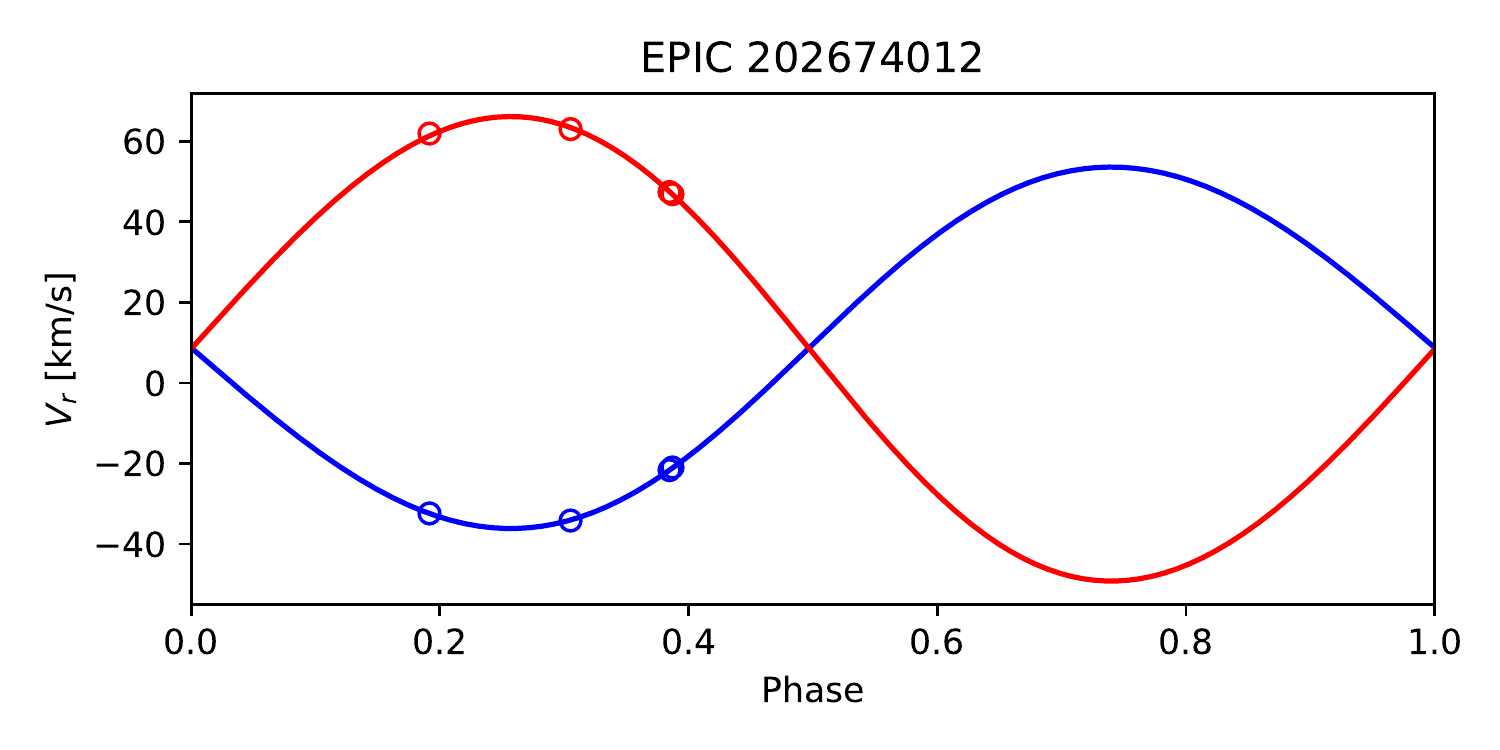} 
\includegraphics[width=0.49\textwidth]{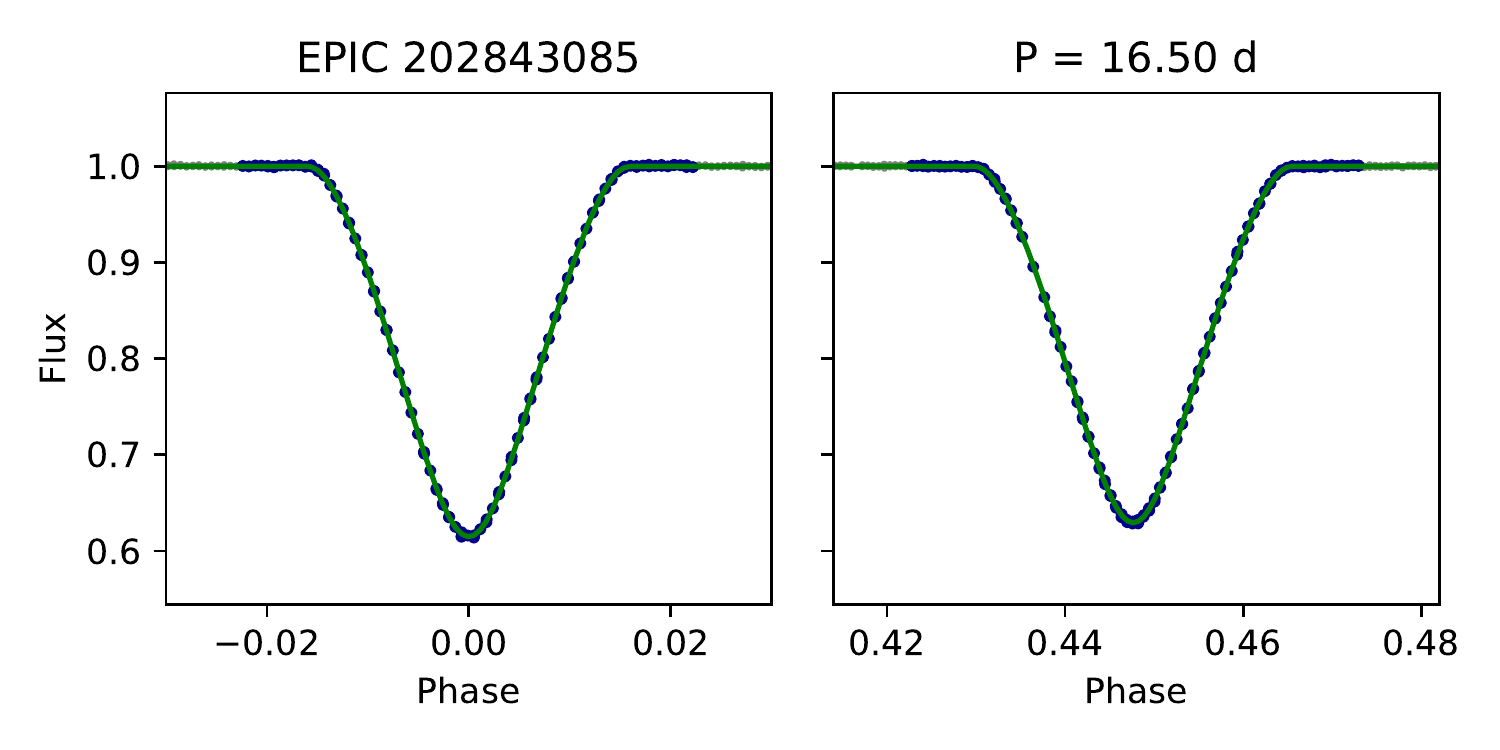} 
\includegraphics[width=0.49\textwidth]{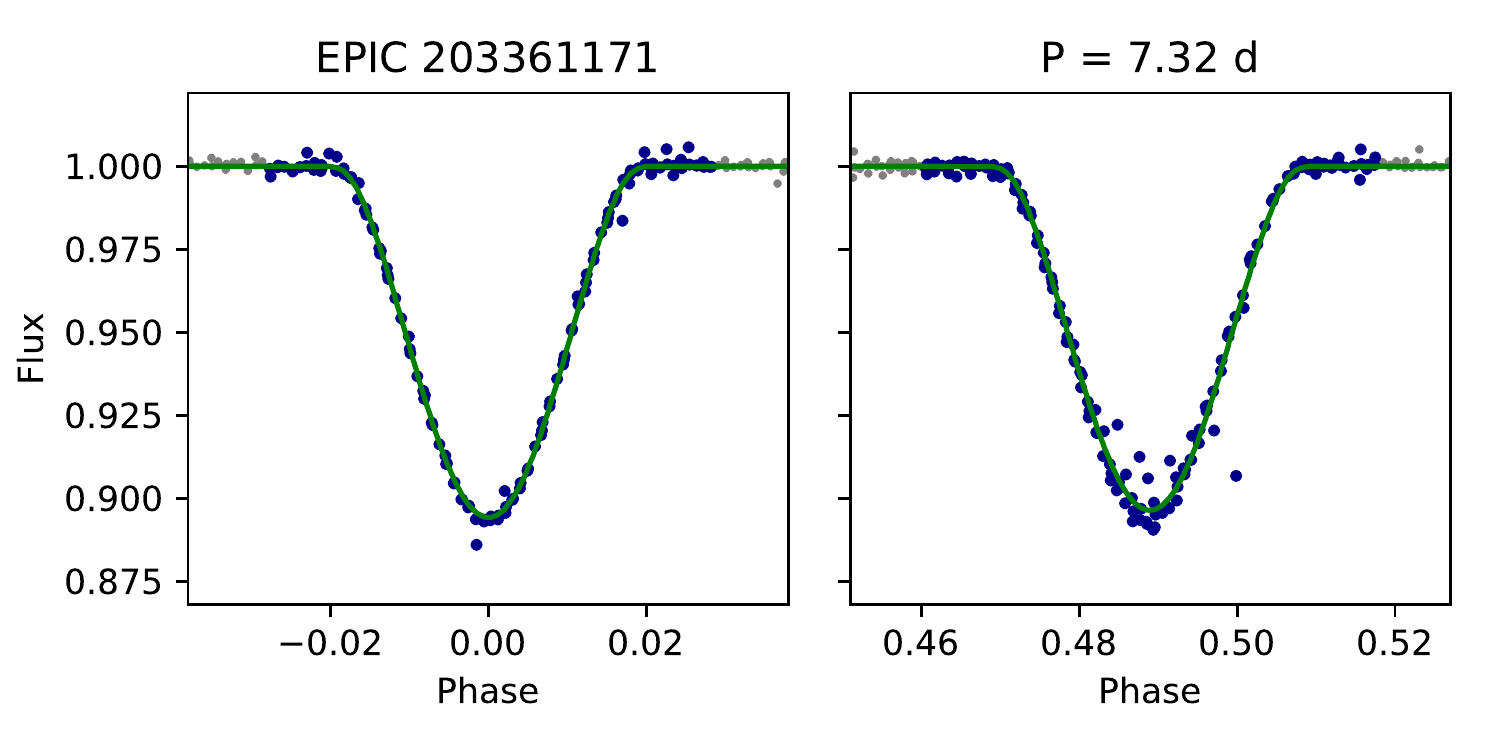} 
\includegraphics[width=0.49\textwidth]{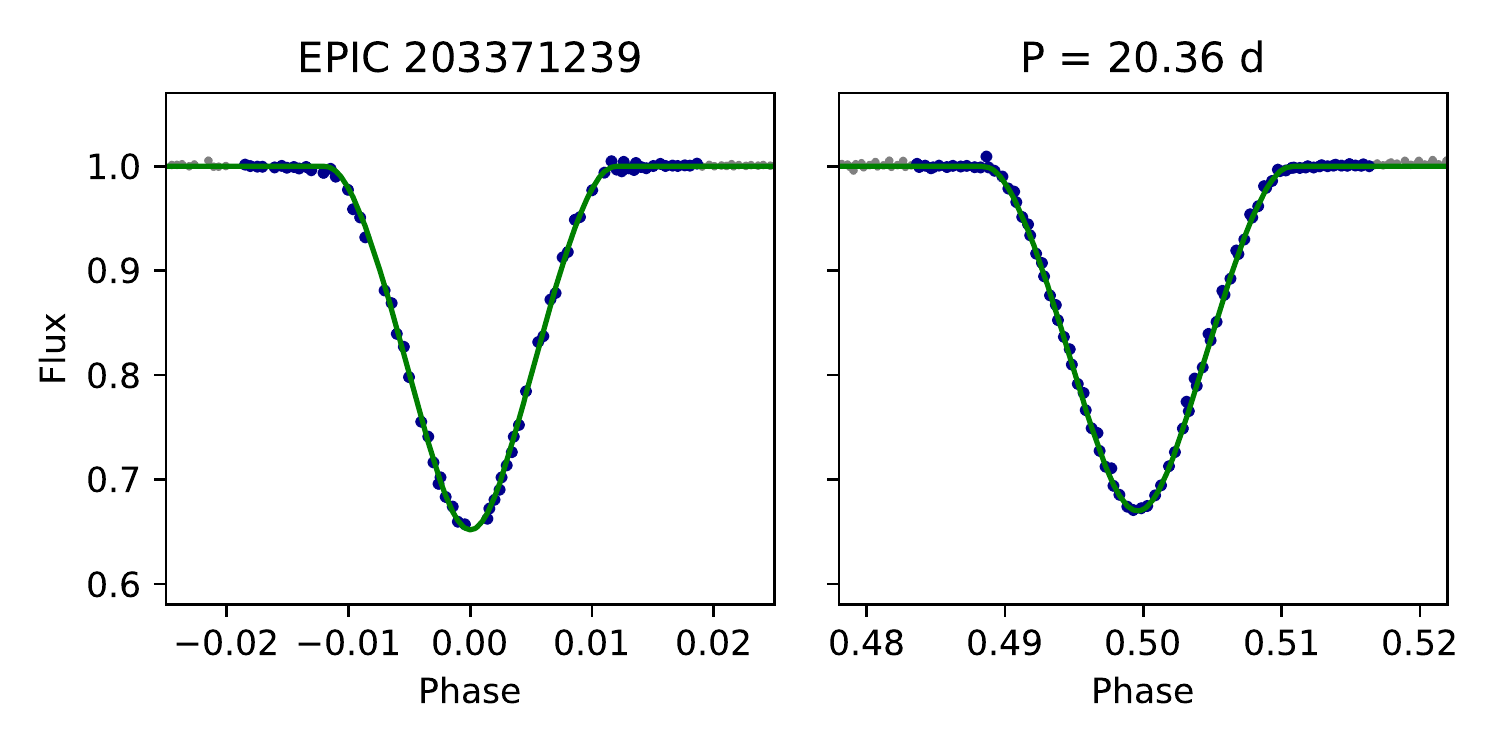} 
\includegraphics[width=0.49\textwidth]{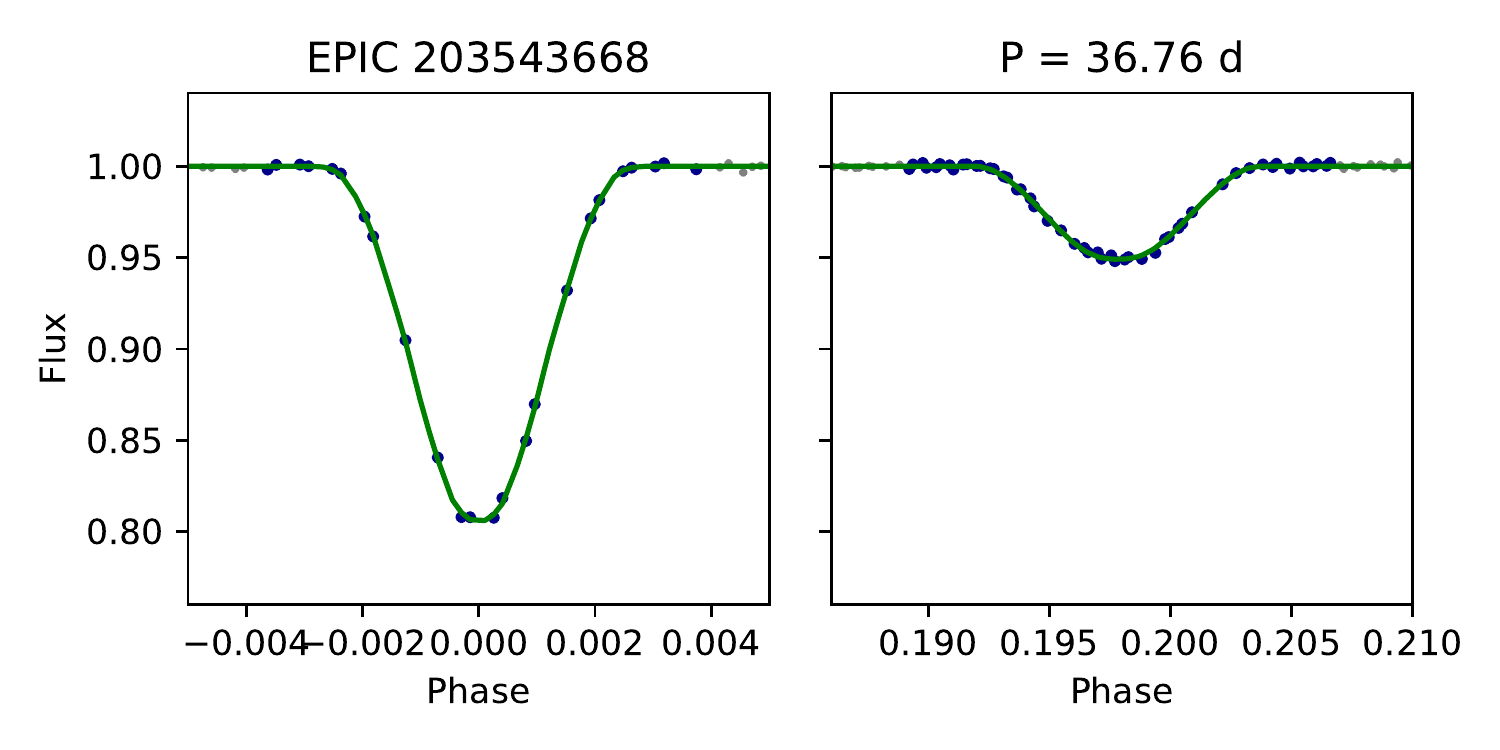} 
\includegraphics[width=0.49\textwidth]{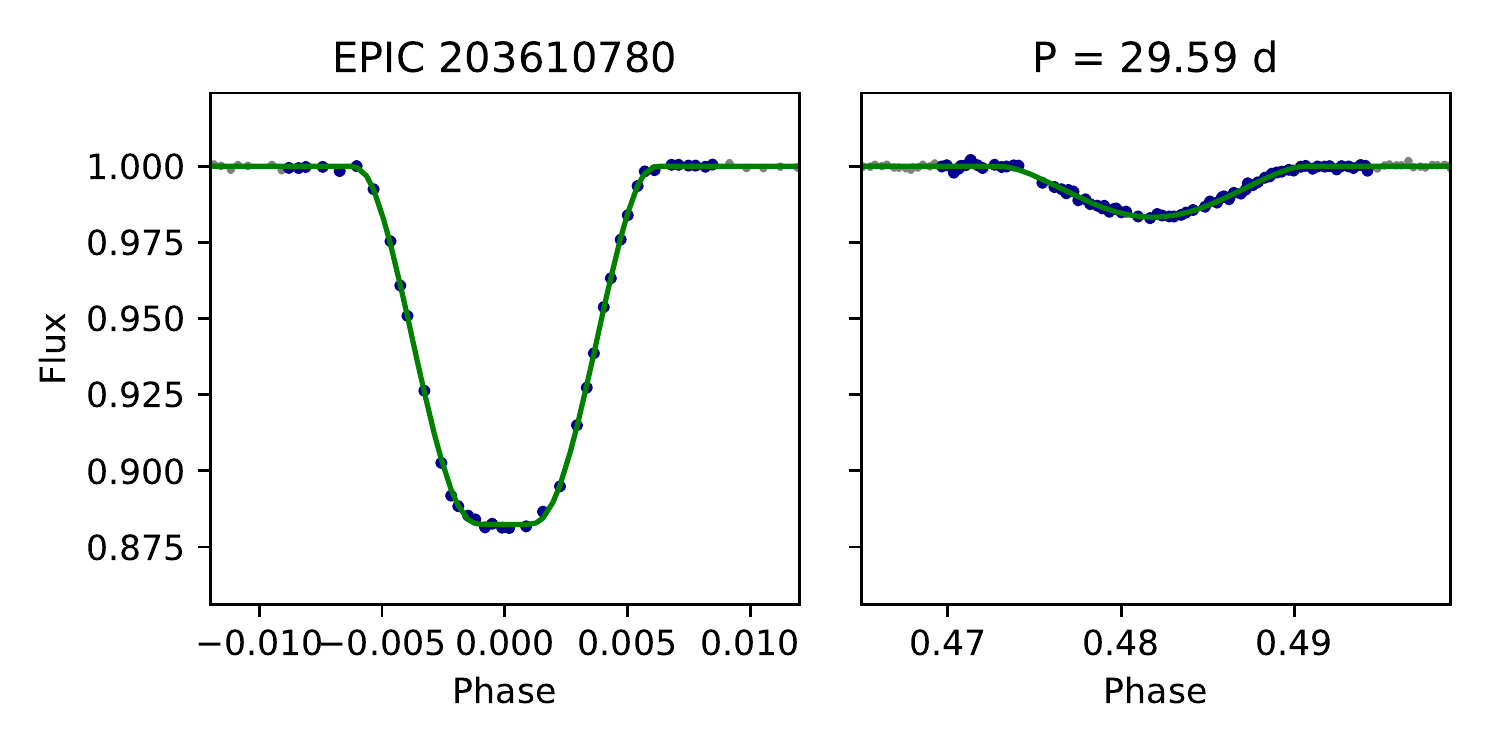} 
\includegraphics[width=0.49\textwidth]{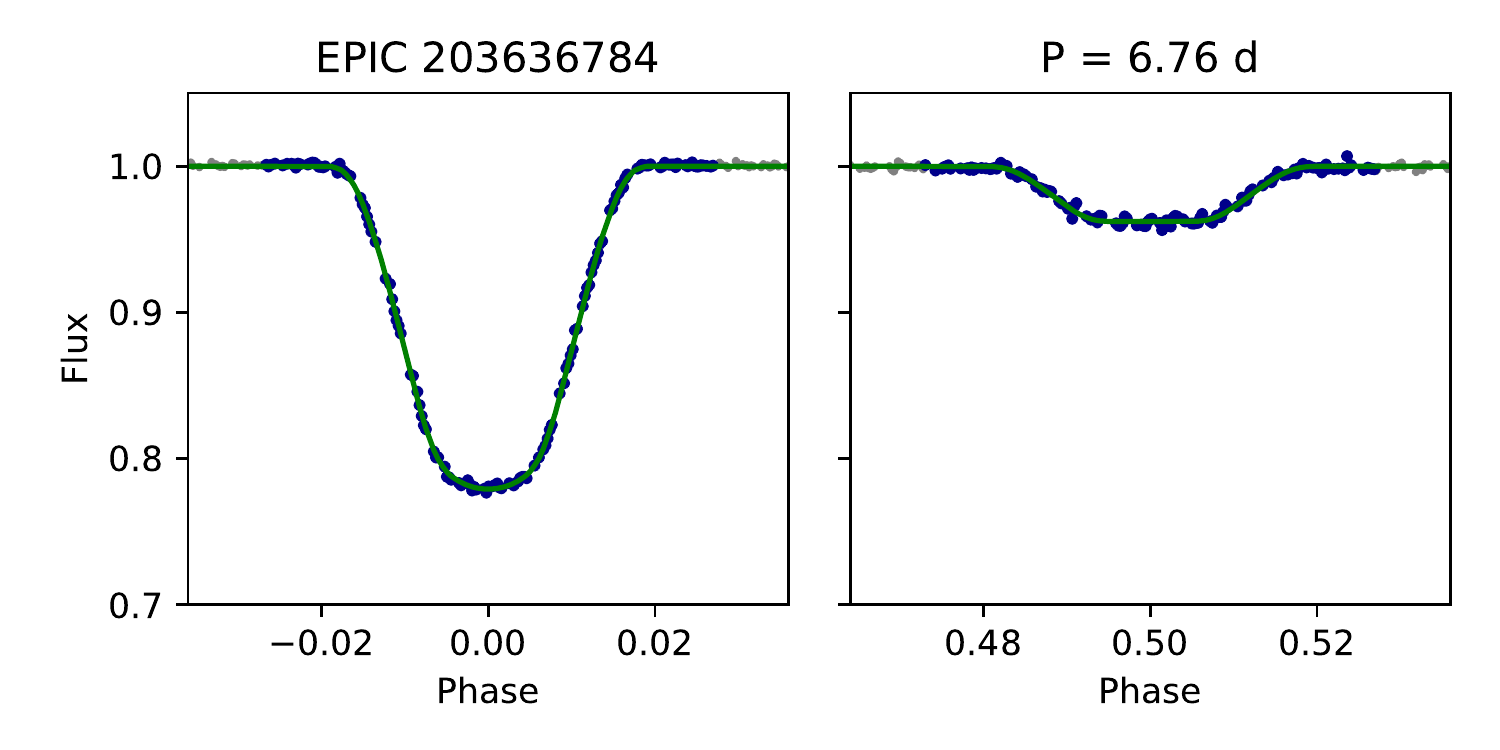} 
  \caption{K2 light curves with the best-fit \texttt{ellc} model. Data not
  included in the fit are plotted using small grey points. For EPIC 202674012
  we also show our best-fit Keplerian orbit to the measured radial velocities
  as a function of the orbital phase relative to the time of mid-primary
  eclipse. 
  \label{lcfit2}}
\end{figure*}

\begin{figure*}
\includegraphics[width=0.49\textwidth]{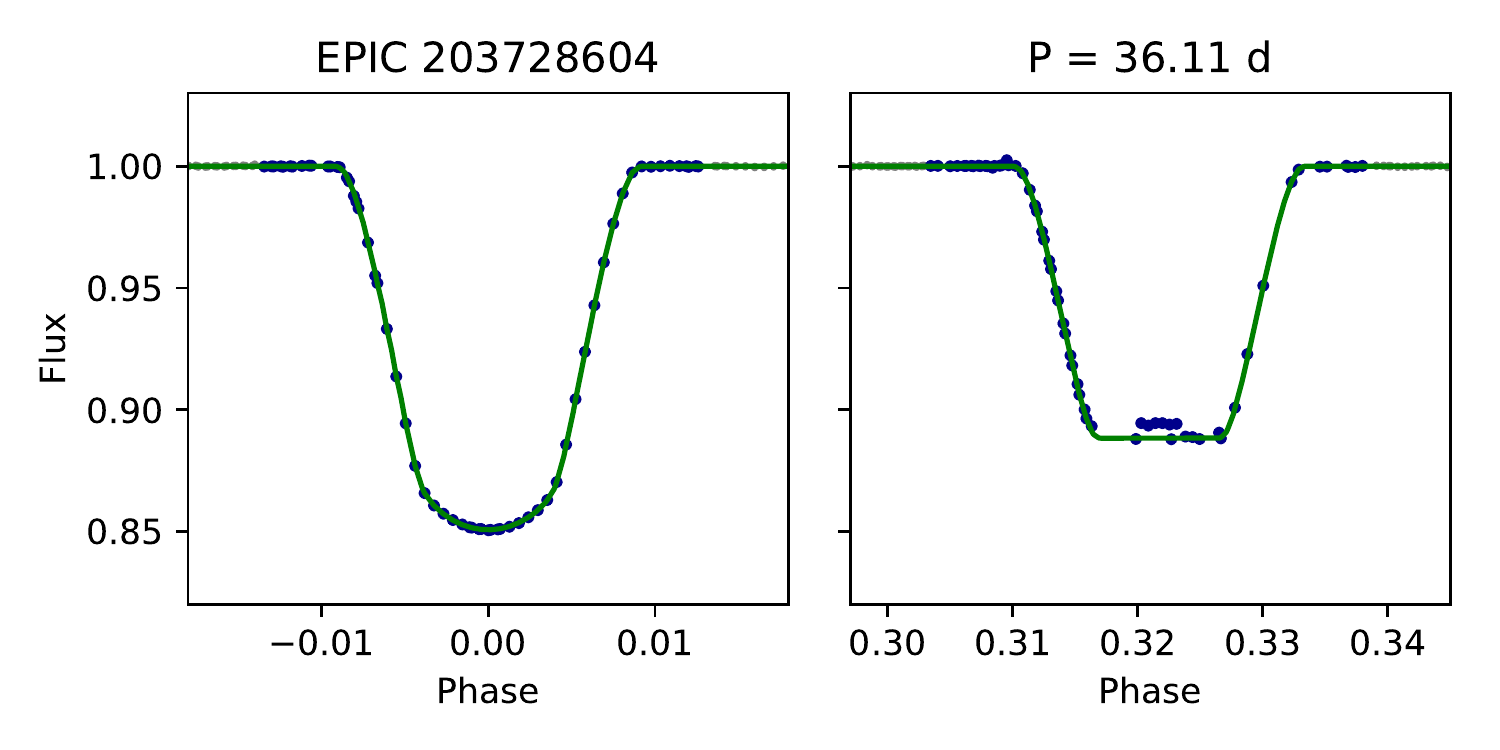} 
\includegraphics[width=0.49\textwidth]{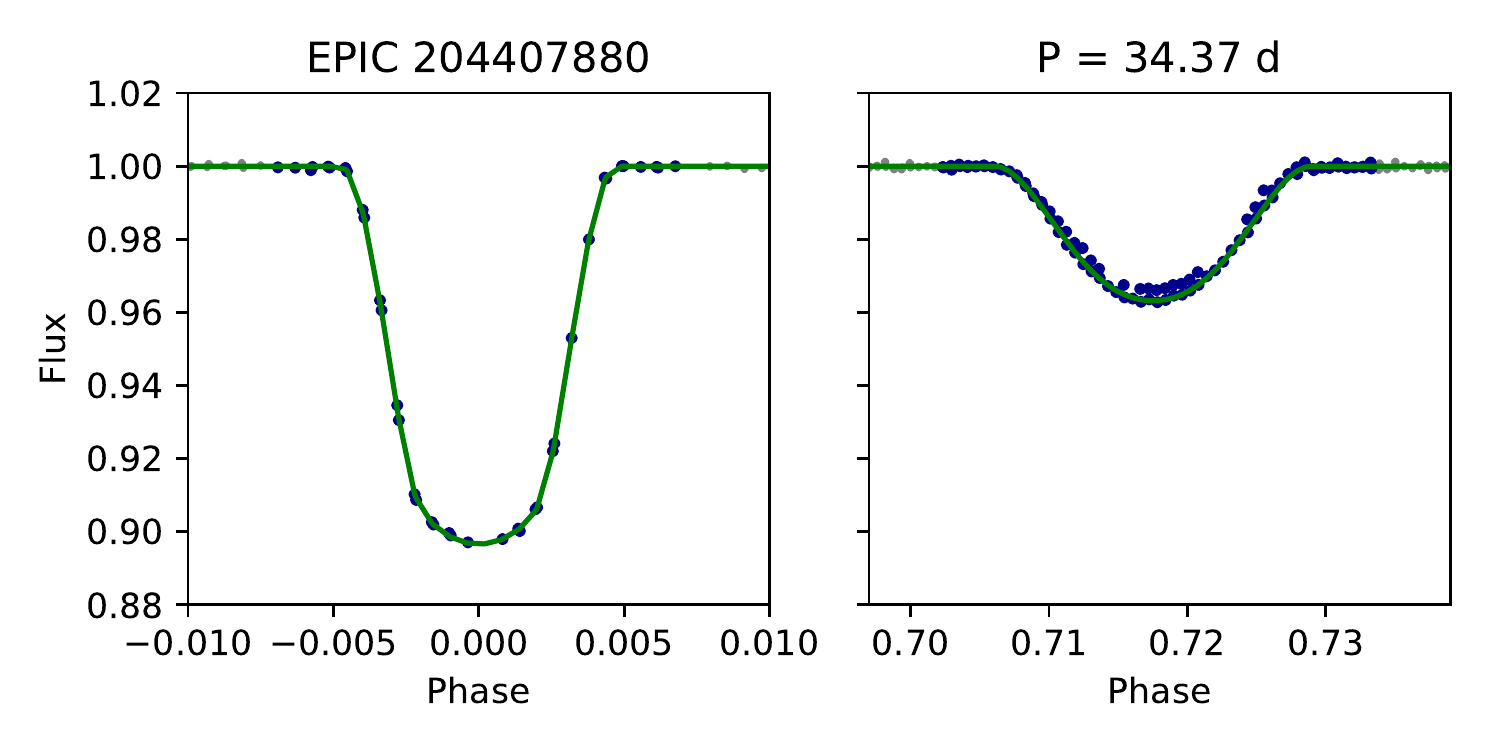} 
\includegraphics[width=0.49\textwidth]{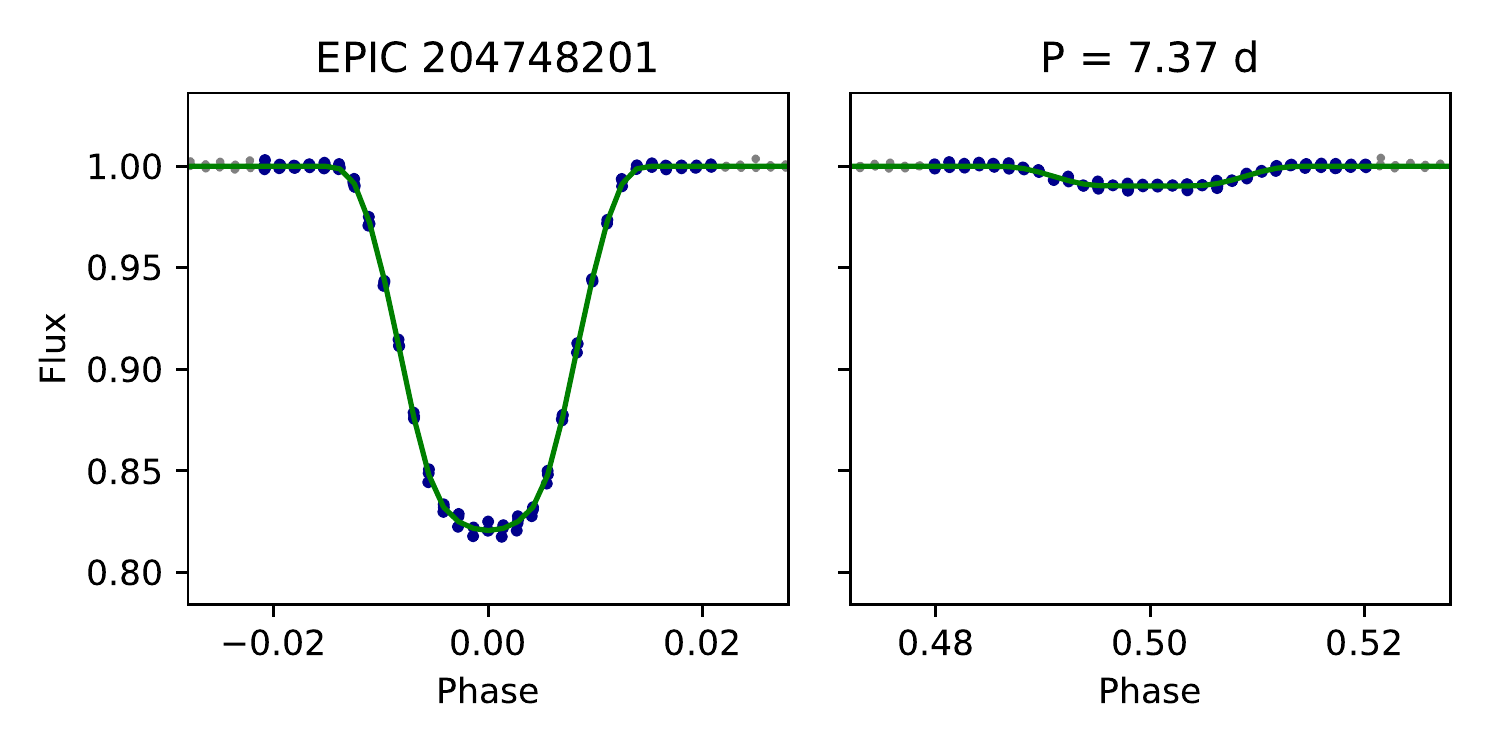} 
\includegraphics[width=0.49\textwidth]{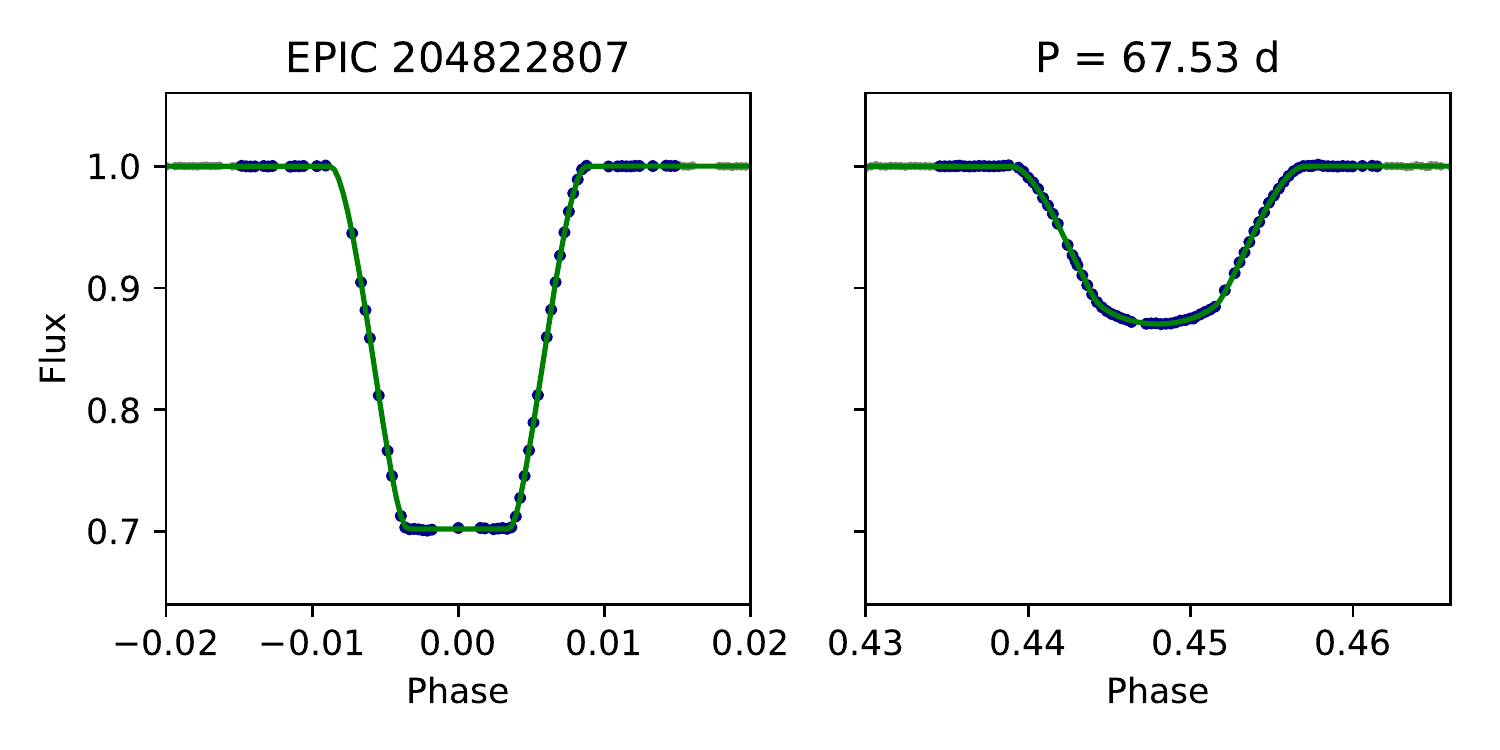} 
\includegraphics[width=0.49\textwidth]{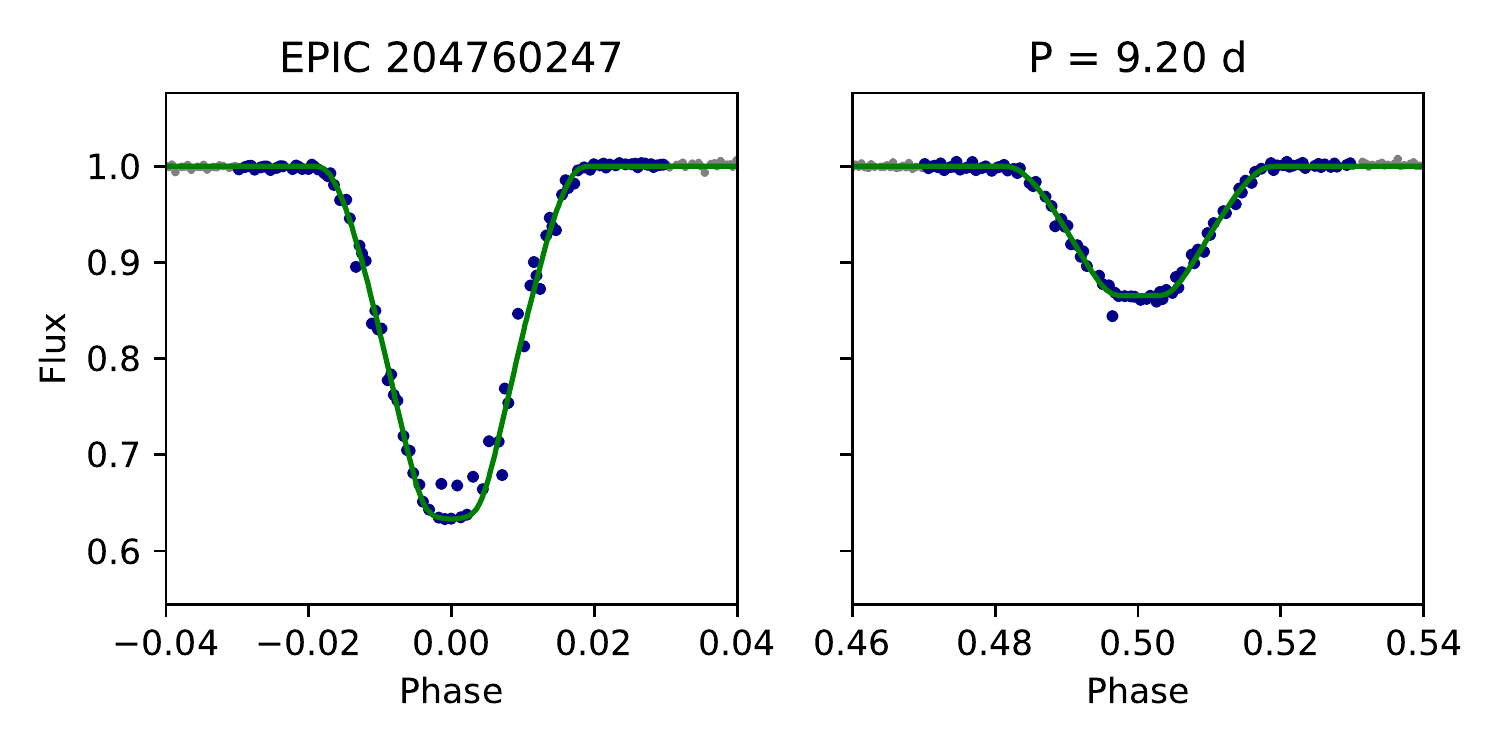} 
\includegraphics[width=0.49\textwidth]{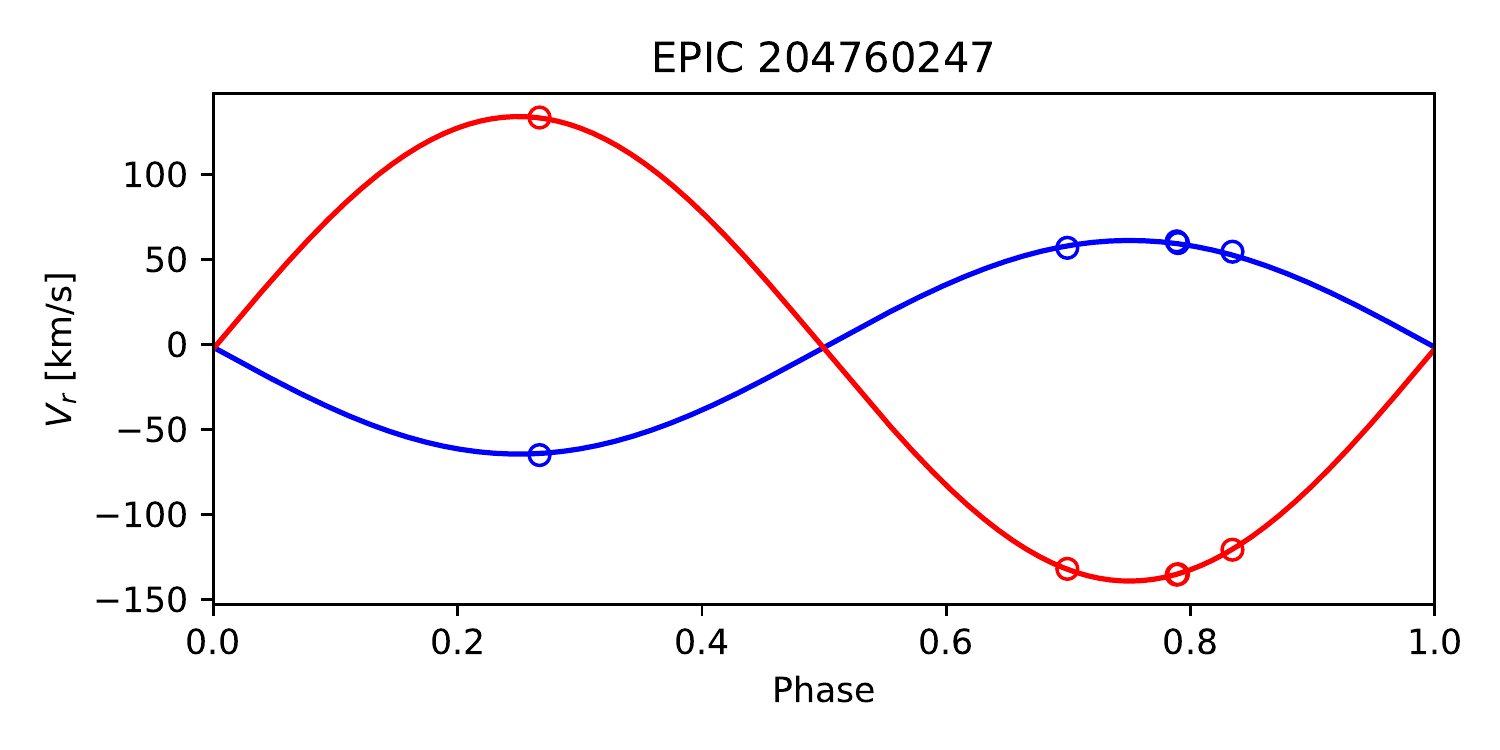} 
\includegraphics[width=0.49\textwidth]{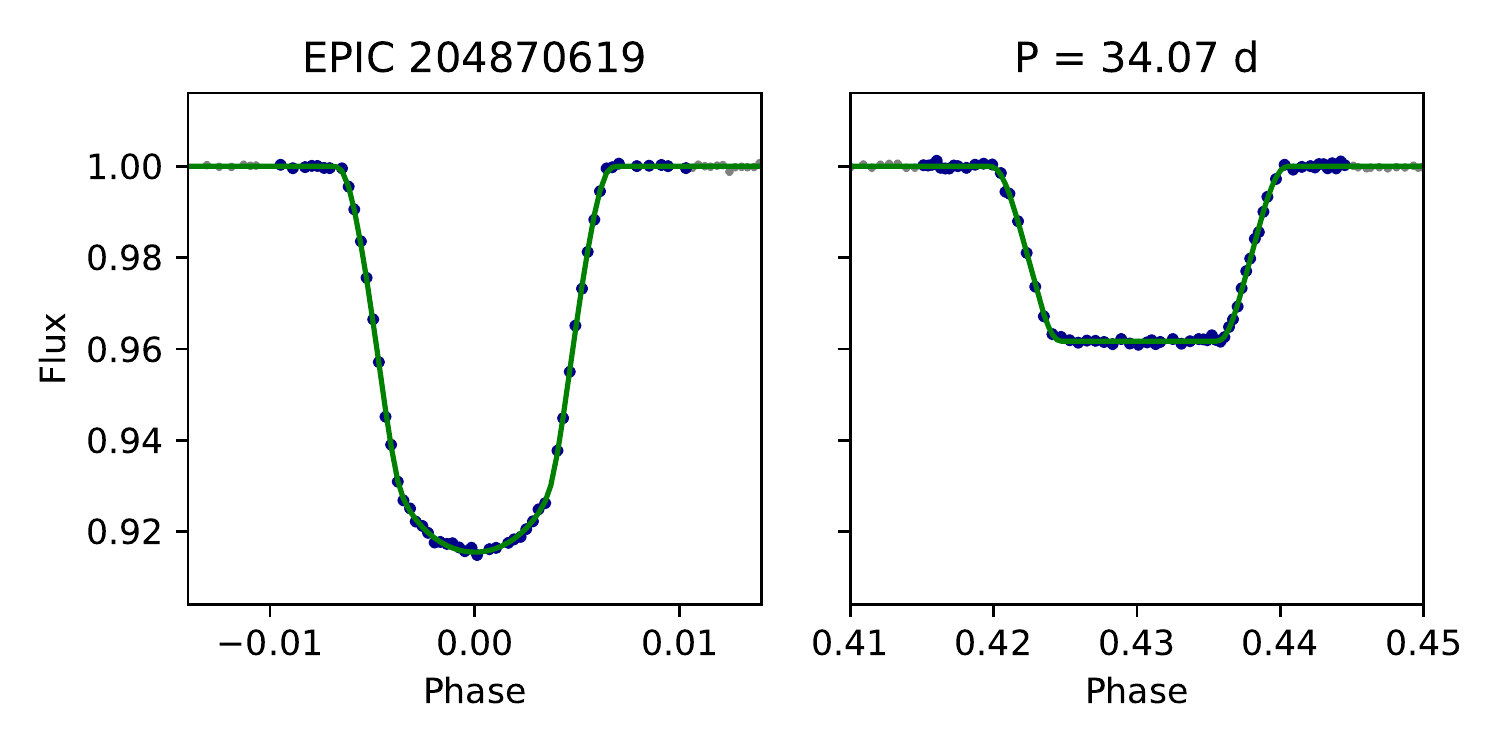} 
\includegraphics[width=0.49\textwidth]{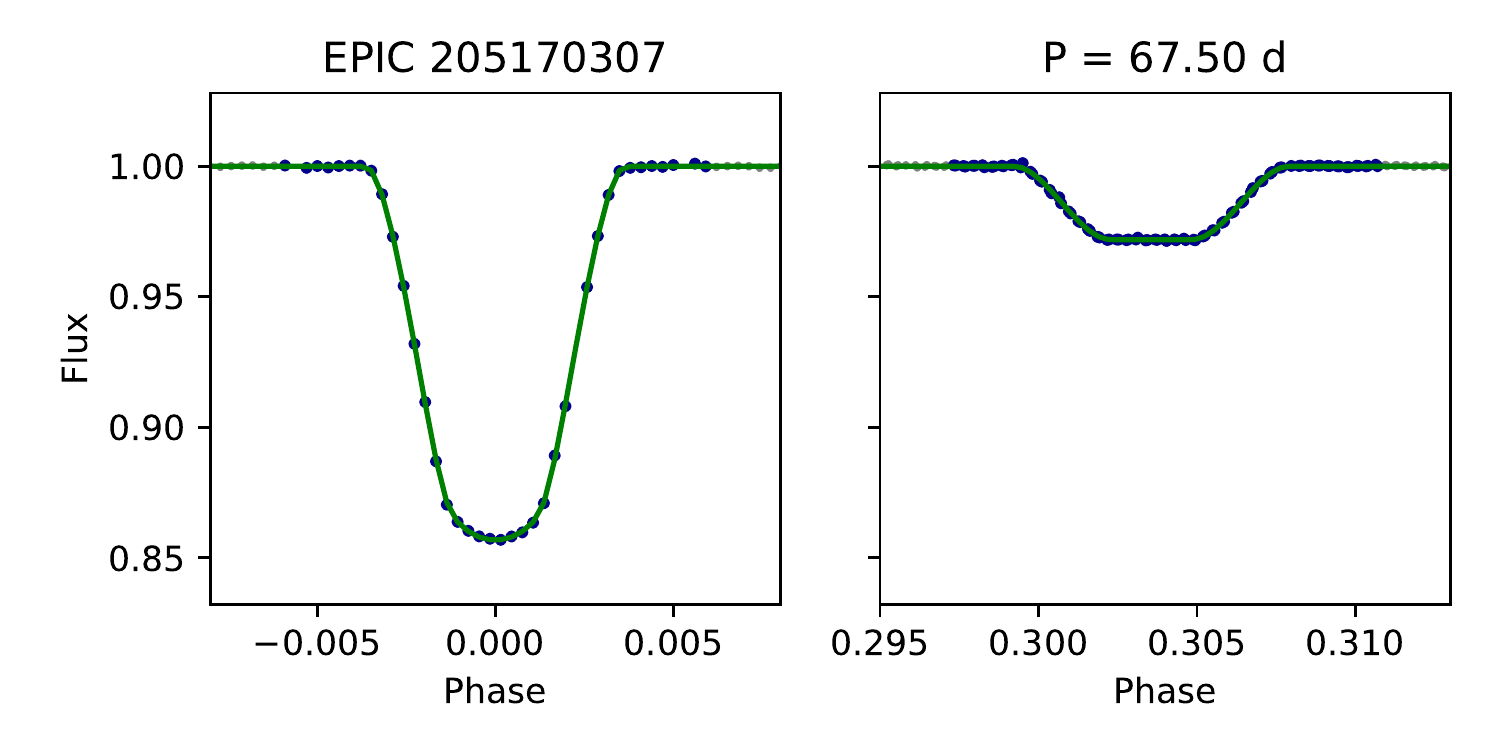} 
\includegraphics[width=0.49\textwidth]{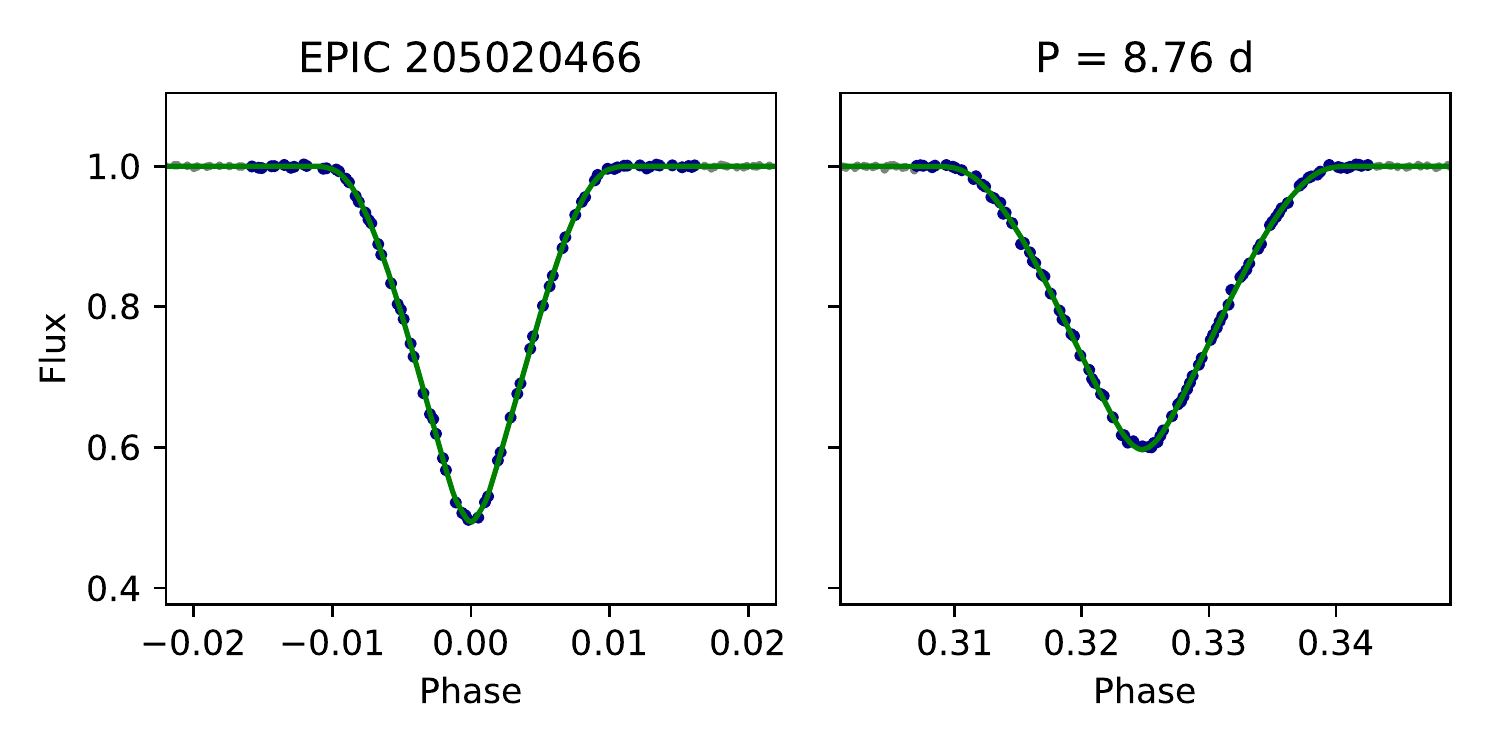} 
\includegraphics[width=0.49\textwidth]{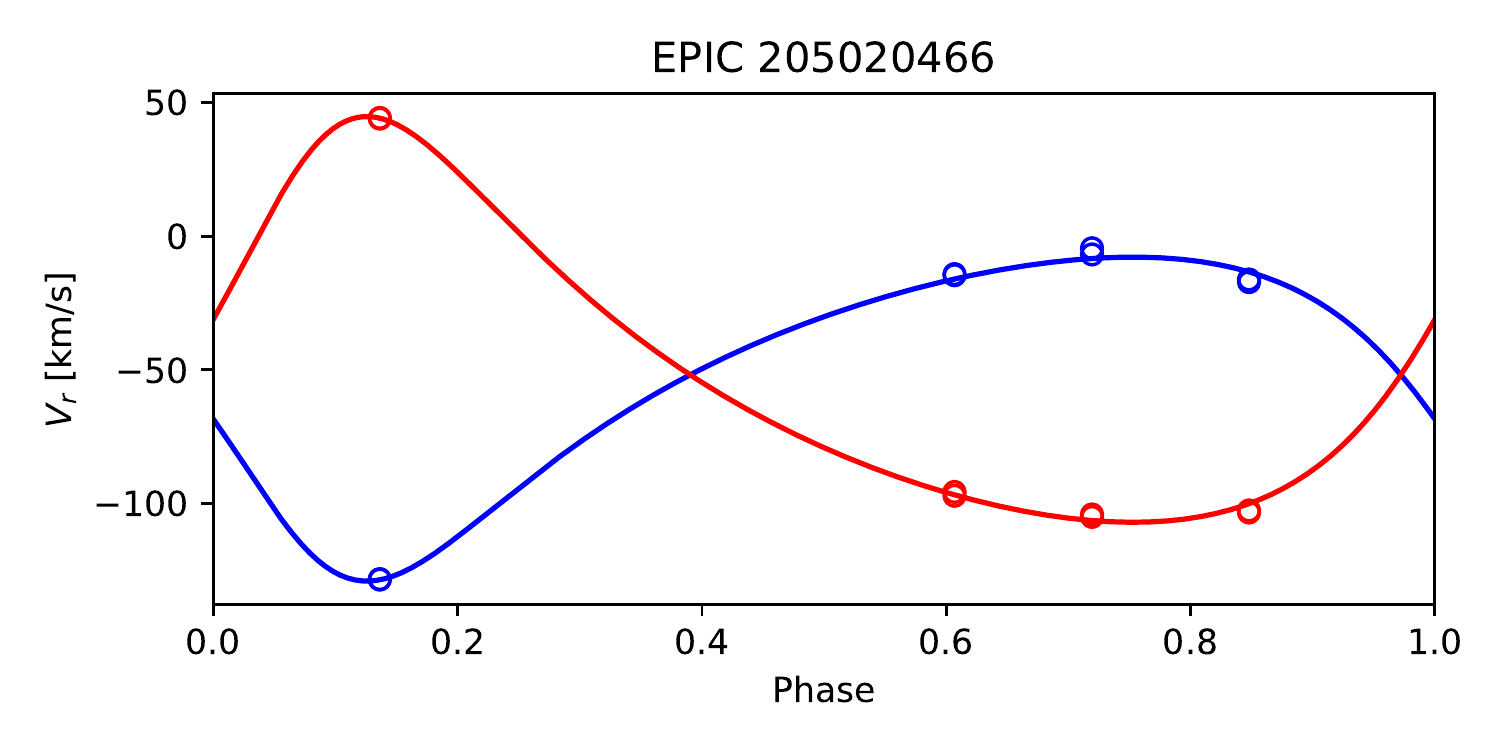} 
  \caption{K2 light curves with the best-fit \texttt{ellc} model. Data not
  included in the fit are plotted using small grey points. For EPIC~204760247
  and EPIC~205020466 we also show our best-fit Keplerian orbit to the measured
  radial velocities as a function of the orbital phase relative to the time of
  mid-primary eclipse. 
  \label{lcfit3}}
\end{figure*}

\begin{figure*}
\includegraphics[width=0.49\textwidth]{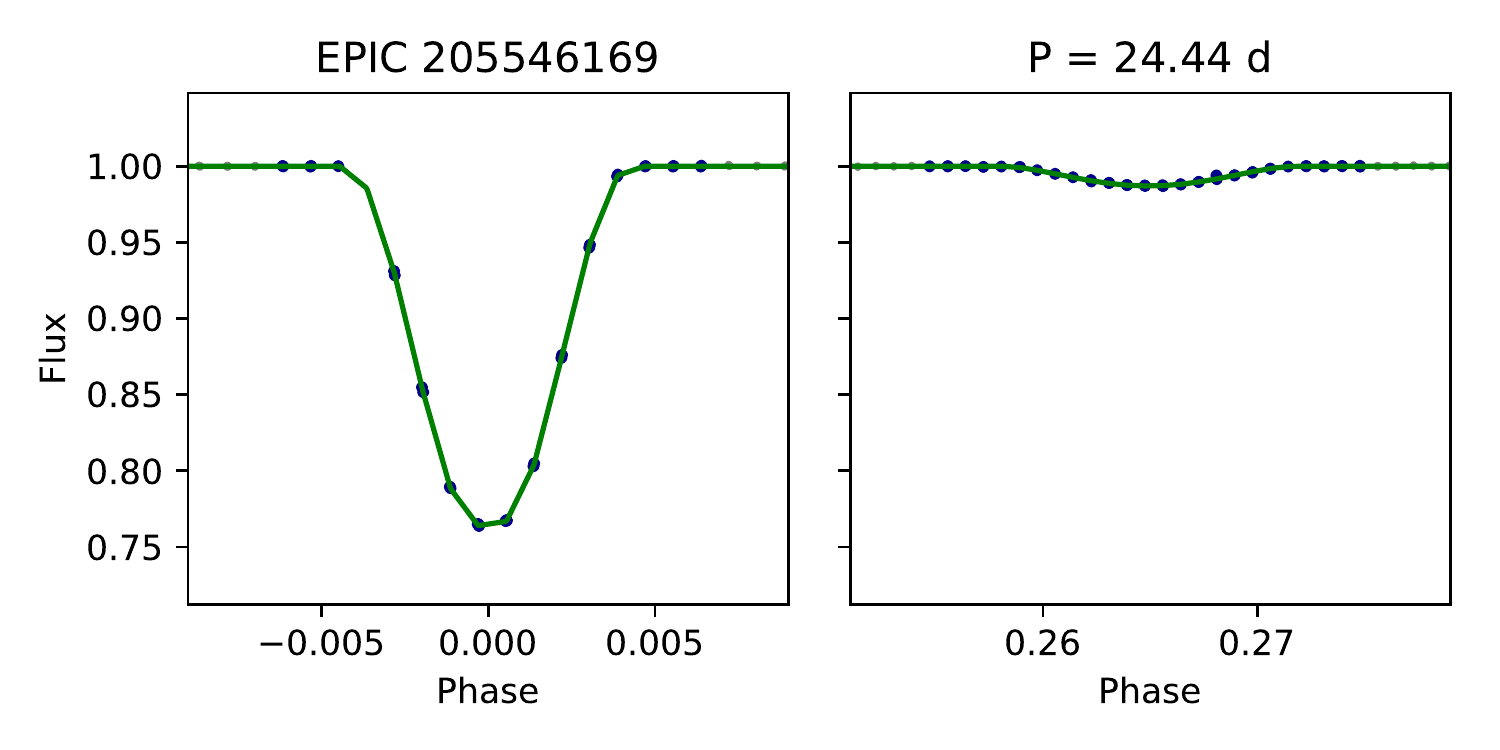} 
\includegraphics[width=0.49\textwidth]{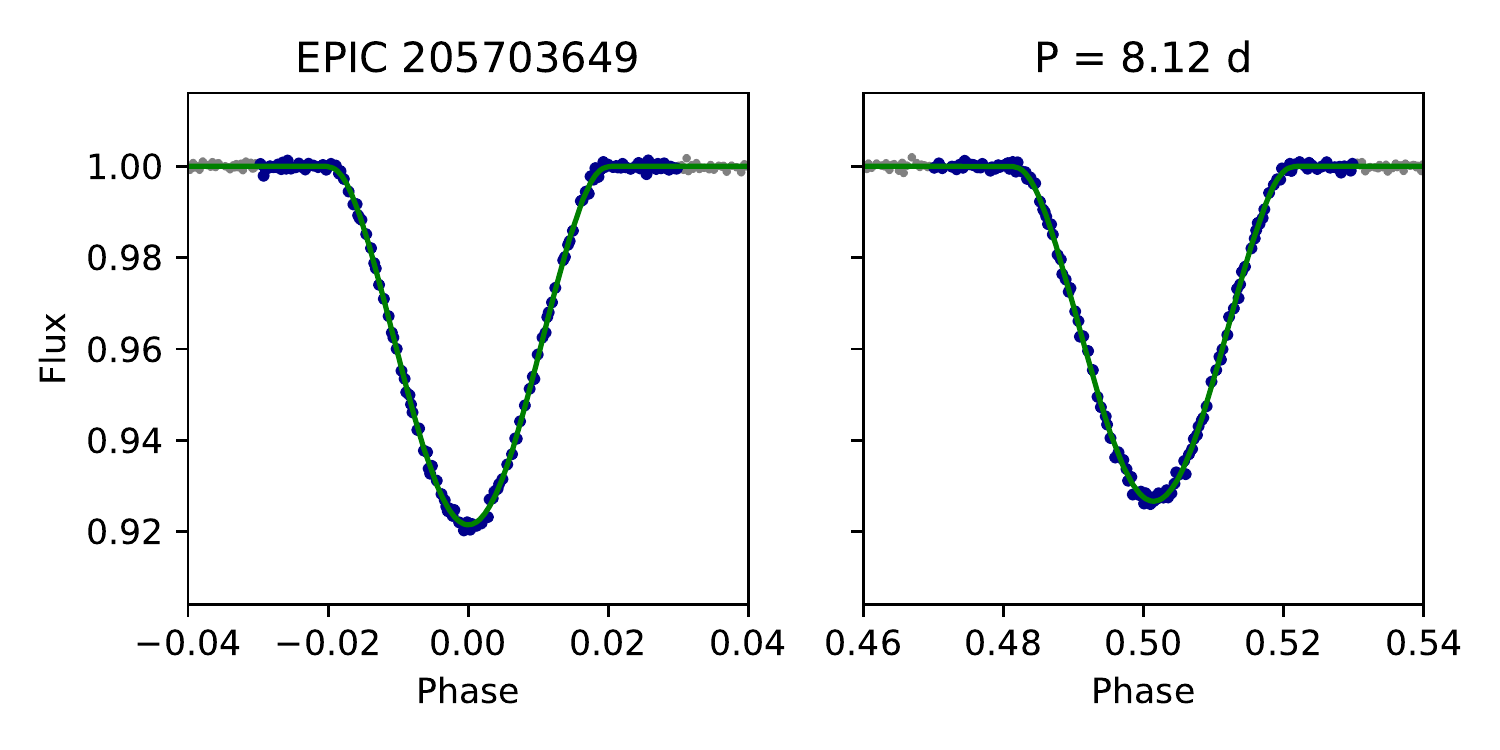} 
\includegraphics[width=0.49\textwidth]{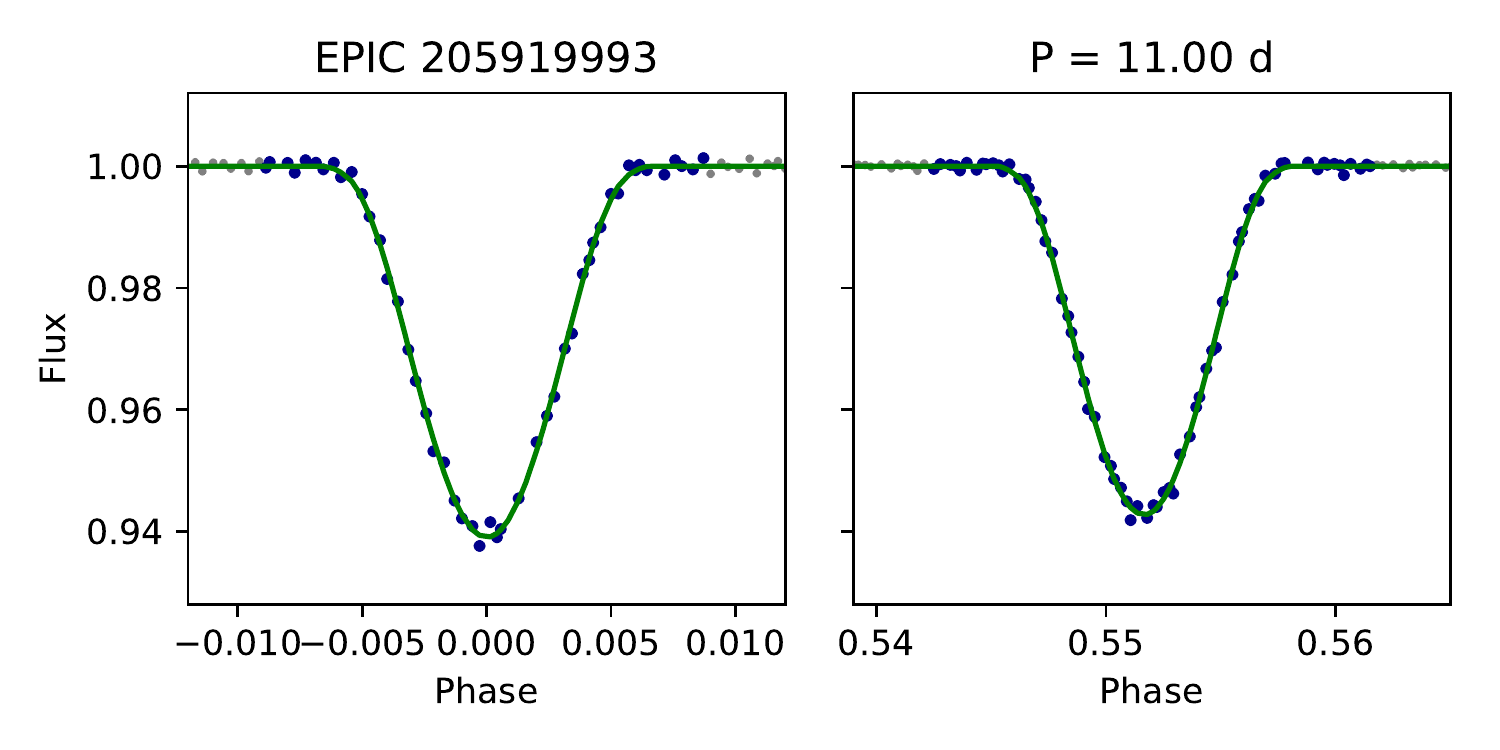} 
\includegraphics[width=0.49\textwidth]{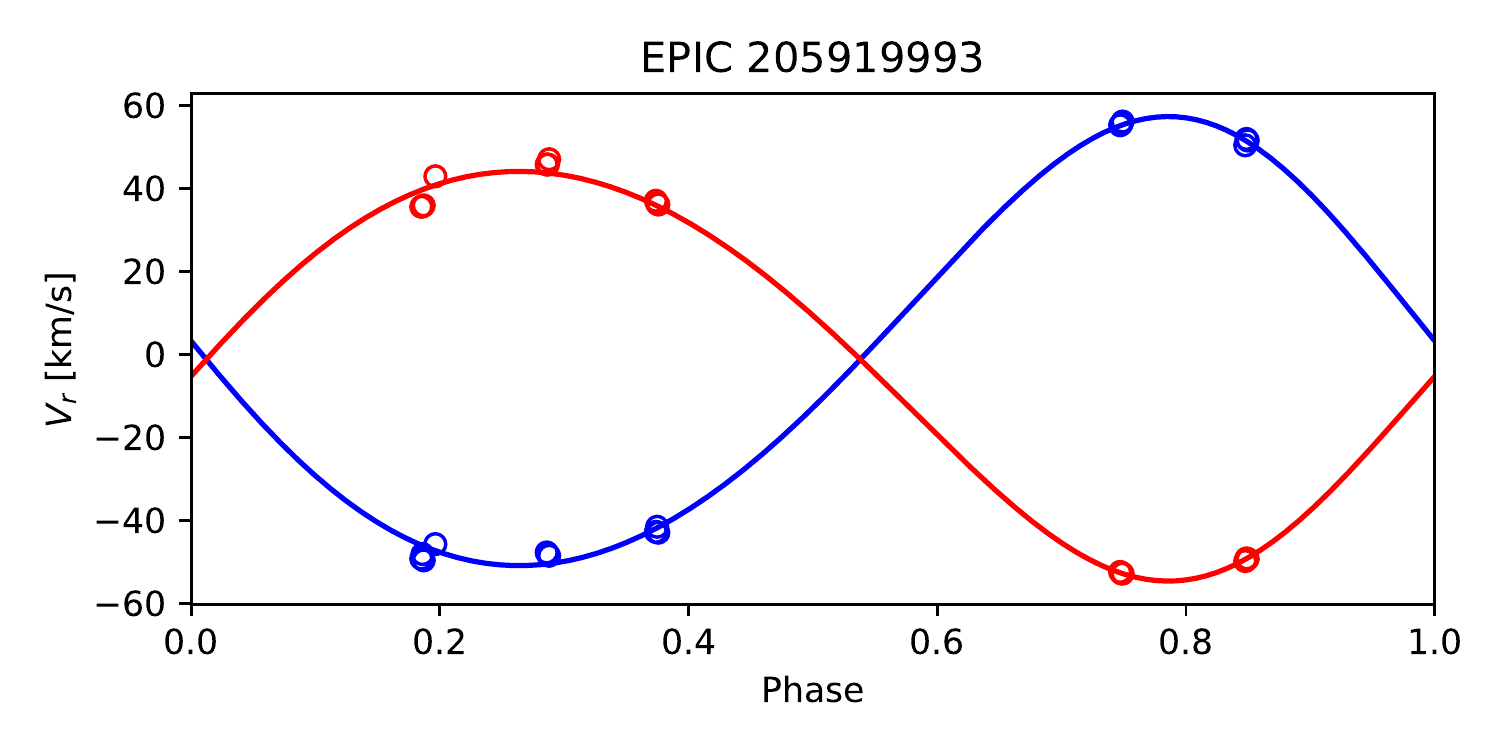} 
\includegraphics[width=0.49\textwidth]{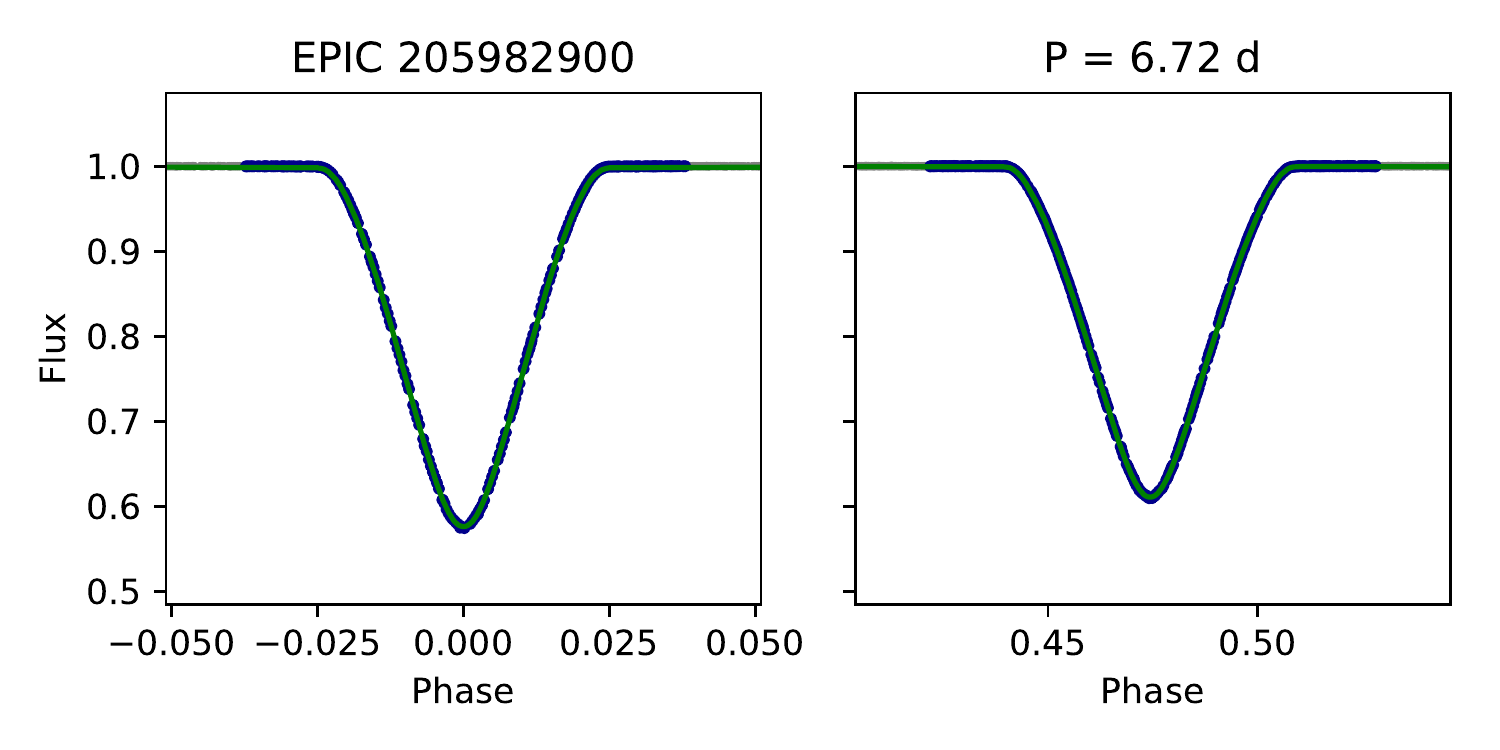} 
\includegraphics[width=0.49\textwidth]{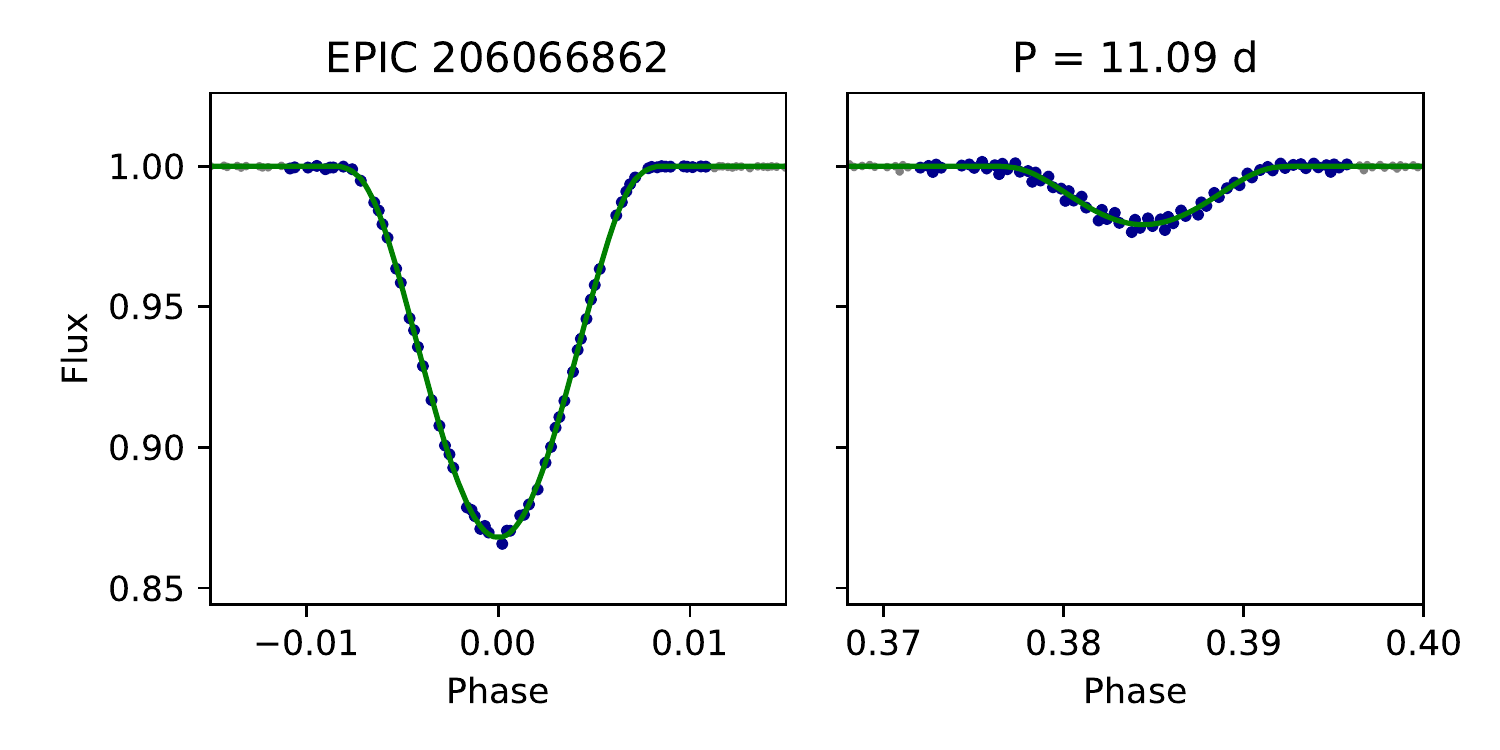} 
\includegraphics[width=0.49\textwidth]{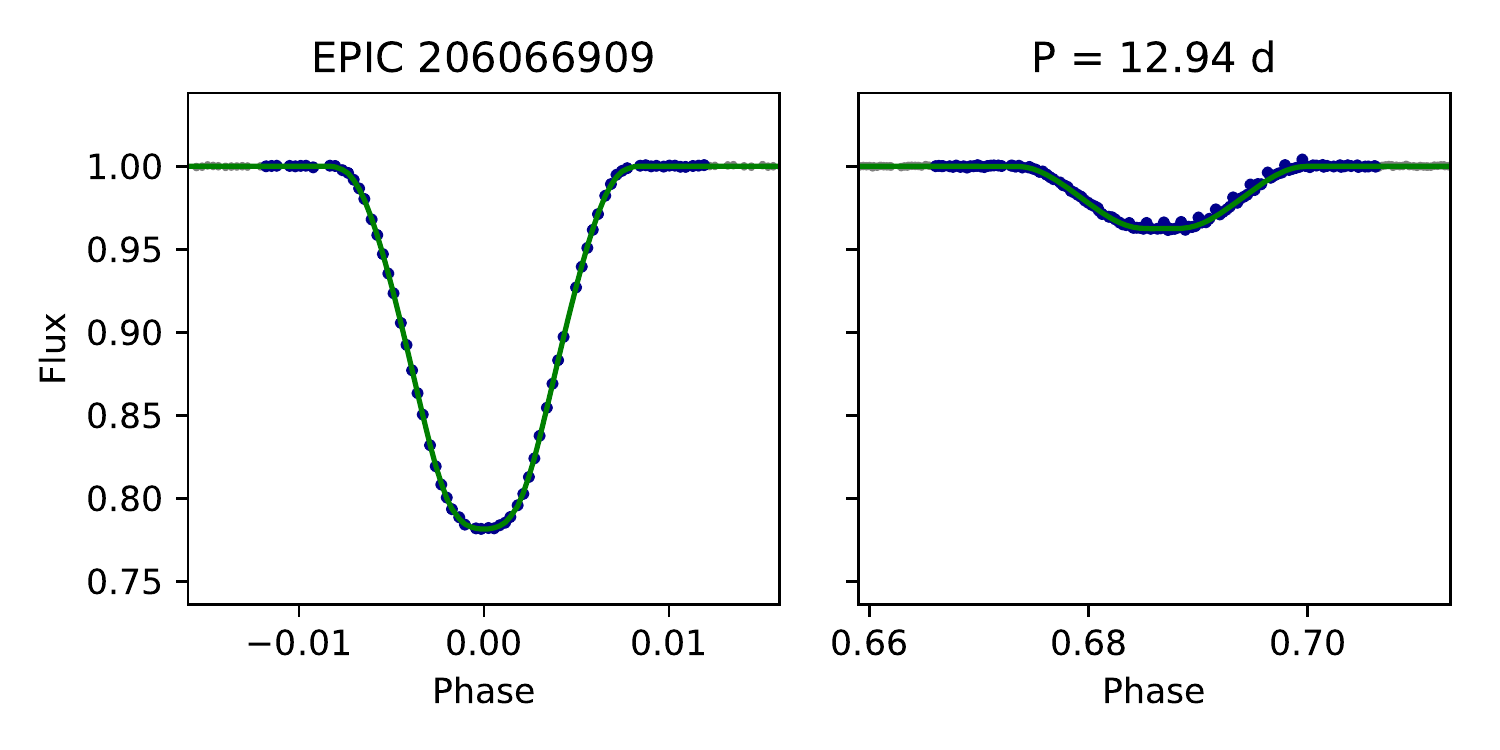} 
\includegraphics[width=0.49\textwidth]{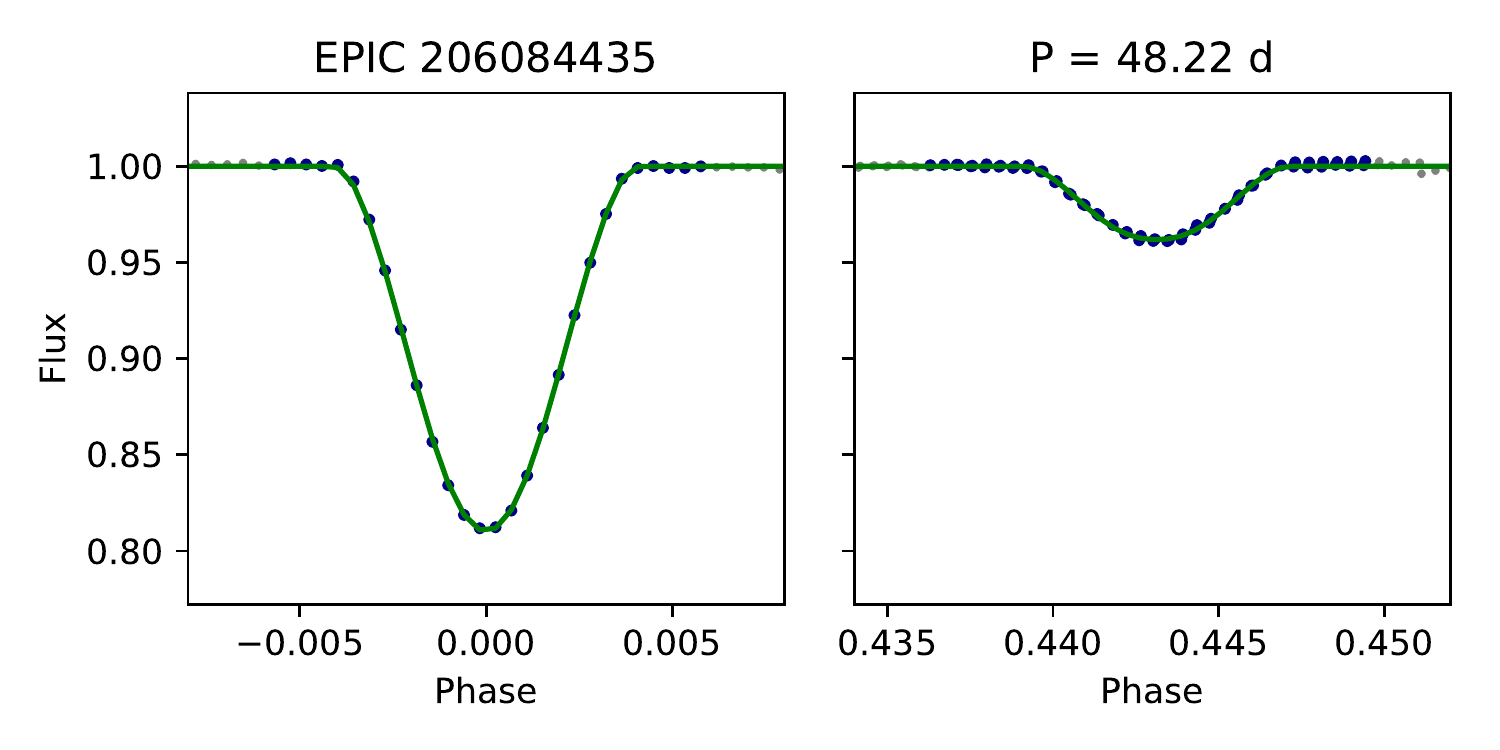} 
\includegraphics[width=0.49\textwidth]{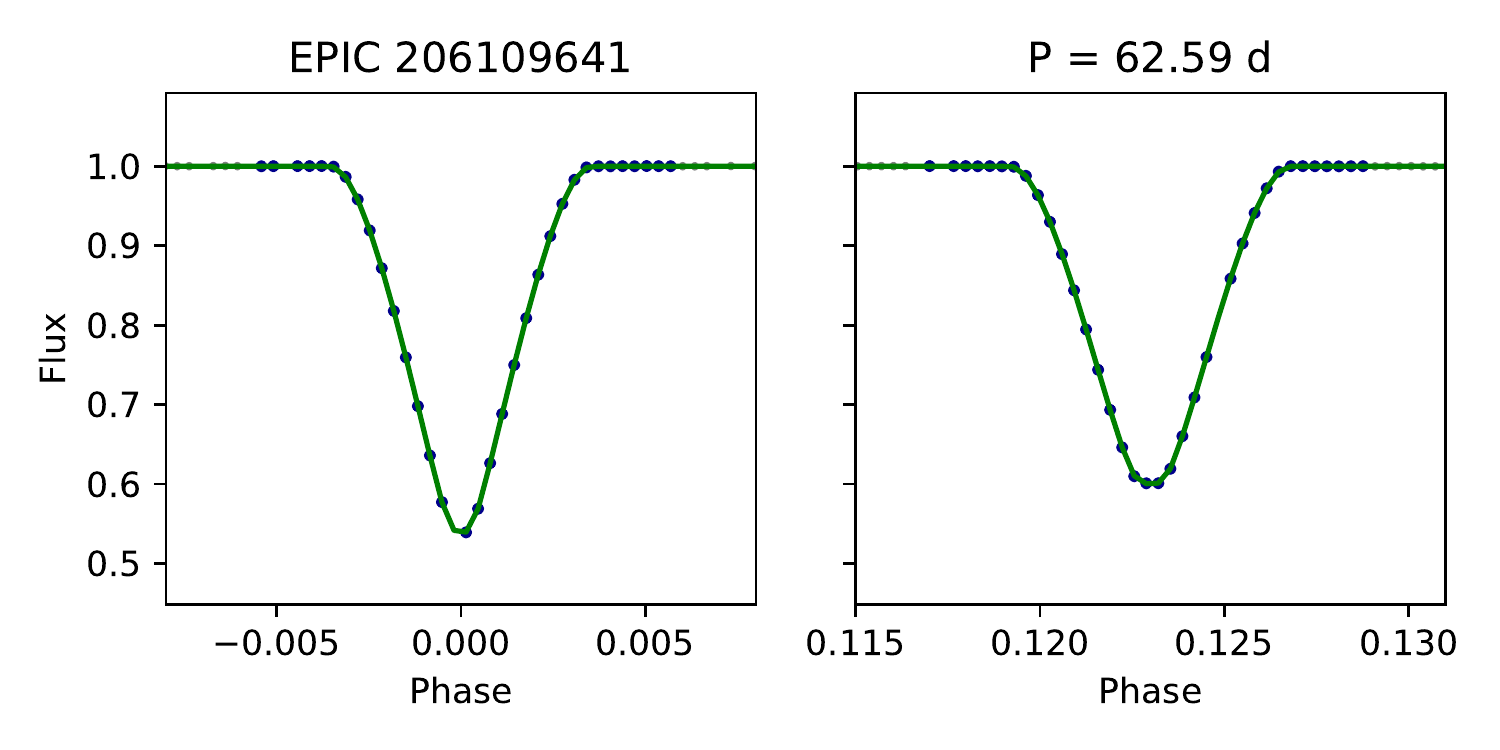} 
\includegraphics[width=0.49\textwidth]{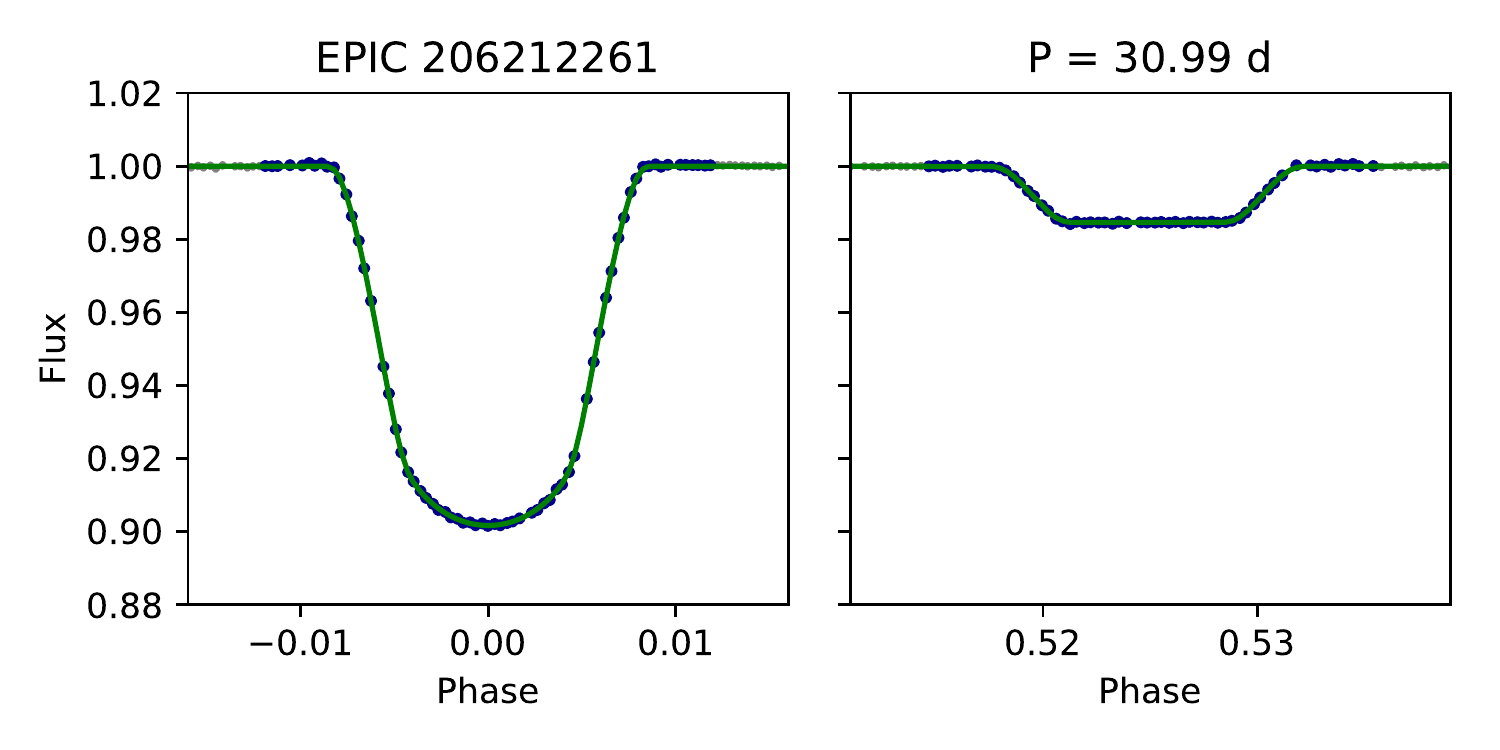} 
  \caption{K2 light curves with the best-fit \texttt{ellc} model. Data not
  included in the fit are plotted using small grey points. For EPIC~205919993
  we also show our best-fit Keplerian orbit to the measured radial velocities
  as a function of the orbital phase relative to the time of mid-primary
  eclipse. 
  \label{lcfit4}}
\end{figure*}

\begin{figure*}
\includegraphics[width=0.49\textwidth]{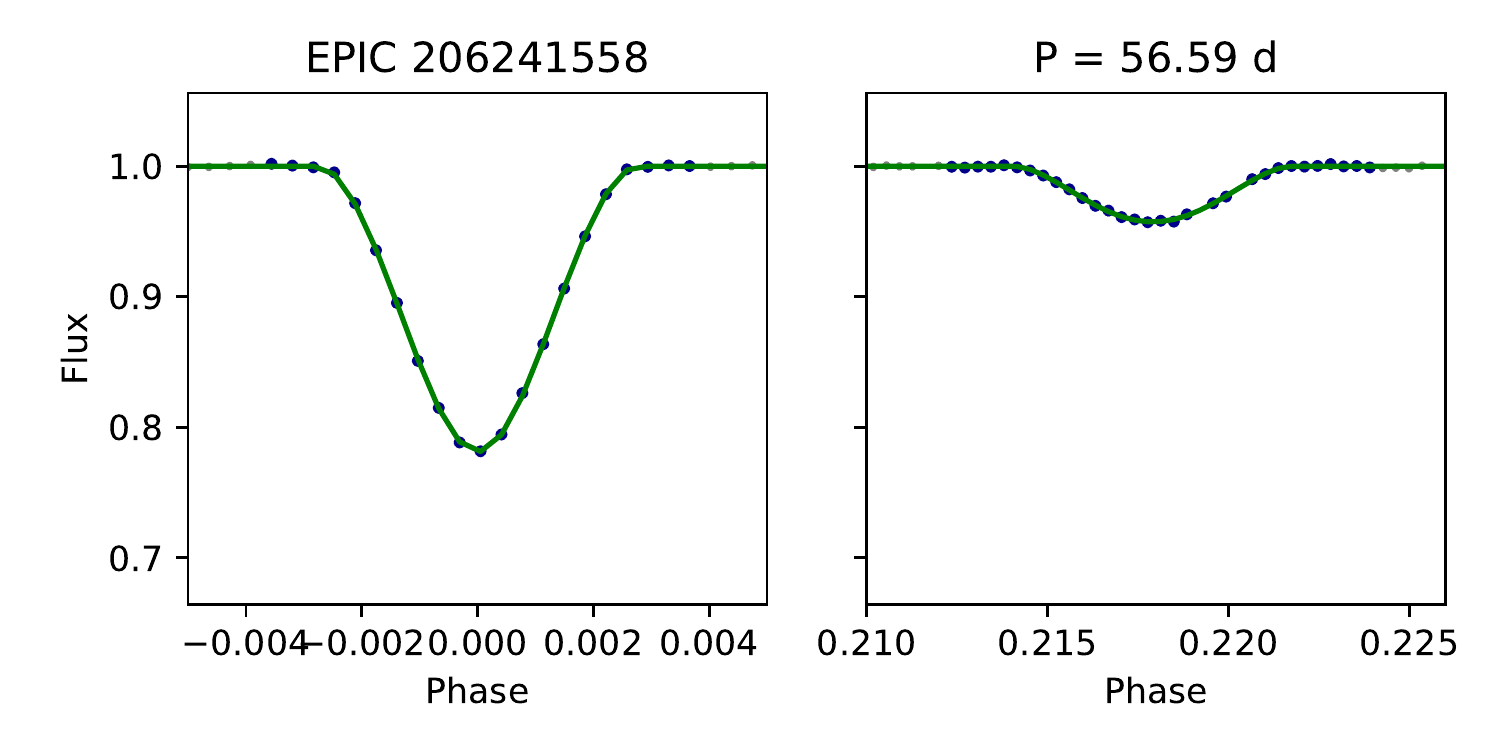} 
\includegraphics[width=0.49\textwidth]{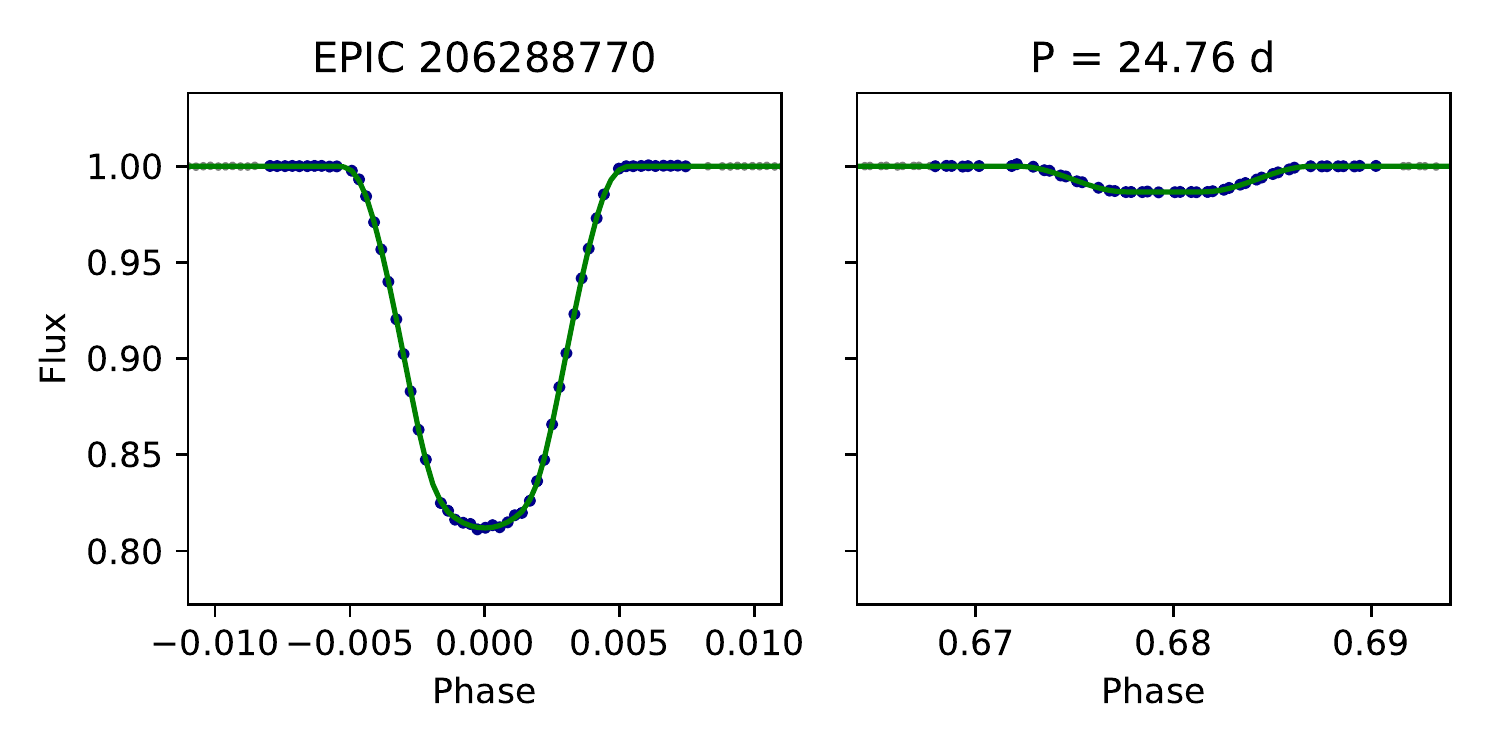} 
\includegraphics[width=0.49\textwidth]{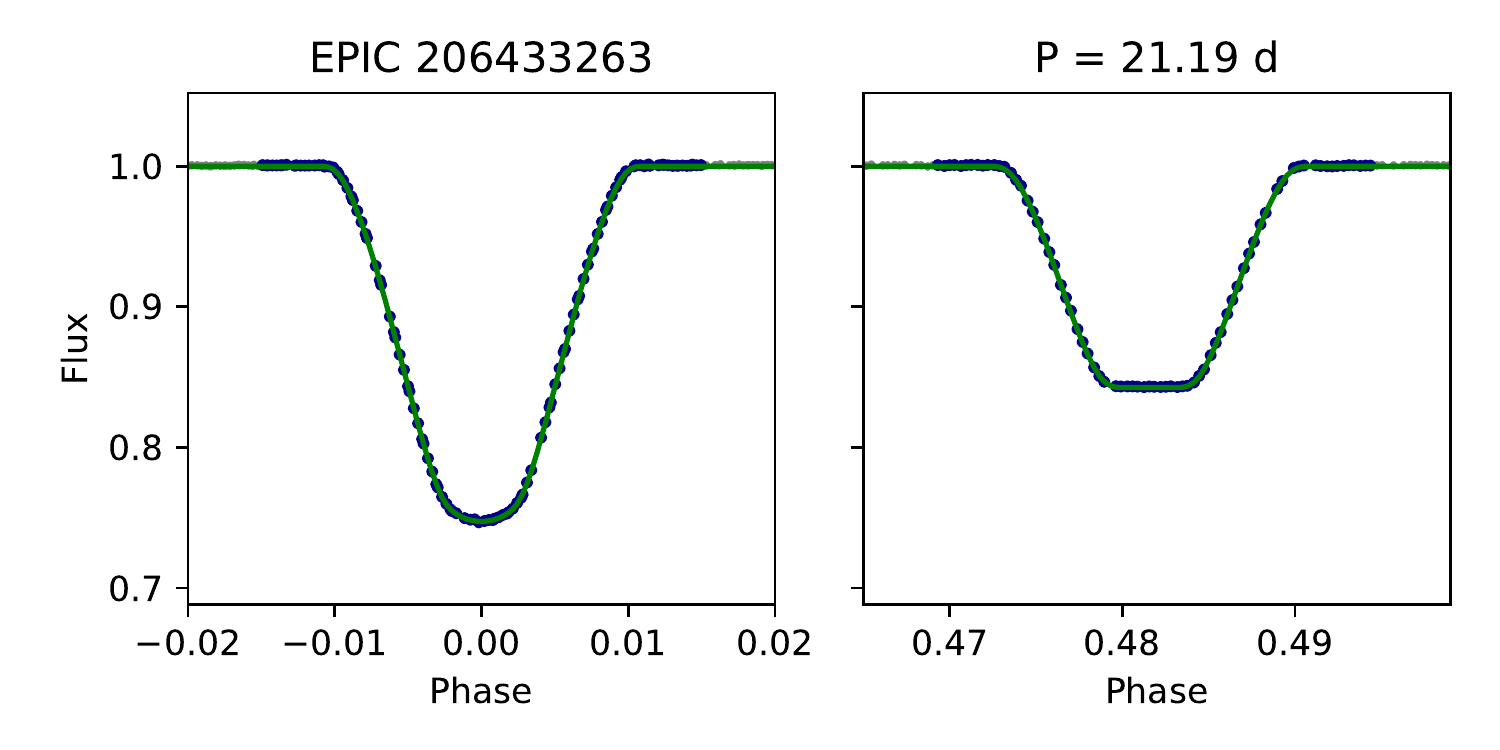} 
  \caption{K2 light curves with the best-fit \texttt{ellc} model. Data not
  included in the fit are plotted using small grey points.
  \label{lcfit5}}
\end{figure*}

\begin{table*}
\caption{Targets selected for further analysis from Kepler K2 campaigns 1, 2
and 3.}
\label{lcinfo}
\begin{center}
\begin{tabular}{@{}rrrrrrrrrll}
\hline
  \multicolumn{1}{@{}l}{EPIC} &
  \multicolumn{1}{l}{C} &
  \multicolumn{1}{l}{Kp} &
  \multicolumn{1}{l}{$P$\,[d]} &
  \multicolumn{1}{l}{$D_1$} &
  \multicolumn{1}{l}{$W_1$} &
  \multicolumn{1}{l}{$\phi_2$} &
  \multicolumn{1}{l}{$D_2$} &
  \multicolumn{1}{l}{$W_2$} &
  \multicolumn{1}{l}{Type} &
  \multicolumn{1}{l}{Notes} \\
\hline
201160323 & 1 & 18.23 & 22.18 & 0.34 & 0.008 & 0.385 & 0.21 & 0.010 & P & Rapid apsidal motion\\
201161715 & 1 & 14.65 & 59.89 & 0.10 & 0.014 & 0.701 & 0.09 & 0.016 & P & \\
201246763 & 1 & 11.93 & 43.68 & 0.35 & 0.008 & 0.258 & 0.31 & 0.014 & P & \\
201253025 & 1 & 12.86 &  6.79 & 0.26 & 0.042 & 0.494 & 0.24 & 0.046 & P & $P_{\rm rot}=6.5,7.2$\,d\\
201379113 & 1 & 14.76 & 21.21 & 0.34 & 0.009 & 0.369 & 0.02 & 0.011 & P & \\
201382417 & 1 & 11.66 &  5.20 & 0.07 & 0.048 & 0.500 & 0.01  & 0.048 & T & $P_{\rm rot}=5.2$\,d\\
201408204 & 1 & 11.85 &  8.48 & 0.42 & 0.031 & 0.611 & 0.39 & 0.036 & T & $P_{\rm rot}=7.4,8.4$\,d \\
201488365 & 1 &  8.81 &  6.73 & 0.36 & 0.048 & 0.500 & 0.36 & 0.048 & P & FM Leo\\
201576812 & 1 & 10.07 &  5.73 & 0.15 & 0.028 & 0.500 & 0.05 & 0.028 & P & TYC 272-458-1, $P_{\rm rot}=6.1$\,d\\
201648133 & 1 & 10.14 & 35.02 & 0.41 & 0.012 & 0.487 & 0.21 & 0.011 & T & Period constraint from WASP data\\
201665500 & 1 & 12.14 &  3.05 & 0.07 & 0.060 & 0.500 & 0.003& 0.060 & T & $P_{\rm rot}=3.2$\,d\\
201705526 & 1 &  9.94 & 18.10 & 0.06 & 0.016 & 0.656 & 0.01 & 0.012 & P & BD +04$^{\circ}$ 2479\\
201723461 & 1 & 14.91 & 22.73 & 0.34 & 0.009 & 0.345 & 0.06 & 0.010 & P & \\
202674012 & 2 &  9.77 & 23.31 & 0.24 & 0.022 & 0.497 & 0.22 & 0.021 & P &  HD 149946. FEROS spectra\\
202843085 & 2 & 11.80 & 16.50 & 0.38 & 0.030 & 0.448 & 0.38 & 0.034 & P & \\
203361171 & 2 & 11.92 &  7.32 & 0.11 & 0.038 & 0.489 & 0.11 & 0.038 & P & \\
203371239 & 2 & 11.74 & 20.36 & 0.35 & 0.025 & 0.500 & 0.35 & 0.022 & P & Pulsations\\
203543668 & 2 & 13.68 & 36.76 & 0.20 & 0.005 & 0.198 & 0.05 & 0.012 & P & $P_{\rm rot}=9.0$\,d\\
203610780 & 2 & 12.24 & 29.60 & 0.12 & 0.012 & 0.482 & 0.02 & 0.017 & T & \\
203636784 & 2 & 12.87 &  6.76 & 0.25 & 0.036 & 0.500 & 0.05 & 0.036 & T & $P_{\rm rot}=7.0$\,d\\
203728604 & 2 & 10.56 & 36.11 & 0.15 & 0.018 & 0.321 & 0.12 & 0.024 & T & $P_{\rm rot}=2.3$\,d (pulsations?)\\
204407880 & 2 & 12.14 & 34.37 & 0.10 & 0.010 & 0.718 & 0.03 & 0.021 & T & $P_{\rm rot}=13.3$\,d\\
204576757 & 2 & 13.67 & 23.28 & 0.19 & 0.010 &  --   & --   &  --   & N & \\
204748201 & 2 & 14.63 &  7.36 & 0.18 & 0.028 & 0.500 & 0.008& 0.028 & T & \\
204760247 & 2 &  5.95 &  9.20 & 0.38 & 0.040 & 0.500 & 0.15 & 0.040 & T & HD
142883. Pulsations. FEROS spectra\\
204822807 & 2 & 11.80 & 67.53 & 0.30 & 0.020 & 0.448 & 0.13 & 0.018 & T & \\
204870619 & 2 & 13.22 & 34.07 & 0.08 & 0.014 & 0.430 & 0.04 & 0.020 & T & \\
205020466 & 2 & 13.44 &  8.76 & 0.52 & 0.022 & 0.325 & 0.40 & 0.024 & P & $P_{\rm rot}=6.7,5.7$\,d, HRS spectra\\
205170307 & 2 & 12.28 & 67.50 & 0.14 & 0.008 & 0.304 & 0.03 & 0.009 & T & $P_{\rm rot}=17$\,d\\
205546169 & 2 & 11.66 & 24.44 & 0.24 & 0.009 & 0.265 & 0.01 & 0.014 & P & \\
205703649 & 2 & 12.49 &  8.12 & 0.08 & 0.040 & 0.500 & 0.07 & 0.040 & P & \\
205919993 & 3 & 10.14 & 11.00 & 0.06 & 0.012 & 0.552 & 0.06 & 0.013 & P & LP 819-72, $P_{\rm rot}=13.4$\,d \\
205982900 & 3 & 10.23 &  6.72 & 0.43 & 0.051 & 0.475 & 0.39 & 0.071 & P & BW Aqr, pulsations\\
206066862 & 3 & 10.28 & 11.09 & 0.13 & 0.015 & 0.384 & 0.03 & 0.016 & P & BD $-13^{\circ}$ 6219. $P_{\rm rot}= 12.2$\,d.\\
206066909 & 3 & 12.37 & 12.94 & 0.22 & 0.016 & 0.686 & 0.04 & 0.027 & T & \\
206075677 & 3 & 12.30 & 31.02 & 0.30 & 0.010 & 0.569 & 0.02 & 0.013 & P & $P_{\rm rot}=9.3,4.3,7.7$\,d. Triple?\\
206084435 & 3 & 14.65 & 48.22 & 0.19 & 0.008 & 0.443 & 0.04 & 0.009 & P & \\
206109641 & 3 & 12.38 & 62.59 & 0.46 & 0.008 & 0.123 & 0.40 & 0.008 & T & Period from K2+WASP data\\
206212261 & 3 & 12.70 & 30.99 & 0.10 & 0.016 & 0.525 & 0.02 & 0.014 & T & \\
206241558 & 3 & 13.38 & 56.59 & 0.28 & 0.005 & 0.218 & 0.05 & 0.008 & P & Period from K2+WASP data\\
206253908 & 3 & 11.18 & 65.45 & 0.10 & 0.005 & --    & --   & --    & N & Period from K2+WASP data\\
206288770 & 3 & 12.45 & 24.76 & 0.19 & 0.011 & 0.679 & 0.01 & 0.015 & T & \\
206433263 & 3 & 12.01 & 21.19 & 0.26 & 0.020 & 0.482 & 0.16 & 0.017 & T & \\
\hline
\end{tabular}
\end{center}
\tablefoot{ The campaign number is given in the column headed ``C''. The
characteristics of the light curve are noted as follows: $P$ -- orbital
period; $D_1$, $D_2$ -- eclipse depths as a fraction of the mean flux between
eclipses; $W_1$, $W_2$  -- eclipse widths (from first to last contact) in phase
units; $\phi_2$ phase of secondary eclipse in phase units. ``Type'' is used to
note light curves with partial eclipses (P), total eclipses (T), or with no
visible secondary eclipse (N). The estimated apparent magnitudes in the Kepler
band, Kp, are taken from the K2 Ecliptic Plane Input Catalog (EPIC)
\citep{2016ApJS..224....2H}. }
\end{table*}

 The two WASP instruments are located at the Observatorio del Roque de
los Muchachos, La Palma and at Sutherland Observatory, South Africa. Both
instruments carry an array of eight wide-field cameras, each with a 2048
$\times$ 2048 pixel CCD detector.  The majority of the survey has been
conducted using 200-mm, f/1.8 lenses combined with a filter that defines a
bandpass covering the wavelengths 400-700 nm \citep{2006PASP..118.1407P}. From
July 2012 the WASP-South instrument has used  85-mm, f/1.2 lenses with  SDSS
r$^{\prime}$ filters \citep{2014CoSka..43..500S}. A dedicated pipeline is used
to perform aperture photometry on the images at the position of catalogued
stars within the images. The data are then processed by a detrending algorithm
that has been developed from the SysRem algorithm of
\citet{2005MNRAS.356.1466T}, as described by \citet{2006MNRAS.373..799C}.

\subsection{Light curve modeling \label{ellc}}

 We used version 1.6.1 of the \texttt{ellc} light curve model
\citep{2016A&A...591A.111M} to determine the geometry and other parameters for
each binary system. Note that the definition of the ``third light'' parameter
used in this version of \texttt{ellc} to account for light from other stars in
the photometric aperture is different to the one described in
\citet{2016A&A...591A.111M}. In the new version, third light is described by
the parameter $\ell_3$. This parameter is used to calculate the flux $ {\cal
F}_3 = \ell_3 \left( {\cal F}_{\rm N,1} + {\cal F}_{\rm N,2}\right)$, where
${\cal F}_{\rm N,1}$ is the flux from star 1 emitted towards star 2 and vice
versa. This value of ${\cal F}_3$ is then used in the calculation of the
observed flux ${\cal F}_{i}$ at time $t_i$  as before, i.e., \[{\cal F}_{i} =
\frac{{\cal F}_{i,1} + {\cal F}_{i,2} + {\cal F}_3}{ {\cal F}_{\rm N,1} +
{\cal F}_{\rm N,2}+ {\cal F}_3},\] where ${\cal F}_{i,1}$ is the flux emitted
by star 1 towards the observer at time  $t_i$ and similarly for ${\cal
F}_{i,2}$. A complete list of changes in \texttt{ellc} version 1.6.1 is
provided in the file {\tt CHANGELOG.rst} provided with the package
distribution.\footnote{\url{https://pypi.python.org/ellc}}

The details of the analysis are not the same for every binary
system because some binary systems have peculiarities that required special
treatment.  Here we outline the main features of the analysis applied to the
majority of the systems analysed. Additional details and differences from
this general approach  for individual systems are described in Section
\ref{notes_sec}. 

The free parameters in the model for each binary were: the sum of radii
of the stars in units of the semi-major axis -- $r_{\rm sum} = (R_1+R_2)/a$,
the ratio of the radii -- $k = R_2/R_1$; the surface brightness ratio in the
Kepler band -- $S_{\rm Kp}$; the orbital inclination, $i$; the time of primary
eclipse -- $T_0$; the orbital period -- $P$;  $f_s = \sqrt{e}\sin(\omega)$ and
$f_c = \sqrt{e}\cos(\omega)$, where $e$ is the orbital eccentricity and
$\omega$ is the longitude of periastron; and ``third light'' -- $\ell_3$.

 We use $f_s$ and $f_c$ as parameters because a uniform prior probability
distribution for these parameters corresponds to a uniform prior probability
distribution for $e$. We use a quadratic limb-darkening law for both stars
with priors on the coefficients calculated using {\sc ldtk}
\citep{2015MNRAS.453.3821P} based on the spherical model atmospheres by
\citet{2013A&A...553A...6H}. To calculate these priors we assume $\log g =4.3
\pm 0.3$ and [Fe/H]$ = 0.0 \pm 0.2$ for all stars and effective temperature
estimates from a preliminary analysis very similar to those derived in below
in Section \ref{TeffSection}. The standard error estimates on the
coefficients inherited from the assumed errors on T$_{\rm eff}$, $\log g$ and
[Fe/H] are likely to be underestimates of the true uncertainties since they do
not account for systematic errors in the models and other issues with
estimating limb darkening coefficients from models
\citep{2011MNRAS.418.1165H}. To allow for this additional uncertainty we add
0.05 in quadrature to the standard error estimates for both coefficients. This
estimate of the systematic error in the coefficients dominates the error
budget for the limb darkening so we did not consider it necessary to
re-calculate these coefficients for the slightly different values of 
$T_{\rm eff}$ derived in Section~\ref{TeffSection} cf. our preliminary solution.
Rather than sampling the limb darkening coefficients $u_1$ and $u_2$ directly,
we use the parameters $q_1 = (u_1 + u_2)^2$ and $q_2 =  0.5u_1/(u_1 + u_2)$
since this makes it easier to uniformly sample the allowed parameter space
\citep{2013MNRAS.435.2152K}. Unless otherwise noted, we used spheres to model
the shape of these well-detached stars so gravity darkening was ignored. There
is little or no information about the geometry of the binary system in the
observations between the eclipses. For the light curve modeling of  most
stars we used only observations over a range 1.5 times the full eclipse width
centered on  each eclipse. This had the advantage of speeding up the
calculation. We used numerical integration of the eclipse model to account for
the exposure time of 1765\,s for data obtained near or during an eclipse.

 It is notoriously difficult to include star spots in the model for an
eclipsing binary star because the number of free parameters required is large
and the constraints on these parameters from the light curve are generally
weak and highly degenerate. We did not attempt to model star spots for any of
the binary systems here since the amplitude of the star spot modulation is
generally quite small ($\la 0.5\%$) so the resulting systematic error in the
parameters derived will, in general, not be large enough to alter our
conclusions regarding the nature of the binary. Instead, we simply divide-out
the time trend due to star spot modulation established from the Gaussian
process fit to the out-of-eclipse data.
 
We used {\sc emcee} \citep{2013PASP..125..306F}, a {\sc python} implementation
of an affine invariant Markov chain Monte Carlo (MCMC) ensemble sampler, to
calculate the posterior probability distribution of the model parameters. We
used an ensemble with at least twice the number samples per chain step
(``walkers'') as there were model parameters and 5,000 or 10,000 steps in the
chain used for the results quoted below. The convergence of the chain was
judged by visual inspection of the parameters and the likelihood as a function
of step number. In cases where we suspected the chain had not sampled the
posterior probability distribution accurately we calculated a new Markov Chain
starting from the best-fit parameters in the previous chain and using an
increased number of chain steps and/or an increased number of walkers with a
large spread of initial parameter values to ensure convergence. The standard
error per observation was either assumed to be constant for all the data, or
assumed to be constant within each of two blocks of data where there is a gap
in the observations. These values of the standard error were included as free
parameters in the MCMC analysis by including the necessary term in the
calculation of the likelihood for each chain step.  Unless otherwise stated,
we only use data within a range of 1.5 times the eclipse width (as listed in
Table~\ref{lcinfo}) centred on each eclipse in this analysis. This ensures
that these standard error estimates (and, hence, the error estimates on the
model parameters) are determined by the scatter in the residuals through the
eclipse, rather than the much lower scatter in the residuals between the
eclipses. From preliminary fits to the complete light curves we found that the
out-of-eclipse level is always very close to the value 1 with a very small
error and is not correlated with the other parameters so we fix this parameter
at 1 for the analysis presented here. 

 The aim of this analysis is to characterise each binary system in order to
identify systems of interest for further study and for comparison to binary
population models. The parameters we have derived are reliable enough for this
purpose  but further work is needed to determine the accuracy of these
parameters. The K2 data  clearly have the potential to produce very precise
parameters for some binary systems, but we have not attempted to characterise
the level of systematic error in these parameters for all the systems studied.
We advise that a careful study of these issues should be done before the
parameters of individual binary systems are used to test stellar evolution
models.

\subsection{Effective temperature estimates\label{TeffSection}}

 We have used empirical colour -- effective temperature and colour -- surface
brightness relations to estimate the effective temperatures of the individual
stars in the binary and triple systems we have studied. We extracted
photometry for each target from the following catalogues -- B$_{\rm T}$ and
V$_{\rm T}$ magnitudes from the Tycho-2 catalogue \citep{2000A&A...355L..27H};
B, V, g$^{\prime}$, r$^{\prime}$ and i$^{\prime}$ magnitudes from data release
9 of the AAVSO Photometric All Sky Survey
\citep[APASS9,][]{2016yCat.2336....0H}; J, H and K$_{\rm s}$ magnitudes from
the Two-micron All Sky Survey \citep[2MASS,][]{2006AJ....131.1163S};
i$^{\prime}$, J and K magnitudes from the Deep Near-infrared Southern Sky
Survey \cite[DENIS,][]{2005yCat.2263....0T}. Not all stars have data in all
these catalogues. Photometry from the Sloan digital sky survey (SDSS) can be
unreliable for these bright stars because they saturate the detectors, but we
have used  g$^{\prime}$-, r$^{\prime}$- and i$^{\prime}$-band ``psfMag''
magnitudes from data release 9 of the SDSS \citep{2012ApJS..203...21A} in some
cases, as noted in Table~\ref{TeffTable}. Magnitudes from the APASS9 catalogue
that are given with a standard error estimate of 0.00 were not included in our
analysis.

\begin{longtab}
\begin{longtable}{lrrrrrrrrrr}
\caption{\label{TeffTable} Effective temperature estimates from empirical
colour -- effective temperature and colour -- surface brightness relations and
constraints from the K2 light curve analysis.}\\
\hline
\hline
  \multicolumn{1}{@{}l}{EPIC} & 
  \multicolumn{1}{l}{g$^{\prime}_{0,1}$} &
  \multicolumn{1}{l}{T$_{\rm eff,1}$} &
  \multicolumn{1}{l}{g$^{\prime}_{0,2}$} &
  \multicolumn{1}{l}{T$_{\rm eff,2}$} &
  \multicolumn{1}{l}{g$^{\prime}_{0,3}$} &
  \multicolumn{1}{l}{T$_{\rm eff,3}$} &
  \multicolumn{1}{l}{E$({\rm B}-{\rm V})$} &
  \multicolumn{1}{l}{$\sigma_{\rm ext}$} &
  \multicolumn{1}{l}{$N_{\rm mag}$}\\
  \multicolumn{1}{l}{} &
  \multicolumn{1}{c}{[mag]} &
  \multicolumn{1}{c}{[K]} &
  \multicolumn{1}{c}{[mag]} &
  \multicolumn{1}{c}{[K]} &
  \multicolumn{1}{c}{[mag]} &
  \multicolumn{1}{c}{[K]} &
  \multicolumn{1}{c}{[mag]} &
  &
  \\
\hline
\endfirsthead
\caption{continued.} \\
\hline
\hline
  \multicolumn{1}{@{}l}{EPIC} & 
  \multicolumn{1}{l}{g$^{\prime}_{0,1}$} &
  \multicolumn{1}{l}{T$_{\rm eff,1}$} &
  \multicolumn{1}{l}{g$^{\prime}_{0,2}$} &
  \multicolumn{1}{l}{T$_{\rm eff,2}$} &
  \multicolumn{1}{l}{g$^{\prime}_{0,3}$} &
  \multicolumn{1}{l}{T$_{\rm eff,3}$} &
  \multicolumn{1}{l}{E$({\rm B}-{\rm V})$} &
  \multicolumn{1}{l}{$\sigma_{\rm ext}$} &
  \multicolumn{1}{l}{$N_{\rm mag}$}\\
  \multicolumn{1}{l}{} &
  \multicolumn{1}{c}{[mag]} &
  \multicolumn{1}{c}{[K]} &
  \multicolumn{1}{c}{[mag]} &
  \multicolumn{1}{c}{[K]} &
  \multicolumn{1}{c}{[mag]} &
  \multicolumn{1}{c}{[K]} &
  \multicolumn{1}{c}{[mag]} &
  &
  \\
\hline
\endhead
\hline
\endfoot
201161715\,\tablefootmark{a}         &  15.36 &  5030 & 17.15  & 5370 &       &      & 0.034 & 0.019 &   14  \\ 
\nopagebreak\multicolumn{1}{c}{$\pm$}&   0.10 &    85 &  0.21  &  120 &       &      & 0.025 & &       \\ 
 \noalign{\smallskip}                                                                          
201246763                            &  13.23 & 5875  & 12.66  & 6225 &       &      & 0.026 & 0.000 &   11  \\
\nopagebreak\multicolumn{1}{c}{$\pm$}&   0.10 &  110  &  0.09  &  125 &       &      & 0.024 & &       \\  
 \noalign{\smallskip}                                                                          
201253025                            &  13.82 & 6065  & 13.92  & 6070 &       &      & 0.031 & 0.009&   9 & \\
\nopagebreak\multicolumn{1}{c}{$\pm$}&   0.17 &  155  &  0.18  &  155 &       &      & 0.023 & &     & \\
 \noalign{\smallskip}                                                                          
201379113\,\tablefootmark{a}         &  15.25 & 5150  & 19.9   & 3900 &       &      & 0.039 & 0.030 &  14 & \\
\nopagebreak\multicolumn{1}{c}{$\pm$}&   0.12 &  115  &  1.6   &  230 &       &      & 0.028 & &     & \\
 \noalign{\smallskip}                                                                          
201382417                            &  13.29 & 6175  & 17.08  & 4480 & 12.54 & 4950 & 0.055 & 0.035 &  12 & \\
\nopagebreak\multicolumn{1}{c}{$\pm$}&   0.16 &  400  &  0.16  &  200 &  0.14 &  140 & 0.030 & &     & \\
 \noalign{\smallskip}                                                                          
201408204                            &  12.84 & 5845  & 12.94  & 5830 &       &      & 0.030 & 0.002 &  10 & \\ 
\nopagebreak\multicolumn{1}{c}{$\pm$}&   0.09 &  105  &  0.09  &  100 &       &      & 0.024 & &     & \\
 \noalign{\smallskip}                                                                          
201488365\,\tablefootmark{c}         &  9.21  & 6355  & 9.40   & 6345 &       &      & 0.023 & 0.004 &  10 & \\ 
\nopagebreak\multicolumn{1}{c}{$\pm$}&  0.10  &  150  & 0.10   &  150 &       &      & 0.024 & &     & \\
 \noalign{\smallskip}                                                                          
201576812                            &  10.49 & 5905  & 13.1   & 4360 &       &      & 0.037 & 0.061 &  13 & \\ 
\nopagebreak\multicolumn{1}{c}{$\pm$}&   0.11 &  190  &  0.5   &  140 &       &      & 0.023 & &     & \\
 \noalign{\smallskip}                                                                          
201648133                            &  10.65 & 6010  & 12.32  & 5250 &       &      & 0.028 & 0.020 &  11 & \\ 
\nopagebreak\multicolumn{1}{c}{$\pm$}&   0.07 &   95  &  0.07  &   70 &       &      & 0.019& &     & \\
 \noalign{\smallskip}                                                                          
201665500                            &  12.30 & 6270  & 19.4   & 3630 &       &      & 0.028 & 0.001 &   9 & \\ 
\nopagebreak\multicolumn{1}{c}{$\pm$}&   0.09 &  120  &  0.9   &   60 &       &      & 0.022 & &     & \\
 \noalign{\smallskip}                                                                          
201705526                            &  11.71 & 6600  & 15.93  & 4320 & 10.24 & 6020 & 0.033 & 0.048 &  11 & \\ 
\nopagebreak\multicolumn{1}{c}{$\pm$}&   0.15 &  430  &  0.16  &  195 &  0.11 &  170 & 0.024 & &     & \\
 \noalign{\smallskip}                                                                          
201723461\,\tablefootmark{a}         &  16.26 & 4450  & 17.0   & 4170 &       &      & 0.050 & 0.023 &  12 & \\ 
\nopagebreak\multicolumn{1}{c}{$\pm$}&   0.23 &  140  &  0.4   &  160 &       &      & 0.030 & &     & \\
 \noalign{\smallskip}                                                                          
202674012\,\tablefootmark{d}         &  10.0  &  6250 & 11.3   & 6150 &       &      & 0.096 & 0.058 &  11 & \\ 
\nopagebreak\multicolumn{1}{c}{$\pm$}&   0.2  &   285 &  0.2   &  285 &       &      & 0.065 & &     & \\
 \noalign{\smallskip}                                                                          
202843085                            &  12.4  & 6300  & 12.2   & 6260 &       &      & 0.172 & 0.002 &  12 & \\ 
\nopagebreak\multicolumn{1}{c}{$\pm$}&   0.2  &  280  &  0.2   &  280 &       &      & 0.060 & &     & \\
 \noalign{\smallskip}                                                                          
203361171                            &  12.4  & 6070  & 12.3   & 6050 &       &      & 0.148 & 0.009 &  14 & \\ 
\nopagebreak\multicolumn{1}{c}{$\pm$}&   0.3  &  290  &  0.3   &  280 &       &      & 0.064 & &     & \\
 \noalign{\smallskip}                                                                          
203371239                            &  11.6  & 6400  & 12.0   & 6300 & 16.3  & 3660 & 0.378 & 0.019 &   9 & \\ 
\nopagebreak\multicolumn{1}{c}{$\pm$}&   0.3  &  380  &  0.3   &  380 &  0.7  &  335 & 0.084 & &     & \\
 \noalign{\smallskip}                                                                          
203543668                            &  14.2  & 5900  & 16.8   & 4600 & 13.92 & 5300 & 0.267 & 0.002 &  11 & \\ 
\nopagebreak\multicolumn{1}{c}{$\pm$}&   0.3  &  550  &  0.3   &  315 &  0.25 &  310 & 0.066 & &     & \\
 \noalign{\smallskip}                                                                          
203610780                            &  14.2  & 6650  & 13.4   & 5600 & 13.3  & 5700 & 0.058 & 0.055 &  8  & \\ 
\nopagebreak\multicolumn{1}{c}{$\pm$}&   0.3  &  350  &  0.3   &  300 &  1.1  &  725 & 0.048 & &     & \\
 \noalign{\smallskip}                                                                          
203636784                            &  13.02 & 5970  & 17.1   & 4350 &       &      & 0.099 & 0.032 &   9 & \\ 
\nopagebreak\multicolumn{1}{c}{$\pm$}&   0.17 &  200  &  0.2   &  100 &       &      & 0.044 & &     & \\
 \noalign{\smallskip}                                                                          
203728604\,\tablefootmark{g}         &  10.98 & 6050  & 13.27  & 5840 &       &      & 0.019 & 0.018 &  11 & \\ 
\nopagebreak\multicolumn{1}{c}{$\pm$}&   0.07 &  100  &  0.07  &   95 &       &      & 0.019 & &     & \\
 \noalign{\smallskip}                                                                          
204407880                            &  12.1  & 5765  & 15.7   & 4900 &       &      & 0.188 & 0.004 &  14 & \\ 
\nopagebreak\multicolumn{1}{c}{$\pm$}&   0.2  &  205  &  0.2   &  145 &       &      & 0.054 & &     & \\
 \noalign{\smallskip}                                                                          
204748201                            &  14.7  & 6100  & 20.8   & 3600 & 17.1  & 5400 & 0.168 & 0.027 &   8 & \\ 
\nopagebreak\multicolumn{1}{c}{$\pm$}&   0.2  &  235  &  0.2   &  110 &  0.6  & 1100 & 0.050 & &     & \\
 \noalign{\smallskip}                                                                          
204822807\,\tablefootmark{b}         &  13.3  & 5625  & 12.9   & 4620 & 15.3  & 4665 & 0.087 & 0.016 &  12 & \\ 
\nopagebreak\multicolumn{1}{c}{$\pm$}&   0.2  &  215  &  0.2   &  135 &  0.9  &  350 & 0.055 & &     & \\
 \noalign{\smallskip}                                                                          
204870619                            &  13.3  & 5435  & 17.0   & 4800 &       &      & 0.162 & 0.005 &  12 & \\ 
\nopagebreak\multicolumn{1}{c}{$\pm$}&   0.3  &  250  &  0.3   &  190 &       &      & 0.070 & &     & \\
 \noalign{\smallskip}                                                                          
205020466                            &  13.3  & 5300  & 13.7   & 5070 &       &      & 0.670 & 0.021 &   9 & \\ 
\nopagebreak\multicolumn{1}{c}{$\pm$}&   0.5  &  480  &  0.5   &  425 &       &      & 0.145 & &     & \\
 \noalign{\smallskip}                                                                          
205170307\,\tablefootmark{b}         &  12.49 & 5620  & 16.79  & 4240 &  16.1 & 4490 & 0.126 & 0.003 &   9 & \\ 
\nopagebreak\multicolumn{1}{c}{$\pm$}&   0.08 &   65 &   0.09 &    40 &   0.2 &   90 & 0.022 & &     & \\
 \noalign{\smallskip}                                                                          
205546169\,\tablefootmark{b}         &  13.1  & 6300  &  12.0  & 6170 &  16.1 & 5050 & 0.100 & 0.003 &  14 & \\ 
\nopagebreak\multicolumn{1}{c}{$\pm$}&   0.2  &  210  &   0.2  &  210 &   1.0 &  500 & 0.045 & &     & \\
 \noalign{\smallskip}                                                                          
205703649                            &  12.9  & 5610  & 12.8   & 5540 &       &      & 0.256 & 0.000 &   8 & \\ 
\nopagebreak\multicolumn{1}{c}{$\pm$}&   0.4  &  425  &  0.4   &  415 &       &      & 0.103& &     & \\
 \noalign{\smallskip}                                                                          
205919993                            &  12.2  & 4025  &  11.8  & 4230 &       &      & 0.027 & 0.077 & 10  & \\ 
\nopagebreak\multicolumn{1}{c}{$\pm$}&   0.1  &   60  &   0.1  &   60 &       &      & 0.021 & &     & \\
 \noalign{\smallskip}                                                                          
205982900\,\tablefootmark{h}         &  11.49 & 6200  &  11.21 & 6045 &       &      & 0.011 & 0.070 &  11 & \\ 
\nopagebreak\multicolumn{1}{c}{$\pm$}&   0.07 &  115  &   0.07 &  110 &       &      & 0.013 & &     & \\
 \noalign{\smallskip}                                                                          
206066862\,\tablefootmark{e}         &  10.8  & 6250  & 12.4   & 5000 &       &      & 0.043 & 0.048 &  11 & \\ 
\nopagebreak\multicolumn{1}{c}{$\pm$}&   0.2  &  300  &  0.7   &  200 &       &      & 0.032 & &     & \\
 \noalign{\smallskip}                                                                          
206066909                            &  12.91 & 6440  &  16.66 & 4535 & 13.95 & 5225 & 0.057 & 0.020 &   9 & \\ 
\nopagebreak\multicolumn{1}{c}{$\pm$}&   0.13 &  290  &   0.12 &  140 &  0.19 &  360 & 0.031 & &     & \\
 \noalign{\smallskip}                                                                          
206084435                            &  14.91 & 5950  &  18.91 & 4300 &       &      & 0.042 & 0.001 &   8 & \\ 
\nopagebreak\multicolumn{1}{c}{$\pm$}&   0.11 &  135  &   0.15 &  140 &       &      & 0.030 & &     & \\
 \noalign{\smallskip}                                                                          
206109641\,\tablefootmark{e}         &  13.18 & 5905  & 13.65  & 5805 &       &      & 0.034 & 0.001 &  11 & \\ 
\nopagebreak\multicolumn{1}{c}{$\pm$}&   0.10 &  120  &  0.10  &  115 &       &      & 0.025 & &     & \\
 \noalign{\smallskip}                                                                          
206212261                            &  13.15 & 5385  & 18.24  & 4010 &       &      & 0.028 & 0.001 &  14 & \\ 
\nopagebreak\multicolumn{1}{c}{$\pm$}&   0.09 &   85  &  0.09  &   50 &       &      & 0.023 & &     & \\
 \noalign{\smallskip}                                                                          
206241558                            &  14.31 & 5330  & 14.74  & 5885 &       &      & 0.052 & 0.006 &   9 & \\ 
\nopagebreak\multicolumn{1}{c}{$\pm$}&   0.15 &  120  &  0.21  &  160 &       &      & 0.031 & &     & \\
 \noalign{\smallskip}                                                                          
206288770                            &  12.45 & 6290  & 18.02  & 3870 &       &      & 0.078 & 0.007 &  12 & \\ 
\nopagebreak\multicolumn{1}{c}{$\pm$}&   0.09 &  120  &  0.09  &   55 &       &      & 0.024 & &     & \\
 \noalign{\smallskip}                                                                          
206433263                            &  12.52 & 6000  & 14.46  & 5525 &       &      & 0.034 & 0.018 &  14 & \\
\nopagebreak\multicolumn{1}{c}{$\pm$}&   0.08 &   90  &  0.08  &   75 &       &      & 0.020 & &     & \\
\noalign{\smallskip}
\hline
\end{longtable}
\tablefoot{
\tablefoottext{a}{SDSS  g$^{\prime}$, r$^{\prime}$ and i$^{\prime}$ magnitudes
included.} 
\tablefoottext{b}{Third-light contribution assumed to come from a
main-sequence star at the same distance as the eclipsing binary pair.}
\tablefoottext{c}{g$^{\prime}$ magnitude estimated from APASS9 B and V
magnitudes included with nominal 0.5 magnitude error.}
\tablefoottext{d}{g$^{\prime}$ magnitude estimated from Tycho-2 B$_{\rm T}$ and
V$_{\rm T}$ magnitudes included with nominal  0.5 magnitude error.} 
\tablefoottext{e}{DENIS data excluded from fit. }
\tablefoottext{f}{SDSS  g$^{\prime}$ and r$^{\prime}$  magnitudes
included.} 
\tablefoottext{g}{APASS9 g$^{\prime}$ magnitude included with nominal 0.5 magnitude error.}
\tablefoottext{h}{2MASS data excluded.}
 The value of $\sigma_{\rm ext}$
is taken from the maximum likelihood solution.
$N_{\rm mag}$ is the number of apparent magnitude measurements used in the
analysis. }
\end{longtab}

 Our model for the observed photometry then has the following  free
parameters that are determined by a least-squares fit to the observed
apparent magnitudes and other data for each system -- g$^{\prime}_{0, i}$,
the apparent g$^{\prime}$-band magnitudes for stars $i=1$, $i=2$ and (for
triple systems) $i=3$, corrected for extinction; $T_{\rm eff,i}$ the effective
temperatures for each star in the binary or triple system; E$({\rm B}-{\rm
V})$, the reddening to the system; $\sigma_{\rm ext}$ the additional
systematic error added in quadrature to each synthetic magnitude to account
for systematic errors in the conversion to observed magnitudes.

 For each trial combination of these parameters we use the empirical colour --
effective temperature relations by \citet{2013ApJ...771...40B} to predict the
apparent magnitudes for each star in each of the observed bands. We used the
same transformation between the Johnson and 2MASS photometric systems as
Boyajian et al.. We used the Cousins I$_{\rm C}$ band as an approximation to
the DENIS Gunn i$^{\prime}$ band and the 2MASS K$_{\rm s}$ band as an
approximation to the DENIS K band \citep[see Fig. 4][]{2005ARA+A..43..293B}.
We used interpolation in Table 3 of \citet{2000PASP..112..961B} to transform
the Johnson B, V magnitudes to Tycho-2 B$_{\rm T}$ and V$_{\rm T}$ magnitudes.
We assume that the extinction in the V band is $3.1\times {\rm E}({\rm B}-{\rm
V})$. Extinction in the SDSS and 2MASS bands is calculated using A$_{\rm r} =
2.770\times {\rm E}({\rm B}-{\rm V})$ from \cite{2003A&A...401..781F} and
extinction coefficients relative to the r$^{\prime}$ band from
\citet{2014MNRAS.440.3430D}.

 We use the transformation from Sloan g$^{\prime}$, r$^{\prime}$ and
i$^{\prime}$ magnitudes by \citet{2011AJ....142..112B} to estimate Kepler
Kp magnitudes for each star in the system. This enables us to include
the flux ratio $\ell_{\rm Kp}$ as a constraint in the analysis of the
published photometry. Another useful constraint is the surface brightness
ratio in the Kepler band, $S_{\rm Kp}$, which we account for by using the
empirical relation between the V-band surface brightness $S_{\rm V}$ and $(\rm
B - \rm K)$ from \citet{2017ApJ...837....7G}. The comparison between the
predicted and observed values is done in terms of the surface brightness
parameter \[ S_i = m_{i,0} + 5\log\phi, \] where $i$ denotes a particular band
(V or Kp), $\phi$ is the angular diameter in milli-arcseconds, and $m_{i,0}$ is
the de-reddened apparent magnitude in a given band, so that $S_{\rm Kp} =
S_{\rm V} + ({\rm Kp}-{\rm V})$.

 We used {\sc emcee} \citep{2013PASP..125..306F} to sample the posterior
probability distribution for our model parameters. We used the reddening maps
by \citet{2011ApJ...737..103S} to estimate the total line-of-sight extinction
to each target, ${\rm E}({\rm B}-{\rm V})_{\rm map}$. This value is used to
impose the following (unnormalized) prior on $\Delta = {\rm E}({\rm B}-{\rm
V}) - {\rm E}({\rm B}-{\rm V})_{\rm map}$: 
\[ P(\Delta) = \left\{ 
\begin{array}{ll}
  1 & \Delta \le 0 \\ 
  \exp(-0.5(\Delta/0.034)^2) & \Delta > 0 \\ 
\end{array} \right. 
\] 
The constant 0.034 is taken from \citet{2014MNRAS.437.1681M} and is based on a
comparison of ${\rm E}({\rm B}-{\rm V})_{\rm map}$ to ${\rm E}({\rm B}-{\rm
V})$ from Str\"{o}mgren photometry for 150 A-type stars.  A least-squares
optimisation algorithm was used to find an initial set of parameters for the
chain and the Markov chains were calculated using 64 walkers and 256 steps
following a burn-in run of 128 steps. An example of the output from the
program used to implement our method is shown in Fig.~\ref{fitmag}.

\begin{figure*}
\begin{verbatim}

Read 17 lines from EPIC201161715.phot

 Calculating least-squares solution...
  g_1     = 15.50
  T_eff,1 = 4931 K
  g_2     = 17.25
  T_eff,2 = 5264 K
  E(B-V)  =   0.00
  chi-squared = 3.44
  Ndf = 12
  sigma_ext = 0.05

 Starting emcee chain of 384 steps with 64 walkers

 Median acceptance fraction = 0.487
 Best log(likelihood) = 42.52 in walker 12 at step 155

 Parameter median values, standard deviations and best-fit values.
  g_1     = 15.362 +/-  0.102   [ 15.460 ]
  T_eff,1 =   5031 +/-   84 K     [ 4958 ]
  g_2     = 17.148 +/-  0.207   [ 17.231 ]
  T_eff,2 =   5366 +/-  120 K     [ 5282 ]
  E(B-V)  =  0.034 +/-  0.025   [  0.011 ]
  sig_ext =  0.027 +/-  0.016   [  0.019 ]
  chi-squared = 9.90

type band value_obs error   source  value_fit value_A  value_B   z
---- ---- --------- ------ -------- --------- -------- -------- ----
 mag    B    15.759  0.208   APASS9   15.7578  15.9591  17.6869 0.01
 mag    H    12.769  0.021    2MASS   12.8287  12.9722  15.0962 2.11
 mag   Ic    13.916   0.03    DENIS   13.9310  14.0947  16.0660 0.42
 mag    J    13.227   0.07    DENIS   13.2451  13.3976  15.4518 0.25
 mag    J    13.283  0.027    2MASS   13.2451  13.3976  15.4518 1.15
 rat  K_p      0.18   0.03 paper-v3    0.1771  14.8825  16.7617 0.10
 sb2  K_p       1.4    0.1 paper-v3    1.3832   5.3211   4.9689 0.17
 mag   Ks    12.602   0.14    DENIS   12.6763  12.8175  14.9610 0.53
 mag   Ks    12.725   0.03    2MASS   12.6763  12.8175  14.9610 1.37
 mag    V    14.881  0.024   APASS9   14.8917  15.0734  16.9224 0.35
 mag    g   15.2956  0.019     SDSS   15.3090  15.5032  17.2736 0.50
 mag    g    15.259   0.11   APASS9   15.3090  15.5032  17.2736 0.45
 mag    i    14.413   0.05   APASS9   14.3643  14.5308  16.4828 0.91
 mag    i   14.3639 0.0192     SDSS   14.3643  14.5308  16.4828 0.02
 mag    r    14.581  0.103   APASS9   14.5961  14.7691  16.6755 0.14
 mag    r   14.6083 0.0133     SDSS   14.5961  14.7691  16.6755 0.53
 Nobs = 17
 Nmag = 14
 Ndf  = 11
 BIC = -68.04

Completed analysis of EPIC201161715.phot
\end{verbatim}
\caption{Example output from our program to estimate T$_{\rm eff}$ for the
  stars in an eclipsing binary from a least-squares fit to the
  observed apparent magnitudes and other constraints.
  \label{fitmag}}
\end{figure*}

 There will be some systematic error in the T$_{\rm eff}$ estimates for stars
in  eclipsing binaries  cooler than 4900\,K because we have extrapolated the
empirical $S_{\rm V}$ -- $(\rm B - \rm K)$ relation in this regime.  The
empirical colour -- temperature relations we have used are valid over the
approximate range T$_{\rm eff} = 3450$\,K to 8600\,K. Our results may be
biased in systems where one of the stars has an effective temperature near
either of these limits because we exclude trial solutions with any T$_{\rm
eff, i}$ value outside this range. Between these limits we use uniform priors
on the values of T$_{\rm eff, i}$. We also use uniform priors for
g$^{\prime}_{0, 1}$ and g$^{\prime}_{0, 2}$.

 In systems where there is evidence of third light from the light curve
analysis and the star appears unresolved in sky survey images we compare
solutions with a uniform prior on g$^{\prime}_{0, 3}$ and with a constraint on
g$^{\prime}$$_{0, 3}$ assuming that the third light is due to a main-sequence
star at the same distance as the eclipsing binary star.  We use the stellar
model from the Dartmouth stellar evolution database
\citep{2008ApJS..178...89D} for solar composition to define the limits of the
main sequence in the T$_{\rm eff}$ -- $M_{\rm g^{\prime}}$ plane, where
$M_{\rm g^{\prime}}$ is the absolute magnitude of star $i$ in the g$^{\prime}$
band. For each trial solution we use interpolation between these model
isochrones to define limits to g$^{\prime}_{0, 3}$ assuming that the fainter
star in the eclipsing binary is a main-sequence star, i.e., we reject
solutions where the combination of T$_{\rm eff, 3}$, T$_{\rm eff, B}$,
g$^{\prime}_{0, 3}$ and g$^{\prime}_{0, B}$ cannot be reproduced by two stars
between the zero-age main sequence and terminal-age main-sequence in the
T$_{\rm eff}$ -- $M_{{\rm g}^{\prime}}$ plane, where  $B = 1$ or 2 is the
index for the star in the eclipsing binary that is fainter in the
g$^{\prime}$-band. Systems where we adopted solutions including this
constraint are noted with a $\star$ symbol in Table~\ref{TeffTable}, together
with the median and standard deviation of the model parameters derived using
{\sc emcee}.

Our method requires an estimate of the apparent g$^{\prime}$ magnitude. In
cases where no such estimate is available from APASS9 we either use the SDSS
g$^{\prime}$ magnitude or infer a value from the Tycho-2 B$_{\rm T}$ and
V$_{\rm T}$ magnitudes using equation (6a) from \citet{2011AJ....142..112B}.
In either case, we assign an nominal standard error of 0.5 magnitudes to this
estimate. We also found for some stars that the magnitudes from the DENIS and
2MASS surveys were significantly different. In general, we used the 2MASS
magnitudes in these cases and excluded the DENIS photometry from the fit --
these cases are noted in Table~\ref{TeffTable}.

\section{Results}
\begin{figure*} 
  \includegraphics[width=0.99\textwidth]{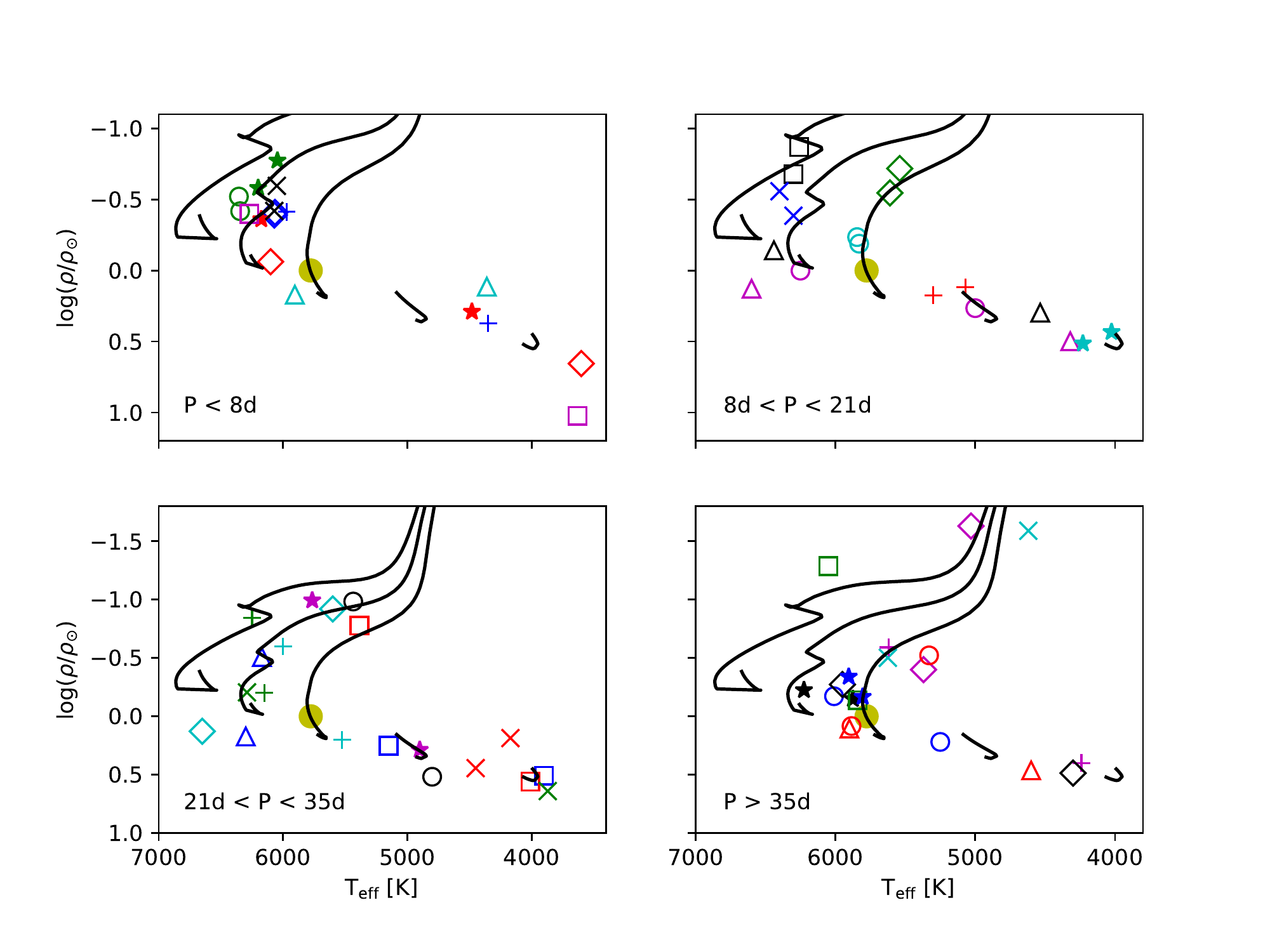}
\caption{Targets in the effective temperature -- mean stellar 
density plane compared to stellar evolution tracks from the  Dartmouth stellar
evolution database for solar composition \citep{2008ApJS..178...89D}. The
evolution tracks are truncated at an age of 13\,Gyr. The location of the Sun in
this plane is shown with a filled yellow circle. Stars from the same binary
system are plotted using the same symbol. Evolution tracks are shown for
stellar masses from 0.6$M_{\sun}$ to 1.4$M_{\sun}$ in steps of 0.2$M_{\sun}$.
Stars are plotted in one of four panels according to the orbital period of the
binary, as noted in each panel.}
\label{TeffDensity} 
\end{figure*}

 The parameters derived from our analysis of the K2  light curves for each
 target  are given in Tables~\ref{ellcpar1} and \ref{ellcpar2}. The best fits
 to the K2 light curves are shown in Figs. \ref{lcfit1}, \ref{lcfit2},
 \ref{lcfit3} and \ref{lcfit4}. The effective temperature estimates for the
 components of each system are given in Table~\ref{TeffTable}. In
 Table~\ref{TrhoTable} we give an estimate of the mean stellar density
 ($\rho_{\star}$) for the two stars in each eclipsing binary calculated using
 the following expression derived from Kepler's third law.
\begin{equation*}
  \mbox{$\rho_{\star}$} =
   \frac{3\mbox{M$_{\star}$}}{4\pi R_{\star}^3} =
   \frac{3\pi}{GP^2(1+q)}
    \left(\frac{a}{\mbox{R$_{\star}$}}\right)^3. 
  \end{equation*}
  Here, $P$ and $a$ are the period and semi-major axis of the Keplerian orbit,
and $q = M_\mathrm{c}/\mbox{$M_{\star}$}$ is the mass ratio for a companion
with mass $M_\mathrm{c}$ to a star with mass $M_{\star}$ and radius
$R_{\star}$. The value of $q$ was estimated by calculating the position of
the stars in Fig.~\ref{TeffDensity} for various values of $q$ and then choosing
the value which is consistent with the approximate masses inferred from the
stellar evolution tracks shown in this figure. 

\begin{table*}
\caption{Geometric and orbital parameters derived from the analysis of the K2
light curves for selected long-period eclipsing binary stars.
\label{ellcpar1}}
\begin{center}
  \begin{tabular}{@{}rrrrrrrr}
\hline
    \multicolumn{1}{@{}l}{EPIC} &
\multicolumn{1}{l}{$(R_1+R_2)/a$} &
  \multicolumn{1}{l}{$R_2/R_1$} &
  \multicolumn{1}{l}{$i$ [$^{\circ}$] }&
  \multicolumn{1}{l}{$T_0$} &
  \multicolumn{1}{l}{$P$} &
  \multicolumn{1}{l}{$f_c$} &
  \multicolumn{1}{l}{$f_s$} \\
\hline
201160323& 0.0276(2)  &  1.10(6)   & 89.80(5) &   2023.026(1)  & 22.200(6)  &$-0.406(2)  $&$ 0.19(1)  $ \\  
201161715& 0.061(1)   &  0.36(1)   & 87.64(8) &  2007.4140(9)  & 59.887(2)  &$ 0.542(4)  $&$ 0.22(2)  $ \\  
201246763& 0.0351(1)  &  1.13(2)   & 89.40(1) &  2014.32552(9) &=43.68281(3)&$-0.504(1)  $&$ 0.514(3) $ \\ 
201253025& 0.1418(7)  &  1.0(1)    & 87.4(2)  &  2011.3371(3)  & 6.78651(9) &$-0.038(2)  $&$ 0.21(1)  $ \\ 
         & 0.1425(8)  &  0.95(8)   & 87.3(2)  &  2011.3367(3)  & 6.78635(8) &$-0.038(2)  $&$ 0.21(1)  $ \\ 
201379113& 0.037(2)   &  0.8(1)    & 88.7(1)  & 1989.2348(2)   & 21.2146(1) &$-0.3(1)    $&$ 0.5(1)   $ \\ 
201382417& 0.1350(5)  &  0.51(1)   & 88.8(2)  & 1983.27151(7)  &5.197721(9) &$-0.008(4)  $&$0.01(3)   $ \\ 
201408204& 0.1067(2)  &  0.97(2)   & 88.95(1) & 2002.45835(4)  &8.48185(1)  &$0.388(1)   $&$0.226(5)  $ \\ 
201488365& 0.1518(1)  & 0.917(6)   & 87.96(1) & 2019.59540(1)  &6.728609(3) &$0.005(2)   $&$0.010(9)  $ \\ 
201576812& 0.106(3)   & 0.9(2)     & 85.7(3)  & 2015.96974(7)  &5.72830(2)  &$0.001(2)   $&$ 0.11(8)  $ \\ 
201648133&0.03513(1)  & 0.687(2)   &89.734(4) & 2015.81404(1)  &=35.02402(1)&$-0.0941(8) $&$-0.192(2) $ \\ 
201665500& 0.171(1)   & 0.2474(9)  & 89.9(5)  & 2017.2344(1)   & 3.05351(2) &$-0.00(2)   $&$0.03(6)   $ \\ 
201705526& 0.0436(3)  & 0.599(8)   & 89.24(4) & 2004.71250(3)  &=18.102928  &$ 0.476(2)  $&$-0.206(7) $ \\ 
201723461& 0.04(1)    & 1.1(3)     & 88(1)    & 1997.146(2)    & 22.731(1)  &$ -0.7(2)   $&$ 0.3(2)   $ \\ 
202674012&0.0718(1)   & 0.562(2)   & 88.65(1) & 2076.36053(5)  &23.30962(5) &$-0.0295(9) $&$-0.162(5) $ \\ 
202843085& 0.1057(2)  & 1.16(1)    & 88.84(2) & 2077.35748(9)  & 16.49843(5)&$-0.259(2)  $&$ 0.184(6) $ \\ 
203361171& 0.148(6)   & 1.1(2)     & 85.0(7)  & 2068.8147(8)   & 7.3216(2)  &$-0.130(8)  $&$-0.02(4)  $ \\ 
203371239&0.0730(4)   &0.88(5)     & 88.98(7) & 2078.9883(3)   & 20.3618(3) &$-0.0034(3) $&$-0.20(1)  $ \\ 
203543668& 0.0278(3)  & 0.68(2)    & 89.7(2)  & 2099.4015(3)   & 36.7623(4) &$-0.583(4)  $&$ 0.516(9) $ \\ 
203610780& 0.0584(8)  & 2.37(3)    & 88.12(5) & 2082.5913(2)   & 29.5937(5) &$-0.044(1)  $&$0.61(1)   $ \\ 
203636784& 0.1140(7)  & 0.461(9)   & 89.1(2)  & 2066.5457(2)   & 6.76465(4) &$ 0.000(1)  $&$0.15(2)   $ \\ 
203728604& 0.0664(1)  & 0.385(2)   & 89.71(9) & 2066.85857(7)  & 36.108(1)  &$-0.5084(4) $&$0.221(2)  $ \\ 
204407880& 0.0542(6)  & 0.316(5)   & 88.56(5) & 2084.7125(1)   &34.36789(2) &$ 0.389(4)  $&$0.664(7)  $ \\ 
204748201& 0.0839(6)  & 0.43(1)    & 89.6(3)  & 2070.5201(1)   & 7.36575(4) &$ 0.00(1)   $&$ 0.00(5)  $ \\ 
204760247& 0.114(1)   & 0.65(1)    & 90.0(4)  & 2079.7559(8)   & 9.2022(6)  &$ 0.007(8)  $&$ 0.01(5)  $ \\ 
204822807& 0.05558(7) & 2.38(1)    & 89.88(9) & 2107.3994(1)   & 67.535(1)  &$-0.2837(3) $&$ 0.051(3) $ \\ 
204870619& 0.0517(2)  & 0.281(4)   & 89.9(2)  & 2073.7038(2)   & 34.0690(3) &$-0.217(1)  $&$0.444(3)  $ \\ 
205020466& 0.0792(2)  & 0.97(2)    & 89.98(9) & 2067.9683(2)   & 8.75903(3) &$-0.467(2)  $&$ 0.35(2)  $ \\ 
205170307& 0.0270(2)  & 0.39(1)    & 89.51(4) & 2112.8240(1)   & 67.5025(8) &$-0.5449(8) $&$ 0.177(5) $ \\ 
205546169& 0.0529(6)  & 1.74(2)    & 87.98(2) & 2125.59047(8)  &24.43581(6) &$-0.393(4)  $&$ 0.696(6) $ \\ 
205703649& 0.152(3)   & 1.1(2)     & 84.7(3)  & 2083.1227(2)   & 8.11699(5) &$ 0.038(7)  $&$ -0.05(2) $ \\ 
205919993& 0.0530(3)  & 0.97(3)    & 87.69(1) & 2182.4928(2)   & 11.00009(6)&$ 0.273(3)  $&$ 0.12(2)  $ \\  
205982900& 0.1787(1)  & 1.196(5)   & 88.68(2) & 2157.46696(2)  & 6.719684(4)&$-0.0942(2) $&$ 0.411(1) $ \\  
206066862& 0.069(1)   & 0.74(9)    & 87.01(9) & 2155.9185(1)   & 11.08666(5)&$ -0.401(8) $&$ 0.21(3)  $ \\  
206066909& 0.0659(2)  & 0.567(6)   & 89.20(4) & 2174.94439(3)  & 12.93712(2)&$ 0.454(1)  $&$ 0.438(3) $ \\ 
206084435& 0.0278(4)  & 0.47(2)    & 89.30(4) & 2182.3183(2)   & 48.221(2)  &$-0.2989(3) $&$ 0.01(2)  $ \\ 
206109641& 0.02915(1) & 0.846(2)   & 89.98(2) & 2178.17503(2)  & 62.58668(6)&$-0.79828(5)$&$ 0.047(1) $ \\ 
206212261& 0.0487(2)  & 0.302(5)   & 89.7(1)  & 2162.0314(1)   & 30.9857(2) &$  0.133(1) $&$-0.260(4) $ \\ 
206241558& 0.0311(3)  & 0.63(4)    & 88.66(1) & 2160.2622(1)   & 56.58934(5)&$ -0.547(4) $&$ 0.536(8) $ \\ 
206288770& 0.0404(3)  & 0.415(4)   & 89.37(4) & 2160.03956(6)  & 24.75656(5)&$ 0.477(3)  $&$ 0.342(8) $ \\ 
206433263& 0.06039(4) & 0.522(2)   & 89.33(1) & 2169.57164(4)  & 21.19385(3)&$-0.0922(6) $&$-0.299(2) $ \\ 
\hline
\end{tabular}
\end{center}
\tablefoot{Symbols used are defined in sections \protect\ref{ellc}  and
\protect\ref{TeffSection}. The time system used for $T_0$ is TDB represented
as ${\rm BJD}-2454833$, i.e., the time system normally used for Kepler K2
observations.  The standard error in the final digit of each parameter is
given in parentheses. Values preceded by ``='' were fixed for this analysis
or (if a standard error is given) imposed as constraints on the solution.  See
text for discussion of possible systematic errors in these parameters.}
\end{table*}

\begin{table*}
\caption{Radiative parameters and other parameters of interest derived from
the analysis of the K2 light curves for selected long-period eclipsing binary
stars. \label{ellcpar2}}
\begin{center}
    \begin{tabular}{@{}rrrrrrrrrrrr}
\hline
      \multicolumn{1}{@{}l}{EPIC} &
  \multicolumn{1}{l}{$S_{\rm Kp}$} &
  \multicolumn{1}{l}{$\ell_3$} &
  \multicolumn{1}{l}{$\ell_{\rm Kp}$} &
  \multicolumn{1}{l}{$R_1/a$} &
  \multicolumn{1}{l}{$R_2/a$} &
  \multicolumn{1}{l}{$e$}    &
  \multicolumn{1}{l}{$\omega$ [$^{\circ}$] }  &
  \multicolumn{1}{l}{$\sigma$ [ppt]  } \\
\hline
201160323&  0.62(1)  &  0.43(5)  & 0.75(7)     & 0.0132(4)  & 0.0145(3)  & 0.202(3)   & 154(2)    & 6.2, 4.8   \\  
201161715&  1.4(1)   &  0.01(3)  & 0.18(3)     & 0.0446(6)  & 0.0161(6)  & 0.345(4)   & 23(2)     & 1.4, 2.7   \\  
201246763&  1.23(2)  &  0.006(6) & 1.58(4)     & 0.0165(2)  & 0.0186(1)  & 0.519(2)   & 134.4(2)  & 0.8, 2.5   \\  
201253025&  1.00(1)  &  0.27(5)  & 1.0(2)      & 0.071(3)   & 0.071(3)   & 0.045(5)   & 100(1)     & 3.2       \\ 
         &  1.01(1)  &   0.16(5) & 0.9(2)      & 0.073(3)   & 0.070(3)   & 0.045(5)   & 100(1)    & 3.5        \\ 
201379113&  0.20(9)  & =0        & 0.1(1)      & 0.021(2)   & 0.016(1)   & 0.35(4)    & 124(14)   & 2.6, 0.5   \\ 
201382417&  0.210(2) & 2.8(2)    & 0.055(3)    & 0.0893(9)  & 0.0457(6)  & 0.001(1)   & 150(65)   & 0.4, 0.4   \\ 
201408204&  0.991(8) & 0.004(4)  & 0.92(2)     & 0.0543(4)  & 0.0524(4)  & 0.202(1)   & 30.3(7)   & 1.2, 1.5   \\  
201488365& 0.9945(7) & 0.0005(8) & 0.84(1)     & 0.0792(2)  & 0.0726(3)  & 0.0001(2)  &  --       & 0.5, 0.4 \\  
201576812& 0.21(4)   & =0        & 0.19(6)     & 0.055(4)   & 0.050(7)   & 0.01(2)    &  --       &  1.9, 0.9  \\  
201648133& 0.542(2)  & 0.006(3)  & 0.2563(8)   & 0.02082(2) & 0.01431(2) & 0.0458(8)  & 243.9(5)  &  0.1, 0.3 \\  
201665500& 0.059(2)  & =0        & 0.0036(1)   & 0.137(1)   & 0.0339(3)  & 0.002(7)   & --        & 2.7, 1.1   \\ 
201705526& 0.128(2)  & 4.5(2)    & 0.046(1)    & 0.0273(2)  & 0.0163(2)  & 0.269(2)   & 336.6(8)  & 0.1, 0.1  \\ 
201723461&  0.4(4)   & =0        &  0.6(5)     & 0.017(7)   & 0.020(5)   & 0.6(2)     & 152(15)   & 24, 22     \\  
202674012& 0.966(3)  & 0.005(5)  & 0.306(2)    & 0.04596(8) & 0.0258(1)  & 0.027(2)   & 259.7(6)  & 0.3,0.4    \\ 
202843085& 0.970(6)  & 0.015(4)  & 1.30(3)     & 0.0490(4)  & 0.0567(3)  & 0.101(1)   & 144(1)    & 1.0, 1.3   \\ 
203361171& 0.98(4)   & 0.3(3)    & 1.2(4)      & 0.069(6)   & 0.079(5)   & 0.018(3)   & 189(18)   & 4.2, 3.6   \\ 
203371239& 0.94(1)   & 0.05(2)   & 0.73(8)     & 0.0388(9)  & 0.034(1)   & 0.039(5)   & 269.0(2)  & 2.7, 7.9   \\ 
203543668& 0.30(1)   & 1.4(1)    & 0.137(5)    & 0.0166(3)  & 0.0112(2)  & 0.606(4)   & 139(1)    & 1.3, 1.2   \\ 
203610780& 0.51(6)   & 1.2(2)    & 2.9(3)      & 0.0173(3)  & 0.0411(5)  & 0.37(1)    &  94.1(2)  & 0.8, 1.0   \\ 
203636784& 0.195(3)  & 0.05(4)   & 0.041(2)    & 0.0780(8)  & 0.0360(5)  & 0.021(6)   & 89.8(5)   & 2.1, 1.8   \\ 
203728604& 0.860(3)  & 0.012(8)  & 0.127(1)    & 0.0480(1)  & 0.01846(6) & 0.3072(4)  & 156.5(2)  & 0.2, 2.4   \\ 
204407880& 0.46(2)   & 0.05(3)   & 0.047(2)    & 0.0412(5)  & 0.0130(2)  & 0.592(6)   & 59.6(5)   & 0.4, 1.8  \\  
204748201& 0.063(2)  & 0.16(6)   & 0.0115(7)   & 0.0588(8)  & 0.0250(4)  & 0.001(3)   &  --       & 1.4, 1.8   \\ 
204760247& 0.389(6)  & 0.06(4)   & 0.167(8)    & 0.0689(7)  & 0.0451(9)  & 0.001(3)   &  --       & 8.7        \\ 
204822807& 0.364(2)  & 0.096(9)  &  2.06(3)    & 0.01644(7) & 0.03914(9) & 0.0831(2)  & 169.7(7)  & 0.5, 0.4   \\ 
204870619& 0.528(4)  & 0.05(2)   & 0.0417(9)   & 0.0403(2)  & 0.0113(1)  & 0.244(2)   & 116.1(2)  & 0.5, 0.7   \\ 
205020466& 0.79(1)   & 0.03(2)   & 0.74(2)     & 0.0402(3)  & 0.0390(4)  & 0.34(1)    & 143(2)    & 2.8, 2.6   \\ 
205170307& 0.219(1)  & 0.17(5)   & 0.034(2)    & 0.0193(2)  & 0.0076(1)  & 0.3282(8)  & 162.0(5)  & 0.3, 0.5   \\ 
205546169& 0.95(7)   & 0.09(5)   &  2.9(2)     & 0.0193(3)  & 0.0336(3)  & 0.639(6)   & 119.4(5)  & 0.5, 0.3   \\ 
205703649& 0.94(2)   & 0.8(2)    &  1.2(4)     & 0.071(5)   & 0.081(5)   &  0.004(2)  & 310(34)   & 0.8, 0.8   \\ 
205919993& 1.35(7)   & =0        & =1.28(5)    & 0.0269(5)  & 0.0261(4)  & 0.088(2)   & 24(3)     & 1.0, 0.9   \\ 
205982900& 0.929(3)  & =0        & 1.329(6)    & 0.0814(2)  & 0.0974(1)  & 0.1775(8)  & 102.92(6) & 0.5, 0.4   \\ 
206066862& 0.36(6)   & =0        & 0.20(5)     & 0.040(2)   & 0.029(2)   & 0.205(8)   & 152(4)    & 1.2        \\ 
206066909& 0.189(2)  & 0.52(2)   & 0.0606(9)   & 0.0420(3)  & 0.0238(1)  & 0.398(2)   & 44.0(3)   & 0.45       \\ 
206084435& 0.200(2)  & 0.04(4)   & 0.044(4)    & 0.0189(2)  & 0.0089(3)  & 0.0896(2)  & 177(2)    &  1.0       \\ 
206109641& 0.932(3)  & 0.0008(7) & 0.6664(9)   & 0.01579(2) & 0.01336(1) & 0.63944(3) &176.65(9)  & 0.20       \\ 
206212261& 0.185(1)  & 0.08(4)   & 0.0169(6)   & 0.0374(3)  & 0.0113(1)  & 0.085(2)   & 297.1(6)  &  0.3       \\ 
206241558& 1.6(1)    & =0        & 0.6(1)      & 0.0191(4)  & 0.0120(5)  & 0.587(4)   & 135.6(7)  & 0.8        \\ 
206288770& 0.0804(7) & 0.02(2)   & 0.0138(2)   & 0.0286(2)  & 0.01186(8) & 0.345(3)   & 35.6(8)   & 0.4        \\ 
206433263& 0.691(3)  & 0.008(7)  & 0.188(1)    & 0.03967(6) & 0.02072(6) & 0.098(1)   &  252.8(2) & 0.4        \\ 
\hline
\end{tabular}
\end{center}
\tablefoot{Symbols used  are defined in sections \protect\ref{ellc} and
\protect\ref{TeffSection}. The standard error in the final digit(s) of each
parameter is given in parentheses. The standard error per observation as
defined in section \protect\ref{ellc} is given in the column $\sigma$ in units
of parts per thousand (ppt). For some light curves we give two values of
$\sigma$ because  we assume different values of the standard error per
observation for different parts of the light curve, as described in section
\protect\ref{ellc} Values preceded by ``='' are either fixed or (if a
standard error is given) imposed as constraints on the solution.   See text
for discussion of possible systematic errors in these parameters.}
\end{table*}

\begin{table*}
\caption[]{\label{TrhoTable} Mean stellar densities, $\rho_1$ and $\rho_2$
calculated assuming a mass ratio $q=M_2/M_1$. }
\begin{center}
\begin{tabular}{crrrc|crrr}
\hline \hline
  \multicolumn{1}{@{}l}{EPIC} & 
  \multicolumn{1}{l}{$P$ [d]} & 
  \multicolumn{1}{l}{T$_{\rm eff,1}$}[K] &
  \multicolumn{1}{l}{$\log(\rho_1/\rho_{\odot})$} &
~~&
  \multicolumn{1}{l}{EPIC} & 
  \multicolumn{1}{l}{$P$ [d]} & 
  \multicolumn{1}{l}{T$_{\rm eff,1}$}[K] &
  \multicolumn{1}{l}{$\log(\rho_1/\rho_{\odot})$} \\

  \multicolumn{1}{@{}l}{Symbol} &
  \multicolumn{1}{c}{$q$} &
  \multicolumn{1}{l}{T$_{\rm eff,2}$}[K] &
  \multicolumn{1}{c}{$\log(\rho_2/\rho_{\odot})$}&
&
  \multicolumn{1}{l}{Symbol} &
  \multicolumn{1}{c}{$q$} &
  \multicolumn{1}{l}{T$_{\rm eff,2}$}[K] &
  \multicolumn{1}{c}{$\log(\rho_2/\rho_{\odot})$}\\

\hline
&&&&&&\\
201253025 & 6.79  &    6065 & $-0.4$ & &  201408204 & 8.48  &     5845 & $-0.2$ \\
 $\diamond$  & 1.0 &   6070 & $-0.4$ & &   $\circ$     & 1.0 &    5830 & $-0.2$ \\[2mm]
201382417 & 5.20  &    6175 & $-0.4$ & &  201705526 & 18.10 &     6600 & $ 0.1$ \\
 $\star$  &    0.6 &  4480 & $ 0.3$ & &   $\triangle$ & 0.5 &    4320 & $ 0.5$ \\[2mm]
201488365\tablefootmark{$\dagger$} 
          & 6.73  &    6355 & $-0.5$ & &  202843085 & 16.50 &     6300 & $-0.7$ \\ 
 $\circ$  & 0.976  &   6345 & $-0.4$ & &   $\square$   & 1.0 &    6260 & $-0.9$ \\ [2mm]
201576812\tablefootmark{$\dagger\dagger$} 
          & 5.73  &    5905 & $ 0.2$ & &  203371239 & 20.36 &     6400 & $-0.6$ \\
 $\triangle$ & 0.66&   4360 & $ 0.1$ & &   $\times$    & 1.0 &    6300 & $-0.4$ \\[2mm]
201665500 &  3.05 &    6270 & $-0.4$ & &  205020466\tablefootmark{$\dagger$} 
                                                    & 8.76  &     5300 & $ 0.1$ \\
 $\square$   & 0.4 &   3630 & $ 1.0$ & &   $+$         &0.80 &    5070 & $ 0.2$ \\[2mm]
203361171 &  7.32 &    6070 & $-0.4$ & &  205703649 &  8.12 &     5610 & $-0.5$ \\
 $\times$    & 1.0 &   6050 & $-0.6$ & &   $\diamond$  & 1.0 &    5540 & $-0.7$ \\[2mm]
203636784 &  6.76 &    5970 & $-0.4$ & &  205919993\tablefootmark{$\dagger$}
                                                    & 11.00 &     4025 & $0.5$  \\
 $+$         & 0.6 &   4350 & $ 0.4$ & &   $\star$     & 1.10&    4230 &  $0.5$ \\[2mm]
204748201 &  7.36 &    6100 & $-0.1$ & &  206066862 & 11.09 &     6250 & $0.0$  \\
 $\diamond$&    0.4 & 3600 & $ 0.7$ & &   $\circ$  &    0.7 &   5000 &  $0.3$ \\[2mm]
205982900 &  6.72 &    6200 & $-0.6$ & &  206066909 & 12.94 &     6440 & $-0.2$ \\
 $\star$     & 1.1 &   6045 & $-0.8$ & &   $\triangle$&    0.5 & 4535 & $ 0.3$ \\[2mm]
\hline                                
&&&&&&\\                            
201379113 & 21.2  &     5150 & $ 0.3$ & &201161715 & 59.89 &     5030 & $-1.6$  \\
  $\square$   & 0.7 &    3900 & $ 0.5$ & & $\diamond$ &    0.8 &  5370 & $-0.5$  \\[2mm]
201723461 & 22.73 &     4450 & $ 0.4$ & &201246763 & 43.68 &     5875 & $-0.1$  \\
 $\times$    & 0.9 &    4170 & $ 0.2$ & & $\star$ &    1.2 &    6225 & $-0.3$  \\[2mm]
202674012\tablefootmark{$\dagger$} 
          & 23.31 &     6250 & $-0.9$ & &201648133 & 35.02 &     6010 & $-0.2$  \\
 $+$         & 0.78&    6150 & $-0.2$ & & $\circ$     & 0.8 &    5250 & $ 0.2$  \\[2mm]
203610780 & 29.59 &     6650 & $ 0.1$ & &203543668 & 36.76 &     5900 & $ 0.1$  \\
 $\diamond$  & 1.2 &    5600 & $-0.9$ & & $\triangle$ & 0.7 &    4600 & $ 0.5$  \\[2mm]
204407880 & 34.37 &     5765 & $-1.0$ & &203728604 & 36.11 &     6050 & $-1.3$  \\
 $\star$     & 0.6 &    4900 & $ 0.3$ & & $\square$   & 0.8 &    5840 & $-0.1$  \\[2mm]
204870619 & 34.07 &     5435 & $-1.0$ & &204822807 & 67.53 &     5625 & $-0.5$  \\
 $\circ$ &    0.7 &    4800 & $ 0.5$ & & $\times$    & 1.1 &    4620 & $-1.6$  \\[2mm]
205546169 & 24.44 &     6300 & $ 0.2$ & &205170307 & 67.50 &     5620 & $-0.6$  \\
 $\triangle$ & 1.1 &    6170 & $-0.5$ & & $+$         & 0.6 &    4240 & $ 0.4$  \\[2mm]
206212261 & 30.99 &     5385 & $-0.8$ & &206084435 & 48.22 &     5950 & $-0.3$  \\
 $\square$   & 0.6 &    4010 & $ 0.6$ & & $\diamond$  & 0.6 &    4300 & $ 0.5$  \\[2mm]
206288770 & 24.76 &     6290 & $-0.2$ & &206109641 & 62.59 &     5905 & $-0.3$  \\
 $\times$    & 0.5 &    3870 & $ 0.6$ & & $\star$     & 0.9 &    5805 & $-0.2$  \\[2mm]
206433263 & 21.19 &     6000 & $-0.6$ & &206241558 & 56.59 &     5330 & $-0.5$  \\
 $+$         & 0.9 &    5525 & $ 0.2$ & & $\circ$     & 1.0 &    5885 & $ 0.1$  \\[2mm]
\hline
\end{tabular}
\end{center}
\tablefoot{
\tablefoottext{$\dagger$}{Mass ratio from spectroscopic orbit.}
\tablefoottext{$\dagger\dagger$}{Mass ratio from \citet{2011AJ....142...50F}.}
The plotting symbol used for each
binary in Fig.~\ref{TeffDensity} is shown for each star and the table is
arranged in the same format as the panels in that figure.}
\end{table*}

\subsection{Notes on individual objects \label{notes_sec}}

\subsubsection{EPIC 201160323}

 This faint star shows rapid apsidal motion. The period measured from the
times of primary and secondary eclipse in the K2 light curve are $P_{\rm pri}
=22.272$\,d and $P_{\rm sec} = 22.300$\,d, respectively. We therefore included
the rate of change of the longitude of periastron in the light curve model as
a free parameter and hence obtained the value $\frac{d\omega}{dt} =
-0.10^{\circ} \pm 0.01^{\circ}$ per anomalistic period. This corresponds to an
apsidal motion period of approximately 220 years if this rate is assumed to be
constant. 

\begin{figure*}
\includegraphics[width=0.99\textwidth]{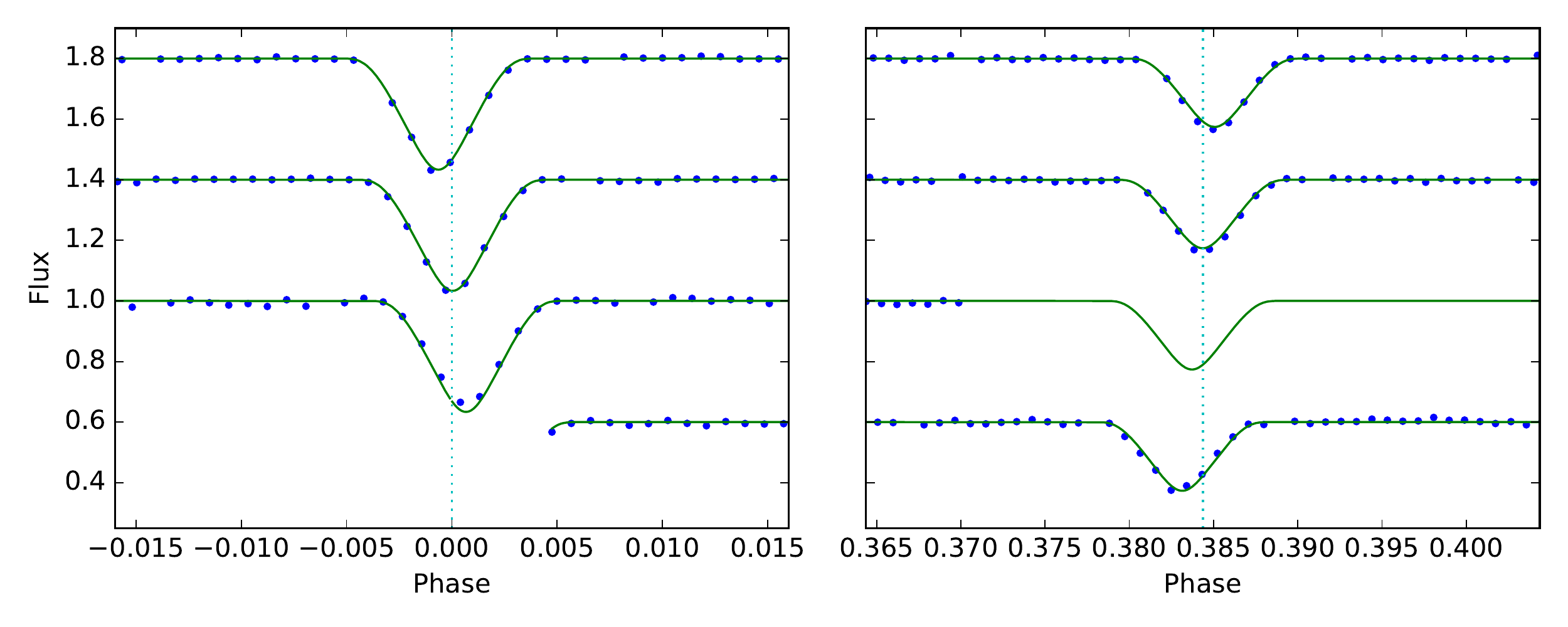} 
\caption{K2 light curve of EPIC~201160323 around primary and secondary
eclipse. The data are shown offset vertically according to cycle number and
have been plotted using a single linear ephemeris to calculate the phase.
Solid lines show our best-fit light curve model.} 
\label{201160323_lcfit} 
\end{figure*}

 Our best-fit model is shown in Fig.~\ref{201160323_lcfit}, where the drift in
eclipse times relative to a single linear ephemeris calculated with the
average period can be clearly seen. There are no nearby stars listed in the
GAIA DR1 catalogue that might explain the large value for the third light
parameter derived from the light curve analysis ($\ell_3 = 0.43 \pm 0.05$).
This suggests that EPIC 201160323 is a triple or multiple star system in which
the gravitational interaction between the eclipsing binary and an additional
body or bodies is causing the rapid change in the orientation of its orbit.

 This star is listed in the K2 Variable Catalogue  \citep{2015A&A...579A..19A}
as an eclipsing binary with a period of 22.299969\,d, which matches closely
our estimate of $P_{\rm sec}$. Note that the orbital period values given in
Tables \ref{lcinfo} and \ref{ellcpar1} are the anomalistic period. We did not 
attempt to estimate the effective temperatures of the stars in this system 
from the published photometry because there are no reliable photometric
measurements at optical wavelengths -- EPIC 201160323 is too faint to appear
in either the APASS9 or Tycho-2 catalogues.

\subsubsection{EPIC 201161715}

 Star 1 is much larger than star 2 but the stars have similar effective
temperatures so we assume that Star 1 is a sub-giant or red giant and $q
=M_2/M_1 < 1$ (since the more massive star will have evolved off the main
sequence first). For any reasonable choice of $q<1$ we find that star 1 is
a red giant with a mass $M_1\approx 1.4\,M_{\sun}$. The evolution tracks for
different masses have similar values of T$_{\rm eff}$ on the red giant branch
so this mass is quite uncertain if we consider the properties of star 1 only.
However, star 2 appears near the main-sequence turn off point so must have a
mass $\ga 0.8\,M_{\sun}$. Both stars are in relatively short-lived
evolutionary phases and the main-sequence life time decreases rapidly with
increasing mass, so the mass ratio cannot be very different from 1. We conclude
that $q\approx 0.8$ such that $M_1\approx 1.2\,M_{\sun}$ and $M_2\approx
0.95\,M_{\sun}$.  From Fig.~\ref{TeffDensity} it can be seen that if
$q\approx 0.8$ then this binary contains a star near the main-sequence
turn-off point (MSTO) and a star at the base of the red giant branch, similar
to the well-known systems AI~Phe \citep{2016A&A...591A.124K} and TZ~For
\citep{2017A&A...600A..41V}. This makes this system an attractive target for
calibrating stellar models. This star is listed in the K2 Variable Catalogue
\citep{2015A&A...579A..19A} as an eclipsing binary with a period of
59.889024\,d, which agrees well with our period estimate. 

\subsubsection{EPIC 201246763}

 The K2 light curve of this star shows one primary eclipse and two secondary
eclipses. The position of the star on the detector during the second of the
secondary eclipses is not well sampled by the other observations of this star
so the detrending corrections applied to some of the data in this eclipse are
extrapolated from the out-of-eclipse data. There are distinct differences
between the shape and depth of this eclipse between the first and second
observation of this feature in the K2 light curve. This makes it difficult to
determine a precise value for the orbital period using the K2 data alone.
Fortunately,  the observations of this star from the WASP photometric archive
have  good coverage of both eclipses of this star that can be used to measure
the orbital period to good precision. 

We used a least-squares fit with the {\sc
jktebop}\footnote{\url{www.astro.keele.ac.uk/~jkt/codes/jktebop.html}}
model \citep{2013A&A...557A.119S} to
664 observations around primary and secondary eclipse from the WASP
photometric archive to measure the orbital period of the binary. The WASP data
cover the minima of two primary eclipses and one secondary eclipse plus a few
observations of the ingress or egress to an eclipse. The first eclipse in the
WASP data occurs on JD 2454881. We included the time of mid-eclipse from a
preliminary fit to the K2 light curve as a constraint in this fit. The
geometric parameters of the binary system were fixed at values from the same
preliminary fit to the K2 light curve. The orbital period value we obtained is
$43.68281 \pm 0.00003 $\, days. We imposed this value as a prior on the
orbital period for our final analysis of the K2 light curve using {\sc emcee}.
From Fig.~\ref{TeffDensity} it can be seen that this binary contains two
main-sequence stars with masses $M_1\approx 1.0\,M_{\sun}$ and $M_2\approx
1.2\,M_{\sun}$. These mass estimates are quite robust because both stars lie
near the evolution tracks with these masses for any reasonable choice of the
mass ratio, $q$.

\subsubsection{EPIC 201253025}
 The aperture used to calculate the light curve is contaminated by another
star approximately 4.7 arcseconds to the west of the main target and 1.6
magnitudes fainter in the G band according to the Gaia DR1 data release
\citep{2016A&A...595A...1G}. We found that we could not get a good fit to the
entire data set using one set of parameters, partly because the level of
contamination from the nearby star is not constant. To deal with this problem
we analysed separately the two parts of the light curve either side of the gap
in the data at BJD 2456849. The results for two subsets of data are both given
in Tables~\ref{ellcpar1} and \ref{ellcpar2}. This approach does improve the
fit to the two parts of the light curve, but residuals of about 0.5\% are
still apparent for some eclipses, presumably as a result of star spots on one
or both stars that are also the cause of the quasi-periodic variations in flux
between the eclipses.  Despite these problems there is very good agreement
between the geometric parameters derived from the two parts of the light
curve. We set $\ell_3 = 0$ for our analysis of the published photometry to
estimate T$_{\rm eff}$ because we assume that the value of $\ell_3$ in
Table~\ref{ellcpar2} is due to the star 4.7 arcseconds to the west of the main
target.

 This star is listed in the K2 Variable Catalogue \citep{2015A&A...579A..19A}
as an eclipsing binary with a period of 6.785544\,d, which agrees well with
our period estimate. High resolution imaging by \citet{2016AJ....151..159S}
did not detect any companions to this star, with the quoted upper limit to the
relative brightness at I-band being  2.02 magnitudes at 0.25\,arcsec.

EPIC 201253025 contains a pair of quite similar stars  so we assume
$q\approx 1$, in which case the stars are towards the end of their main
sequence lifetimes with masses $\approx 1.2M_{\sun}$ (Fig.~\ref{TeffDensity}).
The rotation periods detected in the K2 light curve show that the stars in
this binary system rotate non-synchronously, with one star rotating slightly
faster than predicted for synchronous rotation and one slightly slower. The
values of $R_{\star}/a \approx 0.07$ for these stars put them near the
boundary between synchronous and non-synchronous rotation for stars with
convective envelopes \citep{2010A&ARv..18...67T}. This makes EPIC 201253025 an
interesting test case for theories of the tidal interactions between low mass
stars. 

\subsubsection{EPIC 201379113}

 The secondary eclipse is very shallow (1.5\%) and partial so it is not
possible to determine a reliable value of $\ell_3$ from the K2 light curve
alone. In addition, the observed flux between the eclipses varies by up to 0.4\%
on time scales of 10 days or more. There may be a rotation modulation signal
with a period of about 22\,days in these flux variations, but we are not
confident of this detection. We divided out these slow flux variations so the
observed secondary depth varies systematically by a few parts per thousand. To
derive the parameters in  Tables \ref{ellcpar1} and \ref{ellcpar2} we fixed
the value $\ell_3 = 0$. Even with this restriction, the additional noise in
the eclipse depths results in quite large errors on the light curve model
parameters for this binary. The  precision of these parameters can certainly
be improved using constraints on the luminosity ratio and third-light
contribution from spectroscopy. 

 This star is listed in the K2 Variable Catalogue \citep{2015A&A...579A..19A}
as an eclipsing binary with a period of 21.186043\,d, which is slightly
shorter than the period that we find from our analysis. From the location of
these stars in the T$_{\rm eff}$ -- $\rho_{\star}$ plane we estimate that they
are dwarf stars with masses $M_1\approx 0.8M_{\sun}$ and $M_2\approx
0.6M_{\sun}$.  This conclusion is not affected by the assumed value for
the mass ratio for any reasonable estimate of $q\approx 0.7$.

\subsubsection{EPIC 201382417}

 The light curve between the eclipses shows a quasi-periodic variation that
gradually increases from being barely detectable at the start of the K2
observing sequence up to an amplitude of 0.4\%. We have divided
out this trend rather than trying to fit a model to this variation. As a
result, the depth of the secondary eclipse relative to this ``corrected''
out-of-eclipse level varies from about 1.5\% at the start of the observing
sequence to 1.2\% in the second half of the data set. The best-fit solution to
this corrected light curve has a secondary eclipse depth of 1.27\%. The
parameters in Tables~\ref{ellcpar1} and \ref{ellcpar2} are very precise but
there are certainly systematic errors in these values as a result of the
detrending process, i.e., these parameters are much less accurate than implied
that the quoted precision. To obtain a more accurate solution it will be
necessary to identify and characterise the source or sources of the variation
between the eclipses, i.e., to determine whether it is due to spot modulation
on a third star that dominates the flux from this system (assuming our
estimate of $\ell_3$ is accurate), or from the primary star in the eclipsing
binary system, or a combination of both. Although the error bars quoted in
Tables~\ref{ellcpar1} and \ref{ellcpar2} are underestimates of the current
accuracy in these parameters they do give a useful estimate of the accuracy
that may be possible with a more complete model for this system.

 This star is listed in the K2 Variable Catalogue \citep{2015A&A...579A..19A}
as an eclipsing binary with a period of 5.1976386\,d, which agrees well with
our estimate of the orbital period. From the location of these stars in the
T$_{\rm eff}$ -- $\rho_{\star}$ plane (Fig.~\ref{TeffDensity}) we estimate
that they are dwarf stars with masses $M_1\approx 1.2M_{\sun}$ and
$M_2\approx 0.7M_{\sun}$. Both stars lie near the evolution tracks for
these masses for any reasonable choice of $q=M_2/M_1$.

\subsubsection{EPIC 201408204}

 The stars in this binary system have very similar effective temperatures
and radii so we assume $q\approx 1$. High resolution imaging by
\citet{2016AJ....151..159S} did not detect any companions to this star, with
the quoted upper limit to the relative brightness at I-band being  2.84
magnitudes at 0.25\,arcsec. This star is listed in the K2 Variable Catalogue
\citep{2015A&A...579A..19A} as an eclipsing binary with a period of
8.482149\,d, which agrees well with our estimate of the orbital period. The
rotation periods detected in the K2 light curve suggest that this pair of
main-sequence stars with masses $M\approx 1 M_{\sun}$ (Fig.~\ref{TeffDensity})
rotate non-synchronously with the orbit although one of the rotation periods
is close to the orbital period. However, the orbital eccentricity of this
binary is quite large ($e\approx 0.2$) so in this case it makes more sense to
compare the observed rotation periods to the ``pseudo-synchronisation''
rotation period determined by matching the angular velocity of the star to the
orbital angular velocity at periastron \citep{1981A&A....99..126H}. The
corresponding ratio of the orbital and pseudo-synchronisation rotation periods
is $(1+e)^2/(1-e^2)^{3/2} = 1.54$, suggesting that neither of the stars
rotates pseudo-synchronously. This is another useful system for testing models
of tidal dissipation in solar-type stars.

\subsubsection{EPIC 201488365 = \object{FM Leo}}
 
\begin{table}
\caption[]{\label{tab:fmleo_abspar}
  Absolute astrophysical parameters of FM~Leo (EPIC 201488365).}
\begin{center}
\begin{tabular}{@{}lrr} 
\hline 
\hline\noalign{\smallskip}
 &\multicolumn{1}{l}{Primary} & \multicolumn{1}{l}{Secondary} \\
\noalign{\smallskip} 
\hline 
\noalign{\smallskip} 
Mass   [$M_{\sun}$] & 1.32  $\pm$  0.01 &  1.29   $\pm$ 0.01   \\ 
Radius [$R_{\sun}$] & 1.634 $\pm$ 0.005 &  1.498  $\pm$ 0.006  \\ 
$\log g$ [cm\,s$^{-2}$] & 4.132 $\pm$ 0.002 &  4.197  $\pm$ 0.003  \\
T$_{\rm eff}$ [K]   &  6430 $\pm$ 155   &  6420   $\pm$ 155    \\
$\log(L/L_{\sun})$  &  0.61 $\pm$ 0.04  & 0.54    $\pm$ 0.04   \\ 
M$_{\rm V}$         &  3.24 $\pm$ 0.16  &  3.44   $\pm$ 0.16   \\ 
\noalign{\smallskip} 
Orbital period [d]  & \multicolumn{2}{c}{6.728609  $\pm$ 0.000002} \\ 
Mass ratio & \multicolumn{2}{c}{0.976  $\pm$ 0.005 } \\ 
Distance      [pc]  & \multicolumn{2}{c}{143   $\pm$ 8  } \\ \noalign{\smallskip} \hline
\end{tabular} 
\end{center} 
\tablefoot{Absolute V magnitudes use bolometric corrections from
\citet{1998A+A...333..231B} and the distance is based on the apparent V-band
magnitude inferred from the observed Tycho-2 B$_{\rm T}$ and  V$_{\rm T}$
magnitudes. See text for a discussion of possible systematic errors in these
parameters.} 
\end{table}

 \citet{2010MNRAS.402.2424R} have published spectroscopic orbits for both
components of FM~Leo together with an analysis of the light curves available
to them at that time. We have used the semi-amplitudes $K_1$ and $K_2$ from
Ratajczak et al. together with the parameters from our analysis of the Kepler
K2 light curves with {\sc
jktabsdim}\footnote{\url{www.astro.keele.ac.uk/jkt/codes/jktabsdim.html}} to
derive the absolute parameters for FM~Leo given in
Table~\ref{tab:fmleo_abspar}. The masses derived ($1.29M_{\sun}$ and
$1.32M_{\sun}$) are in reasonable agreement with the estimate $M\approx
1.25M_{\sun}$ implied from the position of the stars in the T$_{\rm eff}$ --
$\rho_{\star}$ plane (Fig.~\ref{TeffDensity}). The precision of the radius
measurements is improved by an order of magnitude compared to what was
possible with the data available to Ratajczak et al..  FM~Leo could be a very
useful system for testing stellar models if more precise estimates for the
metallicity and effective temperature of the stars become available. The
scatter in the residuals through the eclipses is approximately a factor of 2
larger than the residuals between the eclipses so it is likely that there is
additional systematic error in the parameters derived from the K2 light curve
comparable to the quoted error bars.  

 This star is listed in the K2 Variable Catalogue \citep{2015A&A...579A..19A}
as an eclipsing binary with a period of  3.364700\,d, which is approximately
half of the correct orbital period. 

\subsubsection{EPIC 201576812 = \object{TYC 272-458-1}}

 \citet{2011AJ....142...50F} present a detailed analysis of the WASP
light curve and high-resolution spectroscopy of this eclipsing binary. They did
not detect the  secondary star in their spectroscopy and so to estimate the
masses and radii of the stars they adopted the value $M_1 = 0.92 \pm
0.1M_{\sun}$ for the primary star mass based on the values  T$_{\rm eff} =
5483$\,--\,5957\,K and [Fe/H] $= -0.28$ from the analysis of its spectrum
compared to stellar evolution models.
 
 This star is listed in the K2 Variable Catalogue \citep{2015A&A...579A..19A}
as an eclipsing binary with a period of 5.728410\,d, which agrees well with
our period estimate. High resolution imaging by \citet{2016AJ....151..159S}
did not detect any companions to this star, with the quoted upper limit to the
relative brightness at I-band being  2.21 magnitudes at 0.25\,arcsec.

 As there is no evidence for third light in the spectrum of this star and
there are no bright companions within the photometric aperture we have used,
we set $\ell_3 = 0$ in our analysis of the K2 light curve. The geometric light
curve parameters we obtain are not quite consistent with those of
\citet{2011AJ....142...50F} at the 1-$\sigma$ level. This level of
disagreement is not surprising given that the light curve of this star shows a
shallow partial secondary eclipse plus rotational spot modulation visible
between the eclipses with an amplitude $\approx 1\%$.
 
\subsubsection{EPIC 201648133}

 The K2 light curve of this star shows two primary eclipses and two secondary
eclipses, with a gap in the data at the time of a primary eclipse near the
middle of the observing sequence. A least-squares fit of a simple light curve
model to the WASP photometry provides three times of primary eclipse as
follows: HJD 2454852.4432(6), 2454922.4931(4), 2455237.7094(4), where figures
in parentheses denote the standard error in the final digit of these values.
From a fit to these times of mid-eclipse plus one further time of mid-eclipse
from a preliminary fit to the K2 light curve we obtain $P=35.02402(1)$\,d. We
imposed this value of the period with its standard error as a prior for our
analysis of the K2 light curve.
 
 This star is listed in the K2 Variable Catalogue \citep{2015A&A...579A..19A}
as an eclipsing binary, but no period estimate is given. High resolution
imaging by \citet{2016AJ....151..159S} did not detect any companions to this
star, with the quoted upper limit to the relative brightness at I-band being
3.18 magnitudes at 0.25\,arcsec. The location of these stars in the
T$_{\rm eff}$ -- $\rho_{\star}$ plane (Fig.~\ref{TeffDensity}) is consistent
with the assumptions that they are dwarf stars with masses $M_1\approx
1.1M_{\sun}$ and $M_2\approx 0.85M_{\sun}$ for any reasonable estimate of the
mass ratio, $q$.

\subsubsection{EPIC 201665500}

This star is included in our study because we initially assumed the
orbital period is approximately 6.1 days and that there are two similar
eclipses in the light curve. In fact, the orbital period is half this value
and there is a very shallow secondary eclipse visible in the K2 light curve.
The primary eclipse in this light curve is a transit of a solar-type star by a
low mass star. The secondary eclipse is very shallow compared to the star spot
modulation visible between the eclipses (few parts per thousand cf.
peak-to-peak amplitude $\approx 1.5$\%) so there is considerable scatter in
this secondary eclipse depth caused by dividing out the star modulation. As
the secondary eclipse is not well defined we decided to fix the third-light
value at $\ell_3 = 0$.

 High resolution imaging by \citet{2016AJ....151..159S} did not detect any
companions to this star, with the quoted upper limit to the relative
brightness at I-band being  2.37 magnitudes at 0.25\,arcsec. This star is
listed in the K2 Variable Catalogue \citep{2015A&A...579A..19A} as an
eclipsing binary with a period of 3.053723\,d, which agrees well with our
estimate of the orbital period. Star 1 has T$_{\rm eff}\approx 6300$\,K
while star 2 is very cool and much smaller than star 1 so we assume that this
system consists of a solar-type star and a K- or M-dwarf companion. In this
case $q\ll 1$ so the position of the stars in Fig.~\ref{TeffDensity} does not
depend strongly on the assumed value of $q$. From the location of these stars
in the T$_{\rm eff}$ -- $\rho_{\star}$ plane (Fig.~\ref{TeffDensity}) we
estimate that they are dwarfs stars with masses $M_1\approx 1.2M_{\sun}$ and
$M_2\approx 0.5M_{\sun}$.

\subsubsection{EPIC 201705526 = \object{BD +04$^{\circ}$ 2479}}

 The orbital period shown in Table~\ref{ellcpar1} was measured from 85,962
WASP photometric measurements obtained over 1148 days using the {\sc hunter}
algorithm \citep{2006MNRAS.373..799C}. This value is in fair agreement with
the  period of 18.120439\,d given in the K2 Variable Catalogue
\citep{2015A&A...579A..19A}. \citet{2016A&A...594A.100B} include this star in
their table of  planetary candidates. This appears to be based on the depth
and width of the secondary eclipse in the K2 light curve. We speculate that
their outlier rejection algorithm may have removed the narrow primary eclipse
data from the K2 light curve resulting in the misclassification of this
eclipsing binary as a transiting planet candidate. 

A good fit to the K2 light curve is also possible for solutions with a surface
brightness ratio $S_{\rm Kp}\approx$ 7 and $R_2/R_1 \approx 0.9$ but this
leads to estimates of the mean stellar densities and effective temperatures
that are not plausible. In contrast,  the location of these stars in the
T$_{\rm eff}$ -- $\rho_{\star}$ plane for the parameters we have adopted
(Fig.~\ref{TeffDensity}) suggests that they are dwarf stars with masses
$M_1\approx 1.3M_{\sun}$ and $M_2\approx 0.7M_{\sun}$. Both stars appear
near or below the zero-age main sequence for solar-metallicity models of stars
with these masses for any reasonable choice of the mass ratio, $q$.

\subsubsection{EPIC 201723461}

 We decided to fix the third-light parameter at the value $\ell_3 = 0$ since
the eclipses in this light curve are partial and the secondary eclipse is
quite shallow. Even with this assumption, the ratio of the radii is only
weakly constrained by the light curve.   This star is listed in the K2
Variable Catalogue \citep{2015A&A...579A..19A} as an eclipsing binary with a
period of 22.713572\,d, which agrees well with our estimate of the orbital
period. Although the plotted position of the cooler star is less dense than
the hotter star in Fig.~\ref{TeffDensity}, the uncertainty in the radius ratio
is large enough to accommodate solutions where these stars have mean densities
as expected for dwarf stars with masses $M\approx 0.7M_{\sun}$. Changing
the mass ratio from our assumed value of $q=1$ does not alter this
conclusion.

\begin{table}
\caption[]{\label{202674012abspar}
  Absolute astrophysical parameters of HD 149946 (EPIC 202674012).}
\begin{center}
\begin{tabular}{@{}lrr} 
\hline 
\hline\noalign{\smallskip}
 &\multicolumn{1}{l}{Primary} & \multicolumn{1}{l}{Secondary} \\
\noalign{\smallskip} 
\hline 
\noalign{\smallskip} 
Mass   [$M_{\sun}$] & 1.48  $\pm$  0.16 &  1.16   $\pm$ 0.13   \\ 
Radius [$R_{\sun}$] & 2.18  $\pm$ 0.07  &  1.22   $\pm$ 0.04   \\ 
$\log g$ [cm\,s$^{-2}$] & 3.93  $\pm$ 0.02  &  4.33   $\pm$ 0.02   \\
T$_{\rm eff}$ [K]   &  6250 $\pm$ 285   &  6150   $\pm$ 285    \\
$\log(L/L_{\sun})$  &  0.82 $\pm$ 0.08  &  0.29   $\pm$ 0.09   \\ 
M$_{\rm V}$ [mag]   &  2.7  $\pm$ 0.2   &  4.1    $\pm$ 0.2    \\ 
\noalign{\smallskip} 
Orbital period [d]  & \multicolumn{2}{c}{23.30962  $\pm$ 0.00005 } \\ 
Mass ratio & \multicolumn{2}{c}{0.78   $\pm$ 0.05 } \\ 
Distance      [pc]  & \multicolumn{2}{c}{265   $\pm$ 35 } \\ \noalign{\smallskip} \hline
\end{tabular} 
\end{center} 
\tablefoot{Absolute V magnitudes use bolometric corrections from
\citet{1998A+A...333..231B} and the distance is based on the apparent V-band
magnitude inferred from the observed Tycho-2 B$_{\rm T}$ and  V$_{\rm T}$
magnitudes.} 
\end{table}

\subsubsection{EPIC 202674012 = \object{HD 149946}}
 We downloaded four spectra of this star observed with the FEROS spectrograph 
from the ESO science archive. We used cross correlation over the wavelength
range 400\,--\,680\,nm against a numerical mask from an F0-type template star 
in iSpec \citep{2014A&A...569A.111B} to measure the radial velocities given in
Table~\ref{rvtable}. The full widths at half minimum of the dips in
the cross correlation function measured by a simultaneous fit of two Gaussian
profiles are 23\,km\,s$^{-1}$ and 17\,km\,s$^{-1}$ for star 1 and star 2,
respectively. The ratio of depths of these dips is 0.41, which is in
reasonable agreement with the value of  $\ell_{\rm Kp}$ given in
Table~\ref{ellcpar2} if some allowance is made for the different wavelength
range covered by these spectra cf. the Kepler band pass. 

 We used {\sc emcee} to find the best fit Keplerian orbit to these radial
velocity measurements including Gaussian priors on the parameters $f_s$,
$f_c$, $T_0$ and $P$ taken from the values shown in Table~\ref{ellcpar1}. We
assumed a single value for the standard error on these radial velocity
measurements and included this as a free parameter in the analysis by
including the appropriate term in the likelihood function.  The
semi-amplitudes derived from this fit are $K_1 = 45.3 \pm 2.4$\,km\,s$^{-1}$
and $K_2 = 57.8 \pm2.6$\,km\,s$^{-1}$, and the standard error for the
maximum-likelihood solution was 0.26 \,km\,s$^{-1}$.   The absolute
parameters of the stars derived from these values and the data in Tables
\ref{ellcpar1} and \ref{ellcpar2} are given in Table~\ref{202674012abspar}.
The spectral type is F3(V) \citep{1982MSS...C03....0H}, which implies a mean
value of T$_{\rm eff} \approx 6435$\,K \citep{2013ApJ...771...40B}. This agrees
well with our estimates for T$_{\rm eff,1}$ and T$_{\rm eff,2}$ in
Table~\ref{TeffTable}. There is also good agreement between the measured masses
of the stars and their expected masses given their position in
Fig.~\ref{TeffDensity} relative to stellar evolution tracks for solar
composition.

\subsubsection{EPIC 202843085}
 The location of these stars in the T$_{\rm eff}$ -- $\rho_{\star}$ plane
(Fig.~\ref{TeffDensity}) suggests that they are a pair of dwarf stars with
masses $M\approx 1.4M_{\sun}$ near the end of the main sequence. Star 2
is larger than star 1 so $q>1$ but it is very unlikely that both stars would
appear near the  MSTO if they have very different masses so we assume the
value $q=1$ for purposes of plotting these stars in
Fig.~\ref{TeffDensity}.

\subsubsection{EPIC 203361171}

 We used different apertures to calculate the light curve of this star for
images obtained before and after a change in the spacecraft orientation near
BJD 2456936.8. Both apertures include a star approximately 21  arcseconds to
the south-west of the main target and 2.4 magnitudes fainter in the G band
according to the Gaia DR1 data release \citep{2016A&A...595A...1G}. We did not
include the value of $\ell_3$ given in Table~\ref{ellcpar2} in our analysis to
estimate the effective temperatures of the stars because we assume that this
value is dominated by the star 21  arcseconds to the south-west of the main
target whose flux will not be included in the published catalogue photometry.
The location of these stars in the T$_{\rm eff}$ -- $\rho_{\star}$ plane
(Fig.~\ref{TeffDensity}) suggests that they are a pair of dwarf stars both
with masses $M\approx 1.2M_{\sun}$ and near the end of the main sequence. 
This is a similar case to EPIC 205703649 so we again assume $q=1$ for the
purposes of plotting these stars in Fig.~\ref{TeffDensity}.

\subsubsection{EPIC 203371239}
\begin{figure*} 
\includegraphics[width=0.99\textwidth]{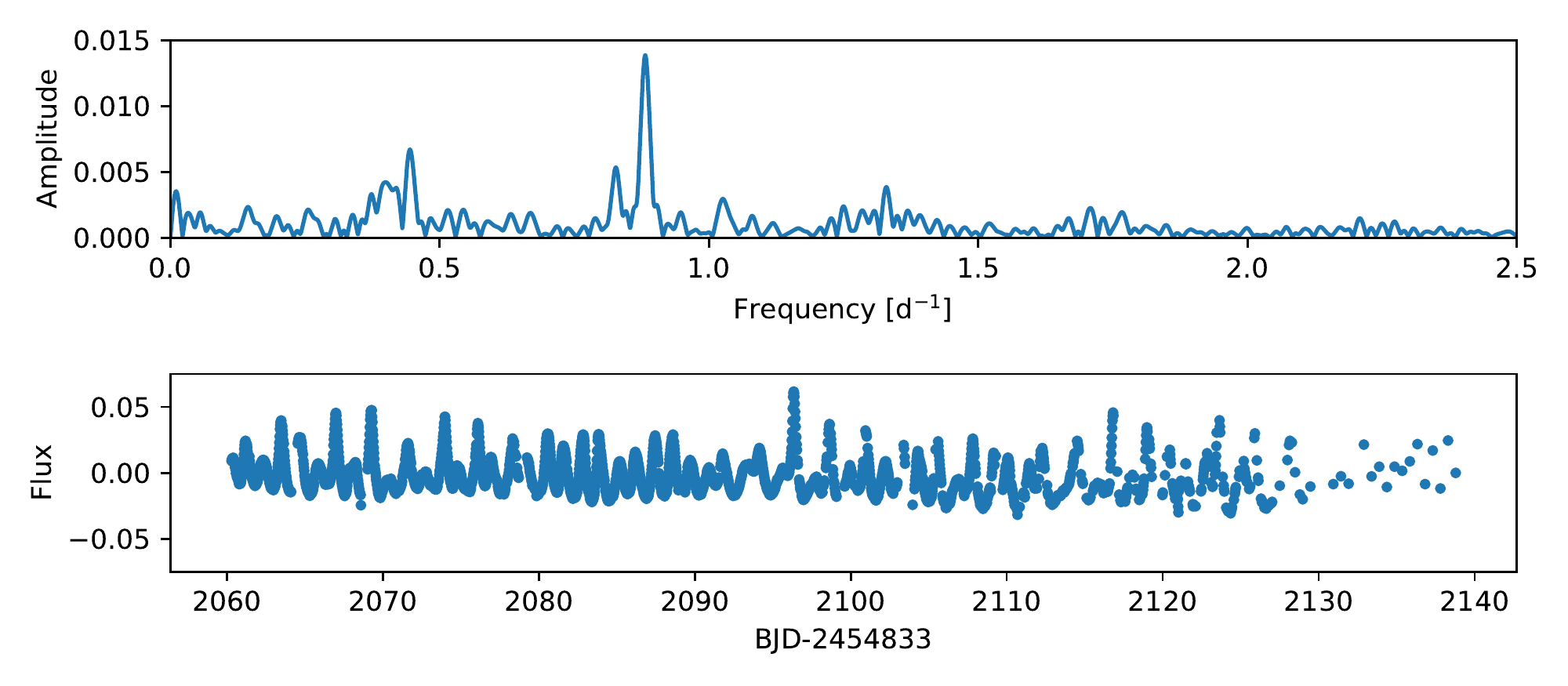}
\caption{Light curve and frequency spectrum of EPIC~203371239 excluding data
obtained during eclipse. A low order polynomial fit by least-squares has been
subtracted from the data shown in the lower panel. The flux is measured
relative to the mean flux between the eclipses.} 
\label{203371239_ft} 
\end{figure*}

 The light curve of this star between eclipses shows a very clear signal due
to multi-periodic pulsations (Fig.~\ref{203371239_ft}) with frequencies near
0.8 cycles/day and 0.4 cycles/day, and amplitudes of about 1\%. These
frequencies and amplitudes taken with the effective temperature estimates
given in Table~\ref{TeffTable} suggest that one or both of the stars in this
binary is a $\gamma$ Dor-type pulsator \citep{2011MNRAS.415.3531B}. The
location of these stars in the T$_{\rm eff}$ -- $\rho_{\star}$ plane
(Fig.~\ref{TeffDensity}) suggests that they are a pair of dwarf stars with
masses $M\approx 1.3M_{\sun}$ and $M\approx 1.2M_{\sun}$. This is a
similar case to EPIC 202843085 and EPIC 202843085 so we again assume $q=1$ for
the purposes of plotting these stars in Fig.~\ref{TeffDensity}.

\subsubsection{EPIC 203543668}
 The photometric aperture we used to construct the K2 light  includes the flux
from some nearby stars, but this is not enough to account for the value of
$\ell_3$ we obtain from the fit to the light curve. The
location of these stars in the T$_{\rm eff}$ -- $\rho_{\star}$ plane
(Fig.~\ref{TeffDensity}) suggests that the primary is a star similar to the
Sun and the secondary is a dwarf star with a mass $M\approx 0.7M_{\sun}$. 
Both stars appear near the zero-age main sequence for solar-metallicity models
of stars with these masses for any reasonable choice of the mass ratio, $q$.

\subsubsection{EPIC 203610780}
 The parameters we have derived for this binary system from the analysis of
the K2 light curve are quite robust because the eclipses are total. Star 2 is
much larger and cooler than star 1 so we can assume $q>1$, but the actual
value of $q=1.2$ used to plot the stars in Fig.~\ref{TeffDensity} is quite
uncertain. The position of the hotter star below the zero-age main sequence
for solar-type stars suggests that this may be a low-metallicity system. This
conclusion is not affected by the exact choice of $q$. The complicating factor
for this interpretation is the large amount of third light in this system that
leads to large uncertainties in the effective temperature estimates. 

\subsubsection{EPIC 203636784}
The rotation signal in the K2 light curve has an amplitude of about 1.5\% at
the start of the observing sequence that gradually decreases to an amplitude
of about 0.5\%. The rotation period is consistent with the assumption of
pseudo-synchronous rotation. Star 1 has T$_{\rm eff}\approx 6000$\,K while
star 2 is much cooler and smaller than star 1 so we assume that this system
consists of a solar type star and a K- or M-dwarf companion. The position of
the stars in Fig.~\ref{TeffDensity} does not depend strongly on the assumed
value of $q$ provided that this value is significantly less than 1. The
location of these stars in the T$_{\rm eff}$ -- $\rho_{\star}$ plane
(Fig.~\ref{TeffDensity}) suggests that the primary is a star near the
main-sequence turn-off with a mass $M\approx 1.1M_{\sun}$ and the secondary is
a dwarf star with a mass $M\approx 0.7M_{\sun}$.

\subsubsection{EPIC 203728604}
\begin{figure*} 
\includegraphics[width=0.99\textwidth]{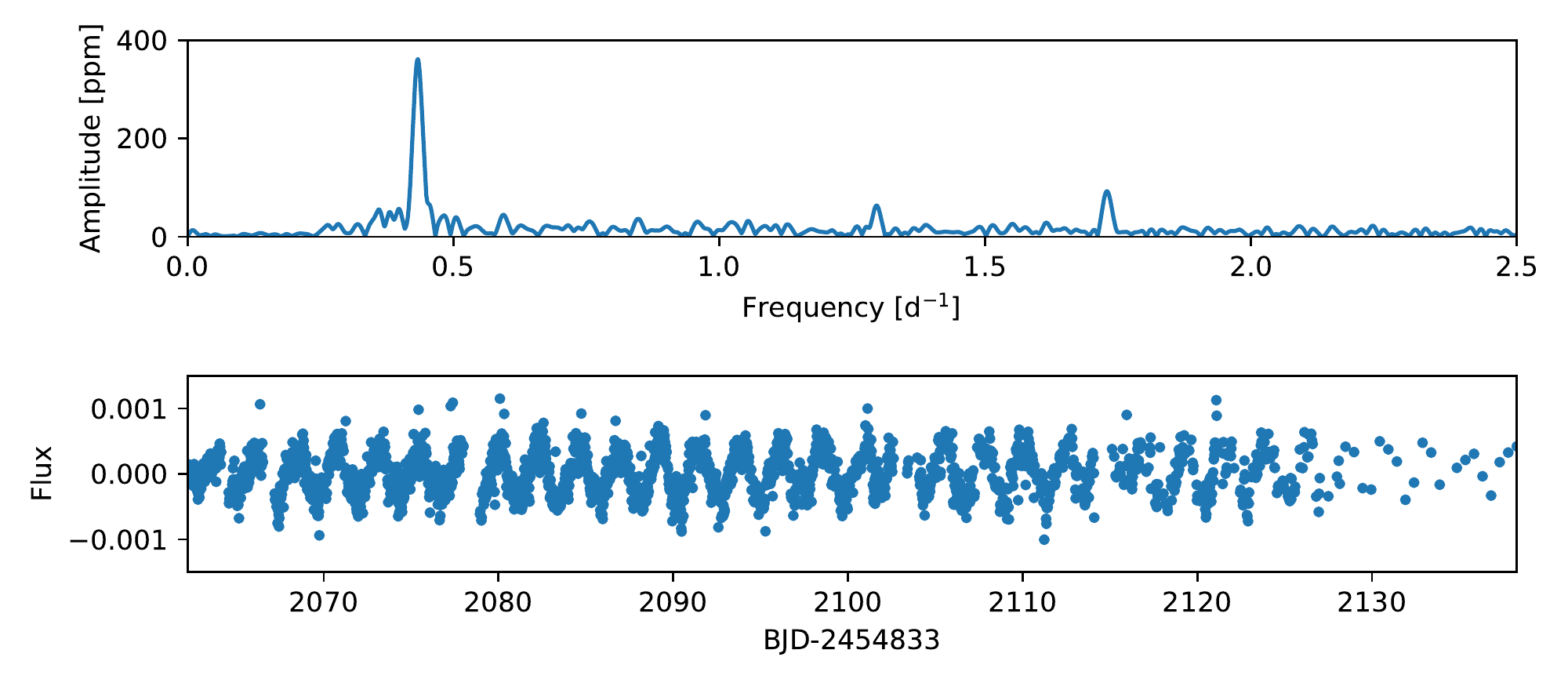}
\caption{Light curve and frequency spectrum of EPIC~203728604 excluding data
obtained during eclipse. A low order polynomial fit by least-squares has been
subtracted from the data shown in the lower panel. The flux is measured
relative to the mean flux between the eclipses.} 
\label{203728604_ft} 
\end{figure*}

 The K2 light curve of this star between the eclipse shows a well defined
periodic signal with a period of 2.306\,d and an amplitude of about 400\,ppm.
The coherence of this signal suggests that this is a pulsation signal rather
than rotational modulation due to star spots, perhaps due to $\gamma$~Dor-type
pulsations in one of the stars. The location of these stars in the T$_{\rm
eff}$ -- $\rho_{\star}$ plane (Fig.~\ref{TeffDensity}) suggests that the
primary is a star near the main-sequence turn-off with a mass  $M\approx
1.5M_{\sun}$ and the secondary is a star similar to the Sun. This conclusion
is not affected by the choice of mass ratio for any value $q\ga 0.8$. The
mass ratio is almost certainly has a value $q>1$ since star 1 has a much
larger radius than star 2. The periodogram of the data between the eclipses
for this star is shown in Fig.~\ref{203728604_ft}.

\subsubsection{EPIC 204407880}
 The WASP 200-mm data for this star include three nights covering the primary
eclipse. To measure the times of mid-eclipse from these data we used a model
with the geometric parameters fixed to the values determined from a
preliminary fit to the K2 light curve. The three times of mid-eclipse and the
surface brightness ratio in the WASP bandpass were free parameters in a fit to
the WASP data using {\sc emcee} to determine the optimum value of these
parameters and their standard errors. The times of mid-eclipse derived using
this method were BJD$_{\rm TDB}-2450000=$  3893.334(2), 4271.385(1),
4649.435(2), where the values in parentheses denote the standard error in the
final digit. From a linear fit to these times of mid-eclipse plus the value
2456917.71237(14) from a preliminary fit to the K2 light curve we find an
orbital period of $34.36789 \pm 0.00002$\,d. This period was included as a
prior in the analysis of the K2 light curve.

\subsubsection{EPIC 204576757}
 This star is listed as a planetary candidate system with a period of
23.277669 days by \citet{2016ApJS..222...14V}, although the estimated radius
of the companion ($\sim 3 R_{\rm Jup}$) is rather large for a planetary-mass
object. Three total eclipses due to the transit of the companion are visible
in the K2 light curve but there is no clear secondary eclipse visible in these
data. This may be because the companion contributes less than about 0.25\% of
the flux at optical wavelengths, or the orbital eccentricity may be large
enough for there to be no secondary eclipse. Given this ambiguity over the
configuration of this binary system we have not attempted any further analysis
of the K2 light curve. 

\subsubsection{EPIC 204748201}

 Although this is a binary with total eclipses, the secondary eclipse is very
shallow so including third light contamination in the analysis results in
parameters that have large uncertainties. We decided to fix the third light
parameter at $\ell_3 = 0$ for this preliminary characterisation of this
system.  Star 1 has T$_{\rm eff}\approx 6100$\,K while star 2 is very cool and
much smaller than star 1 so we assume that this system consists of a solar
type star and a K-dwarf companion. In this case $q$ must be significantly less
than 1, so the position of the stars in Fig.~\ref{TeffDensity} does not depend
strongly on the assumed value of $q$. With these assumptions, the location of
the stars in the T$_{\rm eff}$ -- $\rho_{\star}$ plane
(Fig.~\ref{TeffDensity}) suggests that they are dwarf stars near the zero-age
main sequence with masses $M\approx 1.2M_{\sun}$ and $M\approx 0.5M_{\sun}$.

\subsubsection{EPIC 204760247 = \object{HD 142883}}

 This bright B3V star (V=5.84) is listed in SIMBAD as a Cepheid variable star
-- this is not correct. \citet{1967ApJS...14..263H} found this star to be a
variable with a possible period  0.2872\,days  based on 20 observations in
each of the U and B bands but note that "because of the extremely small
amplitude of the variation \dots this star must be considered a {\it
tentative} $\beta$ Cephei variable."  \citet{1977A&AS...29..309A} noted that
this star is a double-lined spectroscopic binary with the secondary component
``much fainter''.  In a later study \citep{1983A&AS...52..471A} they estimated
a mass ratio for this binary of $0.38\pm0.03$. \citet{1987ApJS...64..487L}
report 8 radial velocity measurements from which they claimed the first
spectroscopic solution for this star with an orbital period near 10\, days,
but with a very large eccentricity that is not consistent with the data
described below. \citet{2002MNRAS.331...45K} note that this is a variable star
on the basis of the  Hipparcos epoch photometry but were not able to classify
the type variability. \citet{2011MNRAS.416.2477W} used observations from the
STEREO mission to correctly identify the variability of this star as being due
to eclipses with a period of 9.20\,days. This star is a member of the Upper
Scorpius OB association \citep{2002A&A...381..446M}. 

\begin{figure*} 
\includegraphics[width=0.99\textwidth]{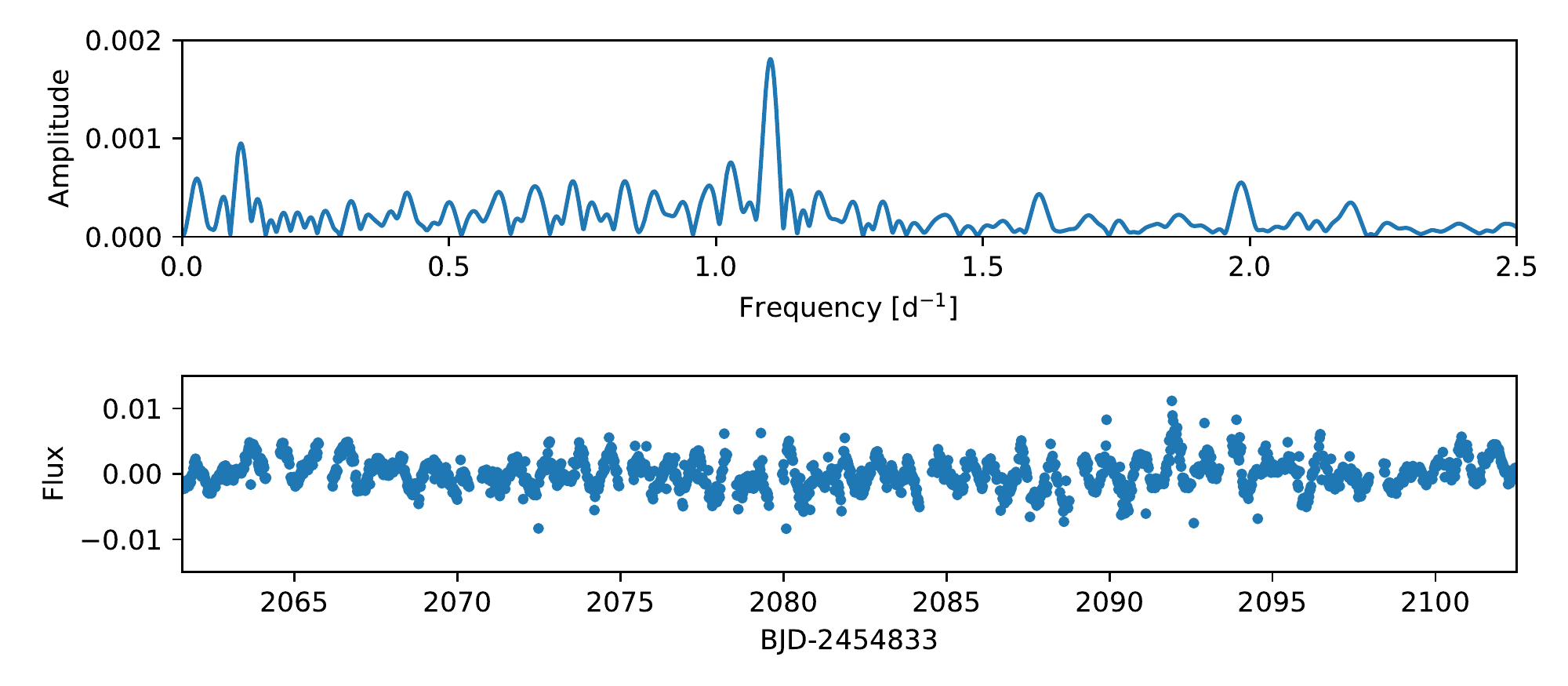}
\caption{Light curve and frequency spectrum of EPIC~204760247 excluding data
obtained during eclipse. A low order polynomial fit by least-squares has been
subtracted from the data shown in the lower panel. The flux is measured
relative to the mean flux between the eclipses.} 
\label{204760247_ft} 
\end{figure*}

 We conducted aperture photometry for this star including the extensive charge
overspill region provided in the K2 target pixel file. This provided useful
photometry for the interval BJD 2456894.5 to 2456935.5. There is a very clear
pulsation signal in the data between the eclipses with an period of 0.908 days
and an amplitude of 0.18\% (Fig.~\ref{204760247_ft}). There is also a periodic
signal in these data with a period close to the orbital period and an
amplitude of 0.1\% that may be due to irradiation of the companion star by the
B3V primary star. We have not included this effect in our model of the light
curve.  The surface brightness ratio from a preliminary light curve solution
combined with an estimate for the primary star effective temperature  $T_{\rm
eff}=18,000$\,K based on its spectral type implies $T_{\rm eff}\approx
10,000$\,K for the secondary star. We used these $T_{\rm eff}$ estimates and
the tabulation by \citet{2011A&A...529A..75C} to estimate the quadratic limb
darkening coefficients  $(a,b) = (0.11,0.24)$ and $(0.21, 0.29)$ for the
primary and secondary, respectively. We assume standard errors of 0.05 on all
these coefficients when imposing them as priors in the light curve analysis.

 We downloaded six spectra of this star observed with the FEROS spectrograph 
from the ESO science archive. We used cross correlation over the wavelength
range 400\,--\,680\,nm against a numerical mask from an A0-type template star
in iSpec \citep{2014A&A...569A.111B} to measure the radial velocities given in 
Table~\ref{rvtable}.  The full widths at half minimum (FWHM) of the
dips in the cross correlation function (CCF)  measured by a simultaneous fit
of two Gaussian profiles are 22\,km\,s$^{-1}$ and 28\,km\,s$^{-1}$ for star 1
and star 2, respectively. A third dip is visible in the CCF with a radial
velocity of $-13$\,km\,s$^{-1}$ and FWHM of 15\,km\,s$^{-1}$ and a strength
approximately half that of the peak for star 2.  The mismatch between the
spectral type of the primary star and the template in this case makes it
difficult to interpret the strength of the dip in the CCF for this star -- no
template is available for spectral type earlier than A0 in the current version
of iSpec. 

 We used {\sc emcee} to find the best fit Keplerian orbit to these radial
velocity measurements assuming a circular orbit ($e=0$). We did not find a
satisfactory fit to these data using the values of  
$T_0$ and $P$ taken from the values shown in Table~\ref{ellcpar1}. Instead we
noted that there is a secondary eclipse visible in the Hipparcos epoch
photometry for this star and used this to estimate an orbital period of
9.199724(4) days. With this orbital period imposed as a prior we find that the
semi-amplitudes are $K_1 = 62.8 \pm 1.7$\,km\,s$^{-1}$ and $K_2 = 136.6
\pm1.4$\,km\,s$^{-1}$. The masses of the stars are $M_1 = 5.18\pm
0.15$\,M$_{\odot}$ and $M_2 = 2.38\pm 0.11$\,M$_{\odot}$ and their radii are
$R_1 = 2.50 \pm 0.04$\,R$_{\odot}$ and $R_2 = 1.63\pm 0.04$\,R$_{\odot}$.

\subsubsection{EPIC 204822807}
The argument that leads to the conclusion $q\approx 1$ for EPIC~201161715
also applies to this binary system, but in this case it is star 2 that is the
cooler and larger star so we assume the value $q= 1.1$ to plot the position of
the stars in Fig.~\ref{TeffDensity}. This system is a bright, totally
eclipsing binary system that contains a star on the red giant branch and a
star with a mass similar to the Sun near the main-sequence turn-off point,
similar to the well-known systems AI~Phe \citep{2016A&A...591A.124K} and
TZ~For \citep{2017A&A...600A..41V}. This makes this system an attractive
target for calibrating stellar models.

\subsubsection{EPIC 204870619} 
 This is a long-period binary in which a sub-giant star with a mass $\approx
1.2\,M_{\odot}$ produces total eclipses of a dwarf star with a mass $\approx
0.8\,M_{\odot}$. The position of the stars in the T$_{\rm
eff}$\,--\,$\rho_{\star}$ plane lie near the stellar evolution tracks for
these masses for any reasonable choice of $q$ so these masses should be quite
accurate. It may be challenging to measure precise radial velocities for the
dwarf star in this binary since it only contributes about 5\% of the flux at
optical wavelengths and the system is quite faint. Nevertheless, this is an
attractive target for follow-up observations to test stellar models given
that, in principle, the masses and radii for these stars can be determined to
an accuracy much better than 1\% and there do not appear to be any
complications in the analysis due to star spots or pulsations.

\subsubsection{EPIC 205020466}

 We obtained 4 spectra of this star using the High Resolution Spectrograph
(HRS) on the Southern African Large Telescope (SALT)
\citep{2014SPIE.9147E..6TC}. We used the medium resolution mode to obtain
spectra at a resolving power $R=43,000$ on the blue arm of the instrument and
$R=40,000$ on the red arm. The exposure time of 577\,s on both arms resulted
in a signal-to-noise per pixel of approximately 10 on the blue arm and 20 on
the red arm. We used spectra reduced automatically using the pipeline
developed by \citet{2017ASPC..510..480K} for our analysis.  We used iSpec
\citep{2014A&A...569A.111B} to measure the radial velocities for both stars
from Gaussian profile fits to the cross-correlation function (CCF) calculated
using a numerical mask based on the solar spectrum.  The results of these fits
are given in Table~\ref{rvtable}. We did not attempt any further analysis of
the HRS spectra because the signal-to-noise is quite low and the reduced
spectra are not corrected for the blaze function of the spectrograph. 

 We used {\sc emcee} to find the best fit Keplerian orbit to these radial
velocity measurements including Gaussian priors on the parameters $f_s$,
$f_c$, $T_0$ and $P$ taken from the values shown in Table~\ref{ellcpar1}. We
assumed a single value for the standard error on these radial velocity
measurements and included this as a free parameter in the analysis by
including the appropriate term in the likelihood function. The semi-amplitudes
derived from this fit are $K_1 = 60.5 \pm 1.3$\,km\,s$^{-1}$ and $K_2 = 75.5
\pm 1.3 $\,km\,s$^{-1}$ and the standard error per observation derived is
2.0\,km\,s$^{-1}$. The absolute parameters of the stars derived from these
values and the data in Tables \ref{ellcpar1} and \ref{ellcpar2} are given in
Table~\ref{205020466abspar}. The masses in this table agree well with the
values that can be inferred from the location of the stars in the T$_{\rm
eff}$ -- $\rho_{\star}$ plane (Fig.~\ref{TeffDensity})  if the errors in
T$_{\rm eff}$ are accounted for.

\begin{table}
\caption[]{\label{205020466abspar}
  Absolute astrophysical parameters of EPIC 205020466.}
\begin{center}
\begin{tabular}{@{}lrr} 
\hline 
\hline\noalign{\smallskip}
 &\multicolumn{1}{l}{Primary} & \multicolumn{1}{l}{Secondary} \\
\noalign{\smallskip} 
\hline 
\noalign{\smallskip} 
Mass   [$M_{\sun}$]     & 1.05  $\pm$  0.04 &  0.84   $\pm$ 0.04   \\ 
Radius [$R_{\sun}$]     & 0.89  $\pm$ 0.015 &  0.86   $\pm$ 0.015  \\ 
$\log g$ [cm\,s$^{-2}$] & 4.56  $\pm$ 0.01  &  4.49   $\pm$ 0.015  \\ 
T$_{\rm eff}$ [K]       &  5300 $\pm$ 480   &  5070   $\pm$ 425    \\
$\log(L/L_{\sun})$      & $-0.25 \pm  0.16$ &$ -0.35   \pm  0.15 $ \\ 
M$_{\rm V}$ [mag]       &  5.5  $\pm$ 0.5   &  5.9    $\pm$ 0.5    \\ 
\noalign{\smallskip} 
Orbital period [d]  & \multicolumn{2}{c}{8.75903   $\pm$ 0.00003 } \\ 
Mass ratio & \multicolumn{2}{c}{0.80   $\pm$ 0.02 } \\ 
Distance      [pc]  & \multicolumn{2}{c}{340   $\pm$ 120} \\ \noalign{\smallskip} \hline
\end{tabular} 
\end{center} 
\tablefoot{Absolute V magnitudes use bolometric corrections from
\citet{1998A+A...333..231B} and the distance is based on the apparent V-band
magnitude from APASS9.}
\end{table}

\subsubsection{EPIC 205170307}  
The analysis of this system is complicated by substantial 3rd light
contamination ($\ell_3\approx 17$\%) but the eclipses are total and
well-defined so a robust determination of the system parameters is possible.
The eclipsing pair are dwarf stars with masses $\approx  0.65\,M_{\odot}$ and
$\approx 1.0\,M_{\odot}$ with the more massive star near the main-sequence
turn-off point in Fig.~\ref{TeffDensity}. The position of the stars in the
T$_{\rm eff}$\,--\,$\rho_{\star}$ plane lie near the stellar evolution tracks
for these masses for any reasonable choice of the mass ratio, $q$.

\subsubsection{EPIC 205546169} 
Based on the parameters we have derived, the eclipsing pair in this system are
both  dwarf stars with masses $\approx 1.2\,M_{\odot}$ with one star near the
zero-age main sequence  and one near the end of the main sequence. This
conclusion does not depend on the assumed mass ratio for any reasonable choice
of $q$. We assume $q=1.1$ to plot these stars in Fig.~\ref{TeffDensity} since
star 2 is apparently more evolved than star 1. This combination is difficult
to reconcile with the very similar effective temperatures for the two stars.
The secondary eclipse in this system is very shallow, the eclipses are partial
and there may be third-light contamination $\ell_3 \approx 9$\% so the
parameters we have derived here may be subject to quite large systematic
error. Spectroscopic observations to determine more robust estimates for
$\ell_3$ and $\ell_{\rm Kp}$ will be very helpful for the analysis of this
system.

\subsubsection{EPIC 205703649} 

 The photometric aperture we used for this star is contaminated by nearby
stars of comparable brightness to the target star. This is accounted for in
the photometric fit by including $\ell_3$ as a free parameter and accounts for
the large value of this parameter. We have assumed that this contaminating
flux does not affect the catalogue photometry for the target star and so we
set $\ell_3=0$ to derive the effective temperature estimates in
Table~\ref{TeffTable}. With this assumption, both stars appear to be dwarf
stars near the MSTO with masses $\approx 1.0\,M_{\odot}$. Star 2 is larger
than star 1 so $q>1$ seems a reasonable choice. However, it is very unlikely
that both stars would appear near the MSTO if they have different masses,
therefore we assume the value $q=1$ for the purposes of plotting these stars
in Fig.~\ref{TeffDensity}.

\subsubsection{EPIC 205919993  = \object{LP 819-72}}
LP 819-72 was identified as an eclipsing binary using data from the WASP
project prior to the start of the K2 mission. The primary eclipse was never
observed with WASP because the orbital period is so close to exactly 11 days.
As such, it was misclassified as an eclipsing binary with a transiting low
mass companion (``EBLM'') with an orbital period of 3.666 days.

 We submitted LP 819-72 to the guest observer program on K2 and also obtained
33 spectra of this system with the fibre-echelle spectrograph on the CTIO
1.5-m telescope operated by the SMARTS Consortium. The spectra were typically
observed in groups of three with an exposure time of 900\,s plus an
accompanying arc spectrum for wavelength calibration. We extracted a single
order from these echelle spectra using the optimal extraction routines and
wavelength calibration routines {\sc pamela} and {\sc molly}
\citep{1989PASP..101.1032M}.  The spectral order selected covers the
wavelength range 660.9\,--\,647.7\,nm and the resolving power of the
instrument is $R\approx 37,000$. The signal-to-noise ratio per pixel at the
centre of the order is typically ${\rm S/N} \approx 20$. We used iSpec
\citep{2014A&A...569A.111B} to measure the radial velocities for both stars in
those spectra where the lines from the two stars are clearly resolved. The
radial velocities were measured using Gaussian profile fits to the
cross-correlation function (CCF) calculated using a numerical mask based on a
K5 spectral-type template.  The results of these fits are given in
Table~\ref{rvtable}. The individual spectra observed on the night JD 2455477
were of lower signal-to-noise than other spectra so we co-added these spectra
for analysis.  We also co-added the group of three spectra with the highest S/N
in order to look for additional dips in the CCF. No such dips were detected so
we estimate that the contribution from any third star in the system does not
exceed about 10\% at these wavelengths, assuming that any such star is a
slowly rotating star with a late-type spectrum.

For simplicity in the analysis below we fixed the third light parameter
$\ell_3=0$  for our analysis of the K2 light curve. We also imposed a prior on
the flux ratio $\ell_{\rm Kp}$ from the ratio of the depths of dips in the
CCF. The mean and standard error in the mean in this ratio is $1.28\pm0.03$
but we use a Gaussian prior on $\ell_{\rm Kp}$ with mean 1.28 and standard
deviation 0.05 to allow for some uncertainty in converting the depth of the
dip in the CCF to a flux in the Kepler bandpass. This information from the
spectroscopy is extremely useful for the analysis of the K2 light curve
because without these priors on $\ell_3$ and the flux ratio the best-fit
solutions tend to imply a flux ratio for the stars that is inconsistent with
the radius ratio. 

\begin{table}
\label{rvtable}
\caption{Radial velocity measurements.}
\begin{center}
\begin{tabular}{@{}lrr} 
\hline 
\hline\noalign{\smallskip}
\multicolumn{1}{l}{BJD (UTC)} & 
\multicolumn{1}{r}{$V_{r,1}$} &
\multicolumn{1}{r}{$V_{r,2}$}  \\
\multicolumn{1}{l}{$-2450000$} & 
\multicolumn{1}{r}{[km\,s$^{-1}$]} &
\multicolumn{1}{r}{[km\,s$^{-1}$]} \\
\noalign{\smallskip} 
\hline 
\noalign{\smallskip} 
  \multicolumn{3}{@{}l}{EPIC 202674012, FEROS} \\
6100.635 &$-34.2 \pm 0.3$ &$ 63.0 \pm 0.3$  \\
6102.545 &$-21.0 \pm 0.3$ &$ 46.9 \pm 0.3$  \\
6517.565 &$-32.4 \pm 0.3$ &$ 61.9 \pm 0.3$  \\
7174.732 &$-21.6 \pm 0.3$ &$ 47.4 \pm 0.3$  \\
\noalign{\smallskip} 
  \multicolumn{3}{@{}l}{EPIC 205020466, SALT HRS} \\
7810.574 &$ -17.1 \pm 2.0 $&$ -103.3 \pm 2.0$ \\
7810.574 &$ -16.2 \pm 2.0 $&$ -102.5 \pm 2.0$\\
7843.498 &$ -14.6 \pm 2.0 $&$  -97.0 \pm 2.0$ \\
7843.498 &$ -14.2 \pm 2.0 $&$  -95.6 \pm 2.0$\\
7844.484 &$  -4.5 \pm 2.0 $&$ -104.9 \pm 2.0$ \\
7844.484 &$  -6.8 \pm 2.0 $&$ -104.0 \pm 2.0$\\
7865.654 &$-128.4 \pm 2.0 $&$   44.2 \pm 2.0$ \\
7865.654 &$-128.2 \pm 2.0 $&$   43.9 \pm 2.0$\\
\noalign{\smallskip} 
  \multicolumn{3}{@{}l}{EPIC 205919993, CTIO 1.5-m}\\
5429.806  &$ 50.35 \pm 0.39 $&$ -49.73 \pm 0.24$  \\
5429.817  &$ 51.88 \pm 0.31 $&$ -49.10 \pm 0.35$  \\
5429.827  &$ 51.52 \pm 0.61 $&$ -49.21 \pm 0.26$  \\
5439.699  &$ 55.18 \pm 0.62 $&$ -52.34 \pm 0.64$  \\
5439.710  &$ 55.40 \pm 0.56 $&$ -52.70 \pm 0.35$  \\
5439.721  &$ 56.07 \pm 1.05 $&$ -52.74 \pm 0.49$  \\
5445.628  &$-47.66 \pm 0.45 $&$  45.75 \pm 0.29$  \\
5445.638  &$-48.10 \pm 0.34 $&$  45.72 \pm 0.36$  \\
5445.649  &$-48.45 \pm 0.32 $&$  47.00 \pm 0.44$  \\
5477.643  &$-45.65 \pm 0.43 $&$  42.89 \pm 0.42$  \\
5490.594  &$-42.71 \pm 0.44 $&$  37.00 \pm 0.35$  \\
5490.605  &$-41.47 \pm 0.55 $&$  36.30 \pm 0.37$  \\
5490.615  &$-42.89 \pm 0.51 $&$  36.10 \pm 0.28$  \\
5510.519  &$-49.12 \pm 0.37 $&$  35.58 \pm 0.33$  \\
5510.530  &$-48.05 \pm 0.31 $&$  35.64 \pm 0.30$  \\
5510.540  &$-49.56 \pm 0.52 $&$  35.88 \pm 0.33$  \\
\noalign{\smallskip} 
  \multicolumn{3}{@{}l}{EPIC 204760247, FEROS}\\
3129.725&$ 60.6\pm 7.1$&$-135.5\pm 1.6 $\\
3129.735&$ 59.6\pm 4.9$&$-135.2\pm 1.7 $\\
4298.510&$ 54.5\pm 2.3$&$-120.8\pm 2.1 $\\
4302.491&$-65.2\pm 2.3$&$ 133.4\pm 1.7 $\\
6523.599&$ 56.8\pm 1.5$&$-132.1\pm 1.8 $\\
\noalign{\smallskip} 
\hline 
\end{tabular} 
\end{center} 
\end{table}

We used {\sc emcee} to find the best fit Keplerian orbit to the radial
velocity measurement in Table~\ref{rvtable} including Gaussian
priors on the parameters $f_s$, $f_c$, $T_0$ and $P$ taken from the values
shown in Table~\ref{ellcpar1}. We find that the semi-amplitudes are $K_1 =
54.1 \pm 0.6$\,km\,s$^{-1}$ and $K_2 = 49.2 \pm 0.6 $\,km\,s$^{-1}$. 
 The absolute parameters of the stars derived from these values and the
data in Tables \ref{ellcpar1} and \ref{ellcpar2} are given in
Table~\ref{205919993abspar}.

\begin{table}
\caption[]{\label{205919993abspar}
  Absolute astrophysical parameters of LP 819-72 (EPIC 205919993).}
\begin{center}
\begin{tabular}{@{}lrr} 
\hline 
\hline\noalign{\smallskip}
 &\multicolumn{1}{l}{Primary} & \multicolumn{1}{l}{Secondary} \\
\noalign{\smallskip} 
\hline 
\noalign{\smallskip} 
Mass   [$M_{\sun}$] & 0.59  $\pm$  0.02 &  0.65   $\pm$ 0.02   \\ 
Radius [$R_{\sun}$] & 0.60  $\pm$ 0.01  &  0.58   $\pm$ 0.01   \\ 
$\log g$ [cm\,s$^{-2}$] & 4.65  $\pm$ 0.02  &  4.72   $\pm$ 0.02 \\
T$_{\rm eff}$ [K]   &  4025 $\pm$  60   &  4230   $\pm$  60    \\
$\log(L/L_{\sun})$  & $-1.07 \pm  0.03$ &$ -1.01   \pm  0.03 $ \\ 
M$_{\rm V}$ [mag]   &  8.38 $\pm$ 0.13  &  8.05   $\pm$ 0.12   \\ 
\noalign{\smallskip} 
Orbital period [d]  & \multicolumn{2}{c}{11.00009  $\pm$ 0.00006 } \\ 
Mass ratio & \multicolumn{2}{c}{1.10   $\pm$ 0.02 } \\ 
Distance      [pc]  & \multicolumn{2}{c}{42.1  $\pm$ 1.1} \\ 
\noalign{\smallskip} \hline
\end{tabular} 
\end{center} 
\tablefoot{Absolute V magnitudes use bolometric corrections from
\citet{1998A+A...333..231B} and the distance is based on the 2MASS 
K$_{\rm s}$-band magnitude transformed to Johnson K-band and the surface 
brightness -- T$_{\rm eff}$ relation by \citet{2004A&A...426..297K}.}
\end{table}
 
 The distance to this system based on the parallax measurement from Gaia DR1 is
$44.9\pm0.5$\,pc. The distance to this system based on the 2MASS K$_{\rm
s}$-band magnitude transformed to Johnson K-band and the surface brightness --
T$_{\rm eff}$ relation by \citet{2004A&A...426..297K} is $42.1\pm 1.0$\,pc,
which is a fair agreement with the Gaia DR1 estimate, particularly if the
suspected systematic error of about 0.22\,mas in Gaia DR1 parallax values for
stars near the ecliptic is taken into account \citep{2016arXiv160905390S}. It
remains to be seen whether a more accurate estimate of $\ell_3$ will lead to
better agreement between these distance estimates.
 
 This high proper motion star has a spectral type of K5V
\citep{1986AJ.....92..139S}. This implies an effective temperature T$_{\rm eff}
\approx 4436$\,K, which agrees reasonably well with the average of our
estimates of T$_{\rm eff}$ for the two stars in Table~\ref{TeffTable}.  

\subsubsection{EPIC 205982900 = \object{BW Aqr}}

 A detailed study of this eclipsing binary has been presented by
\citet{1991A&A...246..397C} based on {\it uvby} light curves by
\citet{1987A&AS...68..323G} and spectroscopic orbits for both components from
\citet{1987A&AS...69..397I}. The times of primary and secondary eclipse show
apsidal motion with a period of approximately 6000 years
\citep{2014CoSka..43..419V}. The projected equatorial rotational velocities of
the stars suggest that they both rotate at about half the rate  expected
assuming pseudo-synchronisation of the rotation with the orbital angular
velocity at periastron. 

 According to linear ephemeris by \citet{2004AcA....54..207K} the 2MASS
observations of this star were obtained during primary eclipse  so we did not
include these data in the analysis to determine the T$_{\rm eff}$ estimates in
Table~\ref{TeffTable}. These T$_{\rm eff}$ estimates are about 250\,K cooler than
the values from \citet{1991A&A...246..397C} based on the dereddened $(b-y)_0$
colour indices, which is significant at the 2-$\sigma$ level. This is a
consequence of the larger reddening to this system derived by Clausen 
from Str\"{o}mgren photometry compared to the value we obtained from
broad-band photometry together with the prior on E($\rm B-\rm V$) from reddening maps.
The position of these stars in the T$_{\rm eff}$ -- $\rho_{\star}$ plane
(Fig.~\ref{TeffDensity})  suggest that the higher effective temperature
estimate is more reliable given that the masses of these stars are known to be
$M\approx 1.4M_{\odot}$.

 The K2 light curve of BW~Aqr shows an ellipsoidal effect with a
semi-amplitude of about 0.1\%. To account for this, we set the period of the
quasi-periodic kernel used to account for the intrinsic variability of the
star in the detrending process to half the orbital period. We then modeled
this light curve including the ellipsoidal effect, i.e., we did not divide-out
this trend. To account for the ellipsoidal effect and the resulting gravity
darkening we used the Roche potential to calculate the shape of the ellipsoids
used to model the two stars assuming a mass ratio $q=1$. The gravity darkening
exponents were set to $y_1=0.267$ and $y_2=0.280$ for star 1 and 2,
respectively, these values being interpolated from the tabulation by
\citet{2011A&A...529A..75C}. 

\begin{figure*}
\includegraphics[width=0.99\textwidth]{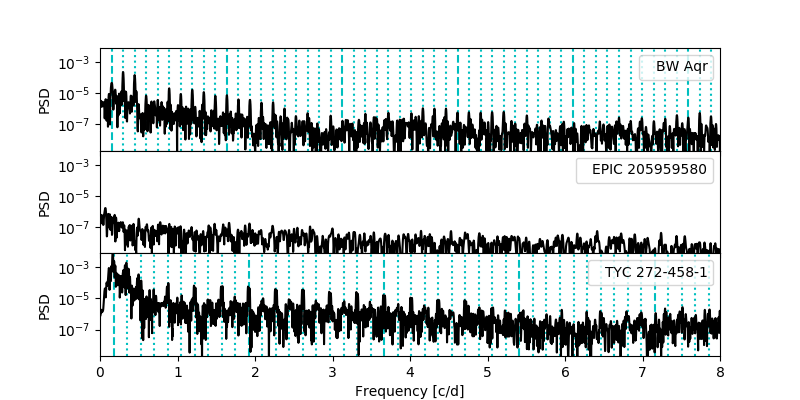} 
\caption{Periodograms of the K2 {\sc everest} light curves for EPIC~205982900
(BW~Aqr) excluding data in eclipse, another star of comparable
brightness observed in the same campaign with the same detector module,
output and channel (EPIC 205959580), and EPIC~201576812 (TYC 272-458-1)
excluding data in eclipse. Vertical lines mark the orbital frequency and
its harmonics for the two eclipsing binaries.}
\label{BWAqr_ft} 
\end{figure*}

 The residuals from the our best-fit model light curve show structure at a
level of about 500 ppm. The periodogram of these residuals shows peaks at
various  harmonics of the orbital period. We used light curves derived using
the {\sc everest} algorithm \citep{2016AJ....152..100L} to check that these
features in the light curve are not a by-product of our data analysis method.
We also checked the  {\sc everest} light curves for 3 other stars of similar
brightness to BW Aqr and observed during the same K2 campaign and using the
same channel of the Kepler instrument. The periodograms of the  {\sc everest}
light curves for BW~Aqr and one of these comparison stars are shown in
Fig.~\ref{BWAqr_ft}. We also analysed the  {\sc everest} light curve for
EPIC~201576812 (TYC 272-458-1). For both  BW~Aqr and TYC 272-458-1 we excluded
data obtained during eclipse from the calculation of the periodogram. The
rotational modulation of TYC 272-458-1 combined with this masking of the
eclipses results in closely separated peaks in the periodogram at harmonics of
the orbital frequency up to approximately the 25$^{\rm th}$ harmonic. In
contrast, the periodogram of  BW~Aqr shows a well defined sequence of single
peaks up to the 50$^{\rm th}$ harmonic of the orbital frequency, at least. A
likely interpretation of these frequencies is that they are tidally induced
pulsations, similar to those seen in the non-eclipsing binary system KOI-54
\citep{2012MNRAS.420.3126F, 2012MNRAS.421..983B}. The strongest frequencies
are at the 2$^{\rm nd}$ and 3$^{\rm rd}$ harmonic of the orbital frequency,
which agrees well with the expectation for a binary system with an
eccentricity e$\approx 0.18$ \citep[][Fig.~3]{2014ARA&A..52..171O}. The
detection of tidally induced pulsations in BW~Aqr open up the possibility of
testing methods to derive stellar parameters using asteroseismology for
non-eclipsing binary stars such as KOI-54, and to investigate the role of these
pulsations in the circularisation and synchronisation of  BW~Aqr's orbit. Such
studies will be aided by the availability of K2 data with a temporal sampling
of 58.8\,s (``SC'' data) for this binary system.

\begin{table}
\caption[]{\label{tab:bwaqr_abspar}
 Absolute astrophysical parameters BW~Aqr.}
\begin{center}
\begin{tabular}{@{}lrr} 
\hline 
\hline\noalign{\smallskip}
 &\multicolumn{1}{l}{Primary} & \multicolumn{1}{l}{Secondary} \\
\noalign{\smallskip} 
\hline 
\noalign{\smallskip} 
Mass   [$M_{\sun}$] & 1.38  $\pm$  0.02 &  1.48   $\pm$ 0.02   \\ 
Radius [$R_{\sun}$] & 1.732 $\pm$ 0.008 &  2.068  $\pm$ 0.009  \\ 
$\log g$ [cm\,s$^{-2}$] & 4.100 $\pm$ 0.003 &  3.977  $\pm$ 0.003 \\
T$_{\rm eff}$ [K]   &  6450 $\pm$ 100   &  6350   $\pm$ 100    \\
$\log(L/L_{\sun})$  &  0.67 $\pm$ 0.03  & 0.80    $\pm$ 0.03   \\ 
M$_{\rm V}$ [mag]   &  3.07 $\pm$ 0.07  & 2.76    $\pm$ 0.07   \\ 
\noalign{\smallskip} 
Orbital period [d]  & \multicolumn{2}{c}{6.719683  $\pm$ 0.000004} \\ 
Mass ratio & \multicolumn{2}{c}{0.931  $\pm$ 0.008 } \\ 
Eccentricity & \multicolumn{2}{c}{$0.1773  \pm 0.0008  $} \\ 
Distance      [pc]  & \multicolumn{2}{c}{411   $\pm$ 21 } \\ \noalign{\smallskip} \hline
\end{tabular} 
\end{center} 
\tablefoot{Absolute V magnitudes use bolometric corrections from
\citet{1998A+A...333..231B} and the distance is based on the apparent V-band
magnitude inferred from the observed Tycho-2 B$_{\rm T}$ and  V$_{\rm T}$
magnitudes. } 
\end{table}

\subsubsection{EPIC 206066862 = \object{BD $-$13$^{\circ}$ 6219}}
 We decided to fix the third light parameter at the value $\ell_3 = 0$ since
the eclipses in this light curve are partial and the secondary eclipse is
quite shallow. \citet{2016A&A...594A.100B} quote an orbital period
P=11.08716751d for this binary, which is in good agreement with our estimate.
These stars appear near the evolutionary tracks for masses of
$0.8\,M_{\sun}$ and $1.2\,M_{\sun}$ for any reasonable choice of the mass
ratio, so we assume $q\approx 0.7$.

\subsubsection{EPIC 206066909}
 The K2 light curve between the eclipses shows variability on timescales of a
few days with an amplitude of about 0.1\%. We were not able to identify a
clear period for this variability and the interpretation of this signal is
complicated by a substantial third-light contribution to the light curve in
this multiple star system.  The position of the stars in the T$_{\rm
eff}$\,--\,$\rho_{\star}$ plane lie near the stellar evolution tracks for
masses $0.7\,M_{\sun}$ and $1.3\,M_{\sun}$  for any reasonable choice of the
mass ratio so we assume $q\approx 0.5$.

\subsubsection{EPIC 206075677}
 There are two partial secondary eclipses visible in the detrended K2 light
curve with depths of 0.4\% and 1.5\% following division by the trends in the
light curve due to star spot modulation. These trends show an
amplitude of 1.5\% and at least three well defined peaks in the periodogram at
the periods listed in Table~\ref{lcinfo}. Given the complexity of this system
and the lack of unambiguous information in the light curve, we decided not to
attempt a model fit to this light curve.
 
\subsubsection{EPIC 206084435} 
 There is only one primary eclipse visible in the K2 light curve for this very
long-period binary ($P\approx 48$\,d) so the period is determined from the two
shallow secondary eclipses. The primary star is a main-sequence star with a
mass $M\approx 1.1M_{\odot}$ and the companion is a dwarf star with a  mass
$M\approx 0.65M_{\odot}$. The position of the stars in the T$_{\rm
eff}$\,--\,$\rho_{\star}$ plane lie near the stellar evolution tracks for
these masses for any reasonable choice of the mass ratio, $q$.

\subsubsection{EPIC 206109641}
 The K2 light curve of this star shows two eclipses with unequal depths
separated by 7.68 days. Only one eclipse of each depth in visible within the
69.16 day span of the K2 observations. From the K2 data alone it is only
possible to establish a lower limit to the orbital period of 41.77 days. Data
from the WASP archive includes one night of 200-mm data covering the minimum
of the secondary eclipse at HJD 2455141.285, another night of 200-mm data
showing the start of the ingress to primary eclipse at HJD 2454632.6 and one
night of 85-mm data covering the egress to the secondary eclipse at HJD
2456144.43. The only period consistent with these observations plus the lack
of visible eclipses on other nights for which we have  WASP data is $P\approx
62.6$\,d. 

\begin{figure*}
  \includegraphics[width=0.99\textwidth]{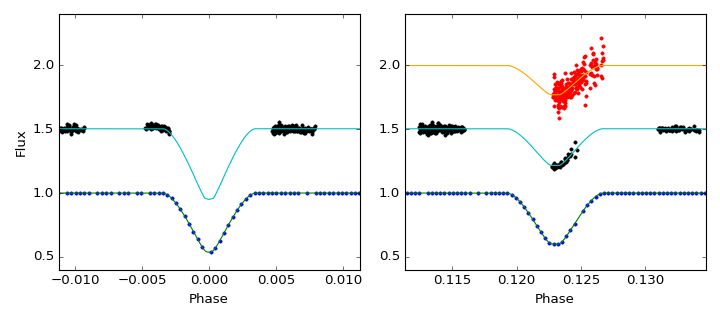} \caption{K2
  and WASP light curves of EPIC~206109641 around primary and secondary
  eclipse. The WASP 200-mm and 85-mm data are shown offset vertically by 0.5
  and 1 units, respectively. Solid lines show our best-fit light curve model.}
\label{206109641_lcfit} \end{figure*}

 We included the WASP photometry from the nights listed above plus data from a
few additional nights to set the out-of-eclipse level in our analysis of the
K2 light curve. The additional parameters required in the fit were the surface
brightness ratio for the 200-mm WASP data,  $S_{\rm 200}$, the surface
brightness ratio for the 85-mm WASP data, $S_{\rm 85}$, the zero-points of the
85-mm and 200-mm data, and the standard deviation of the residuals for the
85-mm and 200-mm data sets, $\sigma_{\rm 200}$ and $\sigma_{\rm 85}$,
respectively. Given the quality of the WASP data we decided to use the same
limb darkening coefficients and values for third light for these data as for
the K2 data rather than adding even more free parameters to the fit. The
best-fit model light curves to the K2 and WASP data are shown in
Fig.~\ref{206109641_lcfit}. The best-fit values of the additional parameters
were found to be $S_{\rm 200}=0.57 \pm0.01 $, $S_{\rm 85}=0.42\pm0.02$,
$\sigma_{\rm 200}=0.016$ and $\sigma_{\rm 85}=0.077$. Both stars lie near
the position of the Sun in the T$_{\rm eff}$\,--\,$\rho_{\star}$ diagram for
any reasonable choice of mass ratio but the relative position of the stars is
inconsistent with star 1 being larger and therefore more evolved unless
$q\approx 0.9$.
  
\subsubsection{EPIC 206212261} 

 The occultation of a dwarf star with a mass $M\approx 0.6M_{\odot}$ by a
1.0-$M_{\odot}$ star near the MSTO results in well-defined total eclipses in
the K2 light curve of this system, so the parameters are determined to good
precision despite the extreme flux ratio in this binary $\ell_{\rm Kp}\approx
1.7$\%. This extreme flux ratio may make it challenging to obtain
spectroscopic observations to determine the masses of these stars. Star 1 has
T$_{\rm eff}\approx 5400$\,K while star 2 is cooler and much smaller than star
1 so we assume that this system consists of a solar type star and a K-dwarf
companion. In this case $q$ must be significantly less than 1, so the position
of the stars in Fig.~\ref{TeffDensity} does not depend strongly on the assumed
value of $q$. 

\subsubsection{EPIC 206241558}
 The K2 light curve of this star shows two eclipses with unequal depths
separated by 12.32 days. Only one eclipse of each depth in visible within the
69.16 day span of the K2 observations. From the K2 data alone it is only
possible to establish a lower limit to the orbital period of 53.1 days. Data
from the WASP archive includes one night of 200-mm data covering the minimum
of the primary eclipse at HJD 2455012.63 and another night of 200-mm data
covering the egress to the secondary eclipse at HJD 2454685.4. The only period
consistent with these observations plus the lack of visible eclipses on other
nights for which we have  WASP data is $P\approx 56.6$\,d. 

 We included the WASP photometry from the nights listed above plus data from
one additional night to set the out-of-eclipse level in our analysis of the K2
light curve. Given the quality of the WASP data we decided to use the same
limb darkening coefficients and values for third light for these data as for
the K2 data. 

\begin{figure*} 
  \includegraphics[width=0.99\textwidth]{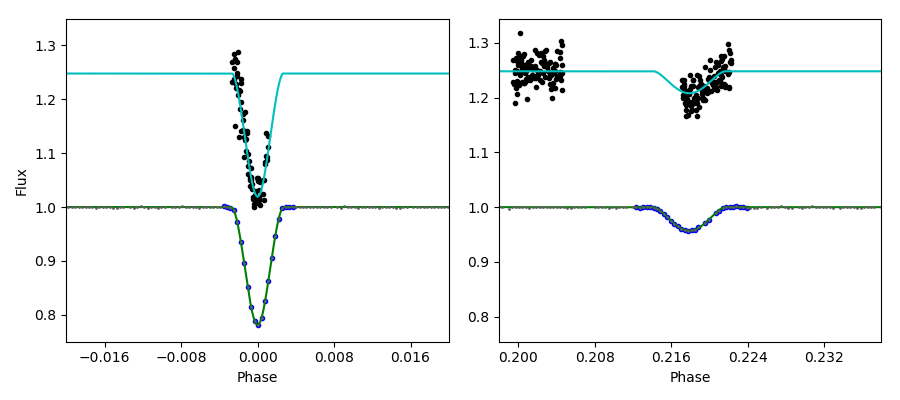}
\caption{K2 and WASP light curves of EPIC~206241558 around primary and
secondary eclipse. Small gray points in the K2 light curve were not included
in the fit. The WASP 200-mm data are shown offset vertically by 0.5
units. Solid lines show our best-fit light curve model.}
\label{206241558_lcfit} 
\end{figure*}

 We made several attempts to find a combination of starting parameters for the
{\sc emcee} analysis that lead to best-fit solutions with both $k < 1$ and
$S_{\rm Kp} <1$ or vice versa, but always found that the Markov chains
converged on solutions with $k < 1$ and $S_{\rm Kp} > 1$ or vice versa,
depending on the initial parameters and whether or not we fixed $\ell_3 = 0$.
We did not find any combination of parameters for which the two stars appear
near the same isochrone in the effective temperature -- mean stellar density
plane so our estimate for the mass ratio $q\approx 1.0$ is very uncertain
in this case. One possibility is that this star is the result of a merger
between two stars in a triple system induced by Kozai-Lidov cycles in a triple
star system \citep{2018arXiv180205854S}. The best-fit model light curves to
the K2 and WASP data are shown in Fig.~\ref{206241558_lcfit}. The surface
brightness ratio in the 200-mm WASP data we found to be $S_{\rm 200}=1.5
\pm0.1  $ and the standard deviation for the residuals is $\sigma_{\rm
200}=0.023$.

\subsubsection{EPIC 206253908}
 The K2 light curve of this star shows only one eclipse, from which it is
possible to establish a lower limit to the orbital period of 62.7 days.
Eclipses are also seen in the WASP archive photometry at HJD 2454758.41 and
2455085.61. We estimate the period of this binary to be 65.45\,d. As there is
no secondary eclipse visible in the K2 or WASP data we  did not attempt any
further analysis of this star. 

\subsubsection{EPIC 206288770} 
The occultation of a dwarf star with a mass $M\approx 0.6\,M_{\odot}$ by a
1.2-$M_{\odot}$ main-sequence star results in well-defined total eclipses in
the K2 light curve of this system, so the parameters are determined to good
precision despite the extreme flux ratio in this binary $\ell_{\rm Kp}\approx
1.4$\%. These mass estimates are quite robust because the position of the
stars in the T$_{\rm eff}$\,--\,$\rho_{\star}$ plane lie near the stellar
evolution tracks for these masses for any reasonable choice of $q$. The
extreme flux ratio in this binary may make it challenging to obtain
spectroscopic observations to determine the masses of these stars.

\subsubsection{EPIC 206433263} 
  This eclipsing binary is a favourable target for detailed study of a
star similar to the Sun. The eclipses of the star with a mass $M\approx
1.0\,M_{\odot}$ are total and the companion with a mass $M\approx
1.2\,M_{\odot}$ is near the MSTO. These mass estimates are quite robust
because the position of the stars in Fig.~\ref{TeffDensity} lie near the
stellar evolution tracks for these masses for any reasonable choice of $q$.
The optical flux ratio is favourable for spectroscopic follow-up to determine
accurate masses for both stars and the star is moderately bright. The mass of
the star near the MSTO will put a tight constraint on the age of the system
and there do not appear to be any complications due to star spots or third
light contamination. 

\section{Summary and conclusions}

 We have used light curves from Kepler K2 to identify 42 long-period
eclipsing binary systems ($P\ga 5.5$\,d) with narrow eclipses and little or no
ellipsoidal effect, i.e., well-detached binaries with very weak tidal
interaction between the stars. This includes systems with periods $P\ga 60$\,d
for which the orbital period cannot be established from the K2 data alone
(EPIC~206109641, EPIC~206241558 and EPIC~206253908). In these cases we have
used data from the WASP project data archive to establish the orbital period
of the binary system.

 For 40 targets we have determined the geometry of the binary system
(fractional radii, inclination, eccentricity, etc.) from the analysis of the
Kepler K2 light curve using the \texttt{ellc} eclipsing binary star model. For
38 of these systems we also estimate the effective temperature of the stars
from an analysis of the observed apparent magnitudes and other data for the
system. For these 38 systems we are able to estimate the mass and evolutionary
state of the stars by comparing their mean stellar densities and effective
temperatures to stellar models. They typically contain main-sequence or
sub-giant stars with masses from 0.6\,$M_{\odot}$ to 1.4\,$M_{\odot}$, with
sub-giant and giant companions being more common among the longer-period
systems.

We have used new radial velocity measurements to make preliminary estimates of
the mass, radius and luminosity of the stars in 3 systems (EPIC~202674012,
EPIC~205020466, EPIC~205919993). We have also re-calculated these absolute
parameters for two systems with spectroscopic orbits that have been previously
studied using light curves from ground-based instruments (EPIC~201576812 =
FM~Leo and EPIC~205982900 = BW~Aqr). We have also estimated the masses and
radii of the eclipsing stars in the early-type triple system EPIC~204760247 =
HD~142883.

 We confirm the presence of variability between the eclipses in HD~142883 due
to $\beta$ Cephei-type pulsations. Variability in the light curve between the
eclipses due to $\gamma$ Doradus-type  pulsations is seen in EPIC~203371239
and perhaps also EPIC~203728604. In BW~Aqr we find variability due to
pulsations which we suspect are induced by dissipation of tidal forces in this
eccentric binary. Variability due to magnetic activity is seen in several
systems and has been used to measure the rotation periods of one or both stars
in 13 cases.

Kepler K2 provides almost continuous observations for each campaign field for
up to 80 days. This makes it possible to identify and characterise long-period
eclipsing binary stars that are very hard to find and study using ground-based
observations. The high quality of the photometry also makes it possible to
identify features in the light curve such as low-amplitude rotational
modulation due to star spots and pulsations that are also very hard to observe
from the ground. The high quality of the photometry also enables the  geometry
of these eclipsing binaries to be measured to very high precision, particularly
for those that show total eclipses. Additional work is needed to establish the
accuracy of the parameters for eclipsing binary stars that can be derived using
Kepler K2 photometry.

 The majority of the eclipsing binary systems we have identified in
the Kepler K2 data with long orbital periods ($P \goa 10$\,d) contain dwarf or
sub-giant stars. The high quality of the K2 photometry makes it possible to
identify binaries where one star has a much lower mass than the other. These
are useful systems for testing stellar evolution models because
free-parameters in the models such as the mixing length parameter will affect
the two stars in the binary in different ways. It can be challenging to obtain
accurate mass estimates for the stars in such binaries because they often have
extreme luminosity ratios as optical/infra-red wavelengths, but there are now
several high-resolution spectrographs available on large telescopes that
should make detailed characterisation for some of these systems feasible.

\begin{acknowledgements}

We thank the anonymous referee for their careful reading of the manuscript
and useful comments that have improved the quality of the paper.

This paper includes data collected by the K2 mission. Funding for the K2
mission is provided by the NASA Science Mission directorate.

This research has made use of the SIMBAD database, operated at CDS,
Strasbourg, France. 

Funding for SDSS-III has been provided by the Alfred P. Sloan Foundation, the
Participating Institutions, the National Science Foundation, and the U.S.
Department of Energy Office of Science. The SDSS-III web site is
http://www.sdss3.org/.

SDSS-III is managed by the Astrophysical Research Consortium for the
Participating Institutions of the SDSS-III Collaboration including the
University of Arizona, the Brazilian Participation Group, Brookhaven National
Laboratory, Carnegie Mellon University, University of Florida, the French
Participation Group, the German Participation Group, Harvard University, the
Instituto de Astrofisica de Canarias, the Michigan State/Notre Dame/JINA
Participation Group, Johns Hopkins University, Lawrence Berkeley National
Laboratory, Max Planck Institute for Astrophysics, Max Planck Institute for
Extraterrestrial Physics, New Mexico State University, New York University,
Ohio State University, Pennsylvania State University, University of
Portsmouth, Princeton University, the Spanish Participation Group, University
of Tokyo, University of Utah, Vanderbilt University, University of Virginia,
University of Washington, and Yale University. 
 
Some  of the observations reported in this paper were obtained with the
Southern African Large Telescope (SALT) under program 2016-2-SCI-001 (PI:
Maxted).

Based on observations made with ESO Telescopes at the La Silla Paranal
Observatory under programme IDs 70.D-0433(A), 091.C-0713(A),  089.D-0097(B)
and 091.D-0145(B).

 We thank Dr Adrian Barker for sharing his expertise in tidally induced
stellar pulsations with us.

\end{acknowledgements}

\bibliographystyle{aa} 
\bibliography{mybib}

\listofobjects

\end{document}